\documentclass[usenatbib]{mnras}

\usepackage{newtxtext,newtxmath}
\usepackage[T1]{fontenc}

\DeclareRobustCommand{\VAN}[3]{#2}
\let\VANthebibliography\thebibliography
\def\thebibliography{\DeclareRobustCommand{\VAN}[3]{##3}\VANthebibliography}

\usepackage{graphicx}	
\usepackage{amsmath}	
\usepackage{color}
\usepackage{makecell}

\newcommand{\MSUN}{{\rm M}_{\sun}}

\newcommand{\RHOST}{R_{\rm 200c}}

\title[Satellites of MW/M31-like galaxies in TNG50]{The abundance of satellites around Milky Way- and M31-like galaxies with the TNG50 simulation: a matter of diversity}

\author[C. Engler et al.]{Christoph Engler$^{1,2}$\thanks{E-mail: c.engler@stud.uni-heidelberg.de},
Annalisa Pillepich$^{2}$, Anna Pasquali$^{1}$, Dylan Nelson$^{3}$,
\newauthor
Vicente Rodriguez-Gomez$^{4}$, Kun Ting Eddie Chua$^{5}$, Eva K. Grebel$^{1}$, Volker Springel$^{6}$,
\newauthor
Federico Marinacci$^{7}$, Rainer Weinberger$^{8}$, Mark Vogelsberger$^{9}$, Lars Hernquist$^{8}$
\\ \\
$^{1}$Astronomisches Rechen-Institut, Zentrum f\"{u}r Astronomie der Universit\"{a}t Heidelberg, M\"{o}nchhofstra\ss e 12-14, 69120 Heidelberg, Germany\\
$^{2}$Max-Planck-Institut f\"{u}r Astronomie, K\"{o}nigstuhl 17, 69117 Heidelberg, Germany\\
$^{3}$Institut für Theoretische Astrophysik, Zentrum f\"{u}r Astronomie, Universit\"{a}t Heidelberg, Albert-Ueberle-Str. 2, 69120 Heidelberg, Germany\\
$^{4}$Instituto de Radioastronom\'ia y Astrof\'isica, Universidad Nacional Aut\'onoma de M\'exico, Apdo. Postal 72-3, 58089 Morelia, Mexico\\
$^{5}$Institute of High Performance Computing, 1 Fusionopolis Way \#16-01, Singapore 138632\\
$^{6}$Max-Planck-Institut für Astrophysik, Karl-Schwarzschild-Straße 1, 85740 Garching bei München, Germany\\   
$^{7}$Department of Physics \& Astronomy ``Augusto Righi'', University of Bologna, via Gobetti 93/2, 40129 Bologna, Italy\\
$^{8}$Center for Astrophysics | Harvard \& Smithsonian, 60 Garden Street, Cambridge, MA 02138, USA\\
$^{9}$Department of Physics, Kavli Institute for Astrophysics and Space Research, Massachusetts Institute of Technology, Cambridge, MA 02139, USA\\}

\date{Accepted 2021 August 23. Received 2021 August 23; in original form 2021 January 28}

\pubyear{2021}

\begin{document}
\label{firstpage}
\pagerange{\pageref{firstpage}--\pageref{lastpage}}
\maketitle

\begin{abstract}
We study the abundance of satellite galaxies around 198 Milky Way- (MW) and M31-like hosts in TNG50, the final instalment in the IllustrisTNG suite of cosmological magnetohydrodynamical simulations. MW/M31-like analogues are defined as disky galaxies with stellar masses of $M_* = 10^{10.5 - 11.2}~\rmn{M}_\odot$ in relative isolation at $z=0$. By defining satellites as galaxies with $M_* \geq 5\times10^{6}~\rmn{M}_\odot$ within $300~\rmn{kpc}$ (3D) of their host, we find a remarkable level of diversity and host-to-host scatter across individual host galaxies. The median TNG50 MW/M31-like galaxy hosts a total of $5^{+6}_{-3}$ satellites with $M_* \geq 8 \times 10^6~\rmn{M}_\odot$, reaching up to $M_* \sim 10^{8.5^{+0.9}_{-1.1}}~\rmn{M}_\odot$. Even at a fixed host halo mass of $10^{12}~\rmn{M}_\odot$, the total number of satellites ranges between $0 - 11$.
The abundance of subhaloes with $M_\rmn{dyn} \geq 5 \times 10^7~\rmn{M}_\odot$ is larger by a factor of more than 10. The number of all satellites (subhaloes) ever accreted is larger by a factor of $4-5$ ($3-5$) than those surviving to $z=0$. Hosts with larger galaxy stellar mass, brighter $K$-band luminosity, more recent halo assembly, and -- most significantly -- larger total halo mass typically have a larger number of surviving satellites. 
The satellite abundances around TNG50 MW/M31-like galaxies are consistent with those of mass-matched hosts from observational surveys (e.g. SAGA) and previous simulations (e.g. Latte). While the observed MW satellite system falls within the TNG50 scatter across all stellar masses considered, M31 is slightly more satellite-rich than our $1\sigma$ scatter but well consistent with the high-mass end of the TNG50 sample. We find a handful of systems with both a Large and a Small Magellanic Cloud-like satellite. 
There is no missing satellites problem according to TNG50.
\end{abstract}

\begin{keywords}
galaxies: dwarf -- galaxies: haloes -- galaxies: luminosity function, mass function -- Local Group -- Galaxy: formation -- Galaxy: evolution
\end{keywords}



\section{Introduction}
\label{sec:intro}

The abundance of dwarf satellite galaxies orbiting the Milky Way (MW) and Andromeda (M31) has helped to inform our understanding of the Universe and the formation of galaxies therein. Yet, these satellite systems -- the closest we can study down to as low as a few thousand solar masses in stars -- continue to challenge the $\Lambda$CDM model of structure formation. 

For about twenty years, the ``missing satellites'' problem \citep{Moore1999, Klypin1999, Bullock2017} has attracted notable interest across the astronomical community by pointing towards a seemingly insurmountable tension between observations and theoretical models. According to its original incarnation, dark matter- or gravity-only simulations of the cosmological assembly of MW-like haloes in a $\Lambda$CDM scenario predict far more satellites \citep[i.e. subhaloes;][]{Springel2008, Diemand2008} than there are actual luminous satellites observed around the Galaxy -- particularly at the low-mass end. 

From an observational perspective, the number of detected satellite galaxies around the MW has in fact continued to grow into the ultra-faint regime in recent years \citep[stellar masses of $\lesssim 10^5\MSUN$, e.g.][]{Zucker2006, Belokurov2006, Sakamoto2006, Willman2010, Drlica-Wagner2015, Drlica-Wagner2020, Koposov2015, Torrealba2016, Torrealba2018}. On the other hand, while the number of bright, classical satellites had been constant for two decades after the discovery of the Sagittarius galaxy \citep{Ibata1994}, additional bright satellites such as Crater~2 and Antila~2 \citep{Torrealba2016, Torrealba2019}, as well as tidal remnants of former bright satellites, such as Gaia-Enceladus \citep{Helmi2018, Myeong2018} and Sequoia \citep{Myeong2019}, have been discovered in recent years. Characterising the satellite populations of similar galaxies within and beyond the Local Volume is essential in order to understand how representative the MW and Andromeda are in a cosmological context. Therefore, the study of the abundance of satellite dwarf galaxies has been extended over the last decade, from the MW \citep{McConnachie2012} and M31 \citep{Martin2016, McConnachie2018} to other nearby galaxies, such as Centaurus A \citep[e.g.][]{Crnojevic2014, Crnojevic2019, Mueller2017, Mueller2019}, M94 \citep{Smercina2018}, and M101 \citep[e.g.][]{Bennet2017, Bennet2019, Carlsten2019}.

However, sample completeness is of the essence when searching for satellite systems to compare to the satellite abundances in the MW and the Local Group. Contamination by foreground and background objects can cause major issues. \cite{Carlsten2020a, Carlsten2021} summarise such comparisons using CFHT/MegaCam data. By utilising surface brightness fluctuations, they obtain reliable distance measurements to confirm satellite candidates around twelve hosts in the Local Volume (i.e. within $12~\rmn{Mpc}$).
The ``Satellites Around Galactic Analogs'' (SAGA) survey extends the search for a ``normal'' satellite system by aiming for a statistical sample of MW-like hosts beyond the Local Volume at distances of $20 - 40~\rmn{Mpc}$. The first stage of SAGA identified 27 satellites around 8 MW-like systems, revealing a significant degree of host-to-host scatter between satellite systems \citep{Geha2017}. This sample was recently expanded to 127 satellites around 36 MW-like hosts in the survey's second stage \citep{Mao2021}. While the satellite abundance of the MW is consistent with those from the SAGA survey, the systems exhibit a remarkable degree of diversity, ranging from MW-like hosts with no satellites whatsoever to systems with up to 9 satellites, down to an absolute $r$-band magnitude of $M_\rmn{r} < -12.3$. Ultimately, SAGA aims to spectroscopically determine the satellite systems of 100 MW-like host galaxies down to satellite stellar masses of about $10^6~\MSUN$.

From a theoretical perspective, the early simulations on which the ``missing satellites'' problem was formulated contained only dark matter subhaloes, as they did not include baryons and therefore did not simulate luminous satellite galaxies. However, the abundance of dark matter subhaloes and luminous galaxies is different. Baryonic effects such as supernova feedback \citep{Larson1974, Dekel1986, Mori2002} and reionisation \citep{Couchman1986, Efstathiou1992, Thoul1996} are thought to reduce the efficiency of star formation and should hence be able to keep visible galaxies from forming in low-mass dark matter haloes. This has been shown repeatedly with both semi-analytic models \citep{Bullock2000, Benson2002a, Benson2002b, Somerville2002, Font2011, Guo2011} and full hydrodynamical simulations \citep{Okamoto2005, Governato2007, Maccio2007, Sawala2016b} over the last two decades. Furthermore, satellites can be destroyed by tidal shocks as they pass through their host galaxy's disk \citep{DOnghia2010, Yurin2015}.

Since galaxy formation is expected to be significantly suppressed at low halo masses, a majority of these (sub)haloes should not host a luminous component -- from theoretical perspectives in general, as well as specifically within a $\Lambda$CDM  ansatz. Therefore, a large body of work in the past years has focused on a line of solutions to the missing satellites problem and considered the idea that luminous dwarf galaxies inhabit only a small fraction of these predicted subhaloes \citep[e.g.][]{Nickerson2011, Shen2014, Sawala2015, Sawala2016a, Benitez-Llambay2020}. On the other hand, even for (sub)haloes that are able to host some star formation, the mapping between dark matter (DM) (sub)halo mass and galaxy stellar mass remains theoretically uncertain -- particularly for systems below $10^{10-11}~\MSUN$ in total mass \citep[e.g.][and references therein]{Sawala2015}. Large systematic variations exist across galaxy formation models, be it numerical simulations, semi-empirical, or semi-analytical models. Finally, a number of cosmological hydrodynamical simulations have shown that galaxy physics affect the survival of subhaloes (luminous or dark) by generally suppressing the total cumulative abundance of low-mass subhaloes at $z=0$, regardless of whether they host a luminous galaxy or not \citep[see e.g.][for a recent discussion based on results from the Illustris simulation]{Chua2017}. However, while different galaxy formation models agree qualitatively on the suppression of subhalo formation and survival, on the shape of the stellar-to-halo mass relation at lower masses, and on the stochasticity of star formation, large quantitative, systematic uncertainties remain across model predictions.

Searching for the ``normal'' satellite system of a MW-like host has also been an open question for simulations. Models for MW-like haloes and their subhalo populations began in DM-only simulations (e.g. Aquarius, \citealp{Springel2008}; Via Lactea II, \citealp{Diemand2008}; Phat ELVIS, \citealp{Kelley2019}) and -- more recently -- have achieved the required numerical resolution to study MW-like galaxies and their abundance of luminous satellite galaxies in cosmological hydrodynamical simulations. So far, these simulations have mostly been performed as zoom-in simulations, focused on either a single or a small sample of MW- or LG-like hosts, with projects such as Latte \citep{Wetzel2016}, FIRE \citep{Hopkins2014, Hopkins2018, Garrison-Kimmel2019}, the DC Justice League simulations \citep{Applebaum2021}, or APOSTLE \citep{Fattahi2016a, Fattahi2016b, Sawala2016b}. Other projects have managed to take steps towards larger samples: the Auriga simulations are comprised of a suite of 30 isolated MW-like galaxies and their satellite systems \citep{Grand2017, Simpson2018}, while \cite{Font2021} have more recently presented satellite abundances of the ARTEMIS simulations, a suite of 45 zoom-in MW-like haloes resimulated with the {\sc eagle} model. 

Whereas all of the aforementioned simulations return satellite systems that are broadly consistent with the observed MW and M31 satellite mass functions, the scatter across hosts is substantial. Both cosmological DM-only simulations \citep[e.g.][]{Gao2004, Boylan-Kolchin2010} and -- more recently -- baryonic galaxy formation hydrodynamical simulations \citep[e.g.][]{Chua2017} have been used extensively to quantify and characterise the scatter of the subhalo mass function at fixed host mass and to show the dependencies of subhalo abundance on host halo properties. However, both the scatter of the {\it satellite} luminosity function around MW-like galaxies, as well as the dependence of {\it satellite} abundance on host properties are yet to be quantified. Therefore, the level of consistency across simulations, as well as between simulations and observations (accounting for such galaxy-to-galaxy variations and for host property dependencies) remains unclear. As a further complication, the aforementioned works all adopt somewhat different definitions or selections for MW-like galaxies, either based on halo mass, host stellar morphology, merger history, isolation criteria, or a combination thereof. Finally, whereas correlations of total satellite abundance with host mass have previously been found in semi-analytic models \citep[e.g.][]{WangWhite2012, Sales2013} and with smaller samples from hydrodynamical simulations \citep{Fattahi2016a, Font2021}, this remains to be confirmed and further quantified with a statistical sample of hosts in a full cosmological context.

In this study, we examine the abundance of satellites around MW- and M31-like hosts using the TNG50 simulation \citep{Nelson2019b, Pillepich2019}. As the final instalment of the IllustrisTNG suite of cosmological magnetohydrodynamical simulations \citep{Marinacci2018, Naiman2018, Nelson2018, Pillepich2018b, Springel2018}, TNG50 bridges the gap between large-scale volumes and the regime of zoom-in simulations. At a volume of $(50~\rmn{comoving~Mpc})^3$ and a baryonic mass resolution of $8 \times 10^4~\rmn{M}_\odot$, TNG50 includes a statistically significant sample of both MW/M31-like galaxies and their satellites. This enables us not only to reliably identify satellite galaxies down to stellar masses of $\sim 5 \times 10^6~\rmn{M}_\odot$ (approximately the stellar mass of Leo I), but also to study the evolution of satellite abundances throughout cosmic time, and to search for statistically significant correlations of satellite abundance with various host galaxy and halo properties. We will analyse specific satellite properties such as their star formation activity and gas fractions, as well as their dependence on host properties and infall times in a future paper (\textcolor{blue}{Engler et al. in prep.}).

This paper is structured as follows: in Section~\ref{sec:methods}, we introduce TNG50 and IllustrisTNG, as well as define our selection of MW/M31-like hosts, satellite galaxies, and subhaloes. We present our results in Section~\ref{sec:results}: the satellite stellar mass function of TNG50 MW/M31-like hosts, comparisons with observational surveys and previous simulations, its evolution with redshift, differences between luminous satellite and dark subhalo populations, as well as baryonic vs. DM-only simulation results. Furthermore, we discuss dependencies of satellite abundances on the adopted selection of host galaxies, as well as various host properties. Finally, we summarise our results in Section~\ref{sec:conc}.

\section{Methods}
\label{sec:methods}

\subsection{The TNG50 simulation}
\label{sec:tng50}

In this study, we analyse satellite abundances in TNG50 \citep{Pillepich2019, Nelson2019b}, the highest resolution flagship run of the IllustrisTNG\footnote{\url{http://www.tng-project.org}} suite of cosmological magnetohydrodynamical simulations of galaxy formation \citep{Marinacci2018, Naiman2018, Nelson2018, Pillepich2018b, Springel2018}. Its galaxy formation model includes physical processes such as primordial and metal-line gas cooling, gas heating by a spatially-uniform, time-dependent UV background, as well as subgrid models for the unresolved structure of the interstellar medium and star formation processes \citep{Springel2003}. Stellar populations are tracked in their evolution and chemical enrichment, which includes ten elements (H, He, C, N, O, Ne, Mg, Si, Fe, Eu) and yields from supernovae Ia, II, and AGB stars \citep{Vogelsberger2013, Torrey2014}. These supernovae can lead to galactic winds as a form of feedback: these winds are injected isotropically and their initial speed scales with the one-dimensional dark matter velocity dispersion \citep{pillepich2018a}. Furthermore, there are two modes of black hole feedback, depending on their accretion: at low accretion rates, black hole feedback occurs purely kinetically, while high accretion rates result in thermal feedback \citep{Weinberger2017}. The implementation of magnetic fields allows for self-consistent amplification from a primordial seed field, following ideal magnetohydrodynamics \citep{Pakmor2013}. IllustrisTNG uses the moving mesh code \textsc{Arepo} \citep{Springel2010}, which solves the equations of hydrodynamics using an adaptive, moving Voronoi tessellation of space. Furthermore, the simulation adopts the $\Lambda$CDM framework with cosmological parameters from Planck data: matter density $\Omega_\rmn{m} = 0.3089$, baryonic density $\Omega_\rmn{b} = 0.0486$, cosmological constant $\Omega_\Lambda = 0.6911$, Hubble constant $h = 0.6774$, normalisation $\sigma_8 = 0.8159$, and spectral index $n_\rmn{s} = 0.9667$ \citep{Planck2016}.

IllustrisTNG encompasses several simulations of varying box sizes and resolution, including TNG300 with a volume of $(300~\rmn{Mpc})^3$ and a baryonic mass resolution of $m_\rmn{b} = 1.1 \times 10^7~\rmn{M}_\odot$, TNG100, a $(100~\rmn{Mpc})^3$ volume with $m_\rmn{b} = 1.4 \times 10^6~\rmn{M}_\odot$, and TNG50, a $(50~\rmn{Mpc})^3$ box with $m_\rmn{b} = 8.5 \times 10^4~\rmn{M}_\odot$. Thus, TNG50 combines a cosmological volume and a statistically significant sample of galaxies with a zoom-in-like level of mass resolution. This work is based almost exclusively on TNG50 and its lower-resolution counterparts (see Appendix~\ref{sec:resolution} for a discussion on the effects of numerical resolution) but we compare to results from TNG100 when possible.

\subsection{Selecting MW/M31-like galaxies in TNG50}
\label{sec:sample_mw}

The very choice of galaxies that can be considered as analogues of our Milky Way and Andromeda is essential in order to compare a simulated galaxy population to these observed systems (see \textcolor{blue}{Pillepich et al. in prep.} for an extended discussion). The abundance of satellite galaxies -- should it follow the abundance of subhaloes -- is expected to depend on at least some properties of their hosts, particularly on host total mass \citep[][]{Gao2004}. Over the next Sections, we quantify this in greater detail. Therefore, it is vital to adopt a clear definition of MW/M31-like hosts. Throughout the paper, we refer to a fiducial sample of MW/M31-like TNG50 analogues as detailed below, as well as to two host samples based on the selection criteria of two observational surveys.

\subsubsection{TNG50 MW/M31-like fiducial sample}
\label{sec:sample_fiducial}
Throughout this paper, we define MW/M31-like galaxies in TNG50 based on their mass and morphology according to \textcolor{blue}{Pillepich et al. (in prep.)}. MW/M31-like galaxies are required to have a stellar mass of $M_* = 10^{10.5} - 10^{11.2}~\rmn{M}_\odot$ within an aperture of $30~\rmn{kpc}$ and to be disky in their stellar shape -- either by having a minor-to-major axis ratio of their 3D stellar mass distribution of $s < 0.45$ (measured between one and two times the stellar half-mass radius) or by visual inspection of synthetic 3-band stellar-light images in face-on and edge-on projection. These visual inspections add 25 galaxies with $s > 0.45$ that display clear disk features or spiral arms to our sample of host galaxies. On the contrary, there are 18 hosts with $s < 0.45$ that display a disturbed morphology, weak or barely visible spiral arms, or a strong bar feature, which are still considered as MW/M31-like candidates. While other works in the literature employ kinematic decomposition as a morphology estimate, the minor-to-major axis ratio was chosen as an observationally motivated and a more readily available indicator. Furthermore, a minimum isolation criterion is imposed at $z=0$. No other massive galaxies with $M_* > 10^{10.5}~\rmn{M}_\odot$ are allowed within a distance of $500~\rmn{kpc}$ of the MW/M31-like candidates and the mass of the candidates' host halo is limited to $M_\rmn{200c} < 10^{13}~\rmn{M}_\odot$. These host haloes are defined using a friends-of-friends (FoF) algorithm and their virial mass $M_\rmn{200c}$ corresponds to the total mass of a sphere around the FoF halo centre with a mean density of 200 times the critical density of the Universe. 

Galaxies that fulfill all of the aforementioned conditions -- a stellar mass of $M_* = 10^{10.5} - 10^{11.2}~\rmn{M}_\odot$, a disky stellar morphology, and a relatively isolated environment -- are considered to resemble the Milky Way and Andromeda to a reasonable degree, at least within the context of the general galaxy population. Note that this selection does \textit{not} require TNG50 MW/M31-like galaxies to be the centrals of their host halo, i.e. they can be satellites of another galaxy. This allows our sample to include also galaxy pairs of Local Group-like systems, as it is not clear which of them -- the MW-like or the M31-like galaxy -- would be considered as the central galaxy of the system. In total, these criteria leave us with a sample of 198 MW/M31-like galaxies in TNG50, eight of which are satellite galaxies. We exclude these satellite hosts in later parts of the paper where we specifically study the infall of satellite galaxy populations or host halo properties.

For the purposes of investigating differences between TNG50 and its dark matter-only (DM-only) analogue TNG50-Dark, we cross-match our sample of MW/M31-like hosts from the baryonic run to their DM-only counterparts \citep{RodGom2015, RodGom2017, Nelson2015} by maximising the number of common DM particles. Therefore, the DM-only host sample consists of 198 MW/M31-like haloes as well.

\subsubsection{SAGA-like host selection}
\label{sec:sample_saga}
This host selection is based on the $K$-band luminosity and the local environment of host galaxies. We adopt the selection criteria from \cite{Geha2017} and \cite{Mao2021}. Potential candidates are required to have a $K$-band luminosity in the range of $-23 > M_\rmn{K} > -24.6$ and to have no bright galaxy within their virial radius (for which we adopt $300~\rmn{kpc}$ as the typical virial radius of a MW-like galaxy). Bright galaxies are defined by a magnitude of at least $K < K_\rmn{host} - 1.6$. Furthermore, galaxies with a host halo mass of $10^{13}~\rmn{M}_\odot$ and above are excluded. Out of all TNG50 hosts that meet these criteria, we match the three most similar ones to each of the 36 observed SAGA hosts based on their $K$-band luminosity. This gives us a sample of 108 TNG50 SAGA-like hosts, which will be used in Figure~\ref{fig:satLF_obs}. Note that this selection does \textit{not} require the TNG50 SAGA-like galaxies to be the centrals of their host halo.

\subsubsection{Local Volume (LV)-like host selection}
\label{sec:sample_lv}
This selection of hosts is based on observations of nearby host galaxies in the Local Volume \citep{Carlsten2021}. As the observational selection is mainly based on luminosity and spatial proximity, we adopt a similar range in $K$-band luminosity of $-22.7 > M_\rmn{K} > 24.5$. As there are no specific observational selection criteria regarding their environment and morphology, we choose our sample candidates to be the central galaxy of their respective dark matter halo and to have a disky stellar shape -- either by having a minor-to-major axis ratio of their 3D stellar mass distribution of $s < 0.45$ (measured between one and two times the stellar half-mass radius) or by visual inspection of synthetic 3-band stellar-light images in face-on and edge-on projection. Out of all TNG50 hosts that meet these criteria, for each of the 6 observed LV hosts (not including the Galaxy and Andromeda), we choose the three TNG50 galaxies with the closest $K$-band luminosity. This gives us a sample of 18 TNG50 LV-like hosts, which will be used in Figure~\ref{fig:satLF_obs}.

\subsubsection{Basic properties of TNG50 MW/M31-like galaxies}

\begin{figure*}
    \centering
    \includegraphics[width=.92\textwidth]{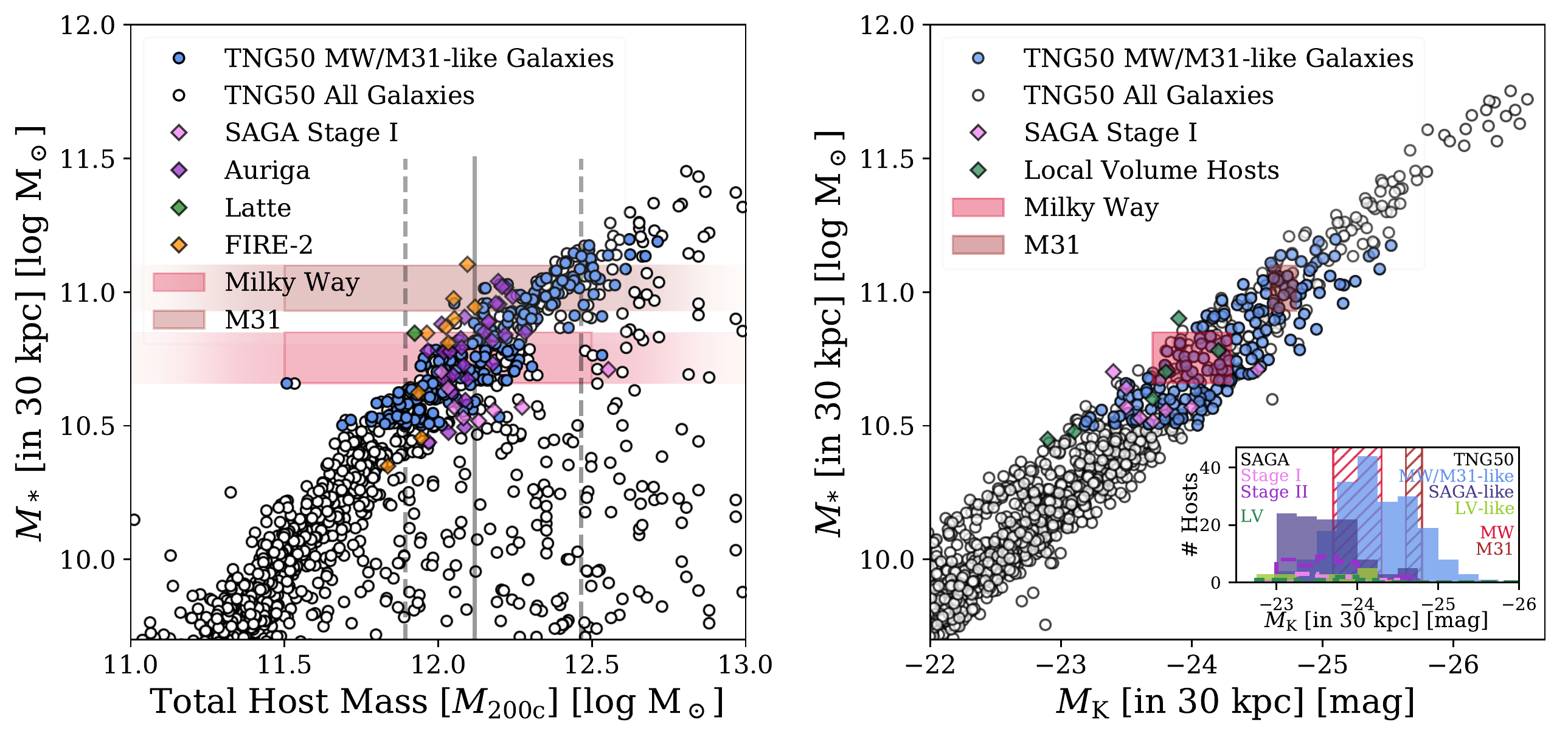}
    \caption{{\bf Properties and basic scaling relations for MW/M31-like hosts selected from the TNG50 simulation.} Blue circles denote the MW/M31 analogues, while open circles indicate all galaxies in the simulation in the depicted parameter space. \textit{Left panel:} stellar-to-halo mass relation, as stellar mass $M_*$ (within 30 physical kpc) as a function of host halo mass $M_\rmn{200c}$: in our selection, MW/M31-like galaxies are not necessarily required to be centrals themselves. We compare our results from TNG50 to the first stage of the SAGA survey \protect\citep[pink diamonds,][]{Geha2017}, previous simulations -- Auriga \protect\citep[purple diamonds,][]{Grand2017}, Latte \protect\citep[green diamond,][]{Wetzel2016}, and FIRE-2 \protect\citep[orange diamonds,][]{Garrison-Kimmel2019} -- as well as the MW \protect\citep[light red box,][]{Licquia2015, Boardman2020} and M31 \protect\citep[light brown box,][]{Sick2015}. As the Latte and FIRE-2 simulations employ different definitions of host halo mass, we convert their masses to $M_\rmn{200c}$ -- note that their actual $M_\rmn{200c}$ masses might be slightly different depending on their employed cosmology. The grey, vertical lines denote the median (solid line, $10^{12.1}~\rmn{M}_\odot$), as well as the $10^\rmn{th}$ and $90^\rmn{th}$ percentiles (dashed lines, $10^{11.9}~\rmn{M}_\odot$ and $10^{12.5}~\rmn{M}_\odot$, respectively) in total host mass for our sample of MW/M31-like galaxies, since our sample is primarily selected based on their stellar mass. \textit{Right panel:} stellar mass as a function of absolute $K$-band luminosity $M_\rmn{K}$, both within $30~\rmn{kpc}$. We compare our TNG50 galaxies to MW-like galaxies from the first stage of the SAGA survey \protect\citep[pink diamonds,][]{Geha2017}, hosts in the Local Volume (green diamonds, \protect\citealp{Carlsten2020b, Carlsten2021}, as available, with additional data from \citealp{Skrutskie2006, Lianou2019}), as well as the MW \protect\citep[light red box,][]{Drimmel2001} and M31 \protect\citep[light brown box,][]{Hammer2007}. \textit{Right panel inset:} distributions of host $K$-band luminosity within $30~\rmn{kpc}$. We compare our fiducial selection of TNG50 MW/M31-like galaxies (blue histogram) to the alternative TNG50 SAGA- and LV-like selections (dark purple and light green histograms, respectively), host galaxies from the SAGA survey's first and second stages (pink and purple histograms, respectively), LV hosts (dark green histogram), as well as the MW (red hatched area) and M31 (brown hatched area).} 
    \label{fig:hostProps}
\end{figure*}

We present our sample of MW/M31-like galaxies and some of their fundamental scaling relations in Figure~\ref{fig:hostProps}. Its left panel shows the TNG50 stellar-to-halo mass relation (SHMR), as stellar mass $M_*$ (within $30~\rmn{physical~kpc}$) as a function of total host halo mass $M_\rmn{200c}$. MW/M31-like galaxies are represented by blue circles and are shown in context of TNG50's general galaxy population, including both central and satellite galaxies (open circles). Vertical grey lines mark specific total host masses that our sample -- which is selected based on stellar mass -- covers. Dashed lines correspond to the $10^\rmn{th}$ and $90^\rmn{th}$ percentiles of our host mass range at $10^{11.9}~\rmn{M}_\odot$ and $10^{12.5}~\rmn{M}_\odot$, respectively. The solid, vertical line denotes the median host mass of our sample of MW/M31-like galaxies at $M_\rmn{200c} = 10^{12.1}~\rmn{M}_\odot$. Furthermore, we compare our sample of host galaxies to the first stage of the SAGA survey \citep[pink diamonds,][]{Geha2017} and previous hydrodynamical simulations of similar-mass hosts: Auriga \citep[purple diamonds,][]{Grand2017}, Latte \citep[green diamond,][]{Wetzel2016}, and FIRE-2 \citep[orange diamonds,][]{Garrison-Kimmel2019}. As both Latte and FIRE-2 employ different measurements of host halo mass -- $M_\rmn{200m}$ (i.e. the total mass of a sphere around the FoF halo centre with a mean density of 200 times the mean density of the Universe) and $M_{\Delta_\rmn{c}}$ (i.e. the total mass of a sphere with a mean density of $\Delta_\rmn{c}$ times the critical density of the Universe, where $\Delta_\rmn{c}$ is derived from the collapse of a spherical top-hat perturbation), respectively -- we convert their host masses into $M_\rmn{200c}$ using the TNG50 relations of these different mass measurements.\footnote{The actual $M_\rmn{200c}$ masses of simulations other than TNG50 might be slightly different depending on their adopted cosmology. We determine the TNG50 relations using least squares minimisation:
\begin{align}
    \log M_\rmn{200c} &= 0.99 \times \log M_\rmn{200m} - 0.02 \label{eq:M200c_vs_M200m}\\
    \log M_\rmn{200c} &= \log M_{\Delta_\rmn{c}} - 0.02
    \label{eq:M200c_vs_Mtophat}
\end{align}}

While both observations and simulations span a similar range in stellar mass, slightly shifted to less massive galaxies, the sample of TNG50 MW/M31-like galaxies is by design allowed to extend towards somewhat more massive host haloes. It is important to note the different approaches that are adopted across analyses in the definition of galaxies such as the MW and M31. While we adopt a host selection based on the host galaxies' stellar mass as an observable property, most previous hydrodynamical works employ a halo mass-based selection instead. Moreover, it is important to note that ours is a volume-limited sample, with more numerous MW/M31-like galaxies towards the lower end of the mass range and fewer galaxies at the high-mass end, i.e. the mass distribution within the sample is not flat. This must be kept in mind when comparing medians and scatters across samples. We summarise the host selection criteria for host galaxies of all other simulations addressed in this paper in Table~\ref{tab:hostSelec} and discuss their impact on satellite populations in Section~\ref{sec:prevSims}. 

Finally, we compare to mass estimates of the MW (light red box) and M31 (light brown box). We show a stellar mass range of $M_* = 10^{10.66} - 10^{10.85}~\rmn{M}_\odot$ for the MW according to \cite{Licquia2015} and \cite{Boardman2020}, as well as a stellar mass range of $M_* = 10^{10.9} - 10^{11.1}~\rmn{M}_\odot$ for M31 according to \cite{Sick2015}. Within these shaded stellar mass bands, we indicate the currently available estimates on the host halo mass of the MW and M31 as darker regions: these lie in the range of $10^{11.5} - 10^{12.5}~\rmn{M}_\odot$ (see \citealp{Callingham2019} for a compilation of estimates). As host halo mass is not an observable, we adopt the same estimate for both the MW and M31. Therefore, our fiducial selection of TNG50 MW/M31-like galaxies returns a range of host halo masses that is well consistent with current inferences for the Galaxy and Andromeda. In fact, the MW's mass estimates lie well in the centre of TNG50's sample of MW/M31-like galaxies. Based on the effective SHMR of TNG50, we would instead expect M31 to reside in a more massive halo than the MW.

The right panel of Figure~\ref{fig:hostProps} depicts the correlation of stellar mass $M_*$ and absolute $K$-band luminosity $M_\rmn{K}$, both measured within an aperture of $30~\rmn{kpc}$. Galaxy luminosities are constructed by assigning broad-band luminosities to each stellar particle using the stellar population synthesis model of \cite{Bruzual2003} according to each particle's age, mass, and metallicity. None of the luminosities include dust attenuation (see \citealp{Vogelsberger2013} for details). As before, we contextualise our sample of MW/M31-like hosts (blue circles) with all galaxies in TNG50 in general (open circles). The two properties are tightly correlated (with a Pearson correlation coefficient of -0.85), depicting the $K$-band luminosity of galaxies as a clear proxy for their stellar mass. We compare the distribution of MW/M31-like hosts in TNG50 to observed systems from the first stage of the SAGA survey \citep[pink diamonds,][]{Geha2017}, as well as hosts in the Local Volume (green diamonds, \protect\citealp{Carlsten2020b, Carlsten2021}, as available, with additional data from \citealp{Skrutskie2006, Lianou2019}), all of which are consistent with the distribution of TNG50 hosts. Furthermore, we include estimates of the actual MW (red box) and M31 galaxies (brown box) using $K$-band luminosity measurements from \cite{Drimmel2001} for the MW and \cite{Hammer2007} for M31. Both galaxies' stellar mass and luminosity estimates agree remarkably well with the relation formed by MW/M31-like galaxies in TNG50.

More specifically, we compare the distributions of $K$-band luminosity (within $30~\rmn{kpc}$) for various samples of host galaxies in the inset panel of the right panel of Figure~\ref{fig:hostProps}. While our fiducial sample of TNG50 MW/M31-like hosts (blue filled histogram) peaks around the luminosity of the MW (red, dashed lines) and has a significant fraction of hosts at brighter luminosities to include analogues of Andromeda (brown, dashed lines), both the SAGA (pink and purple dashed histogram) and Local Volume hosts (dark green dashed histogram) are more concentrated at slightly fainter luminosities than the MW. By construction, the distributions of the observed hosts and of the analogue TNG50 SAGA- and LV-like selections (dark purple and light green histograms) display a good level of compatibility: this ensures an even-handed comparison of their satellite systems, as we do below.

\begingroup
\setlength{\tabcolsep}{3pt} 
\renewcommand{\arraystretch}{1.9} 
\begin{table*}
    
    \centering
    \begin{tabular}{l c c c c}
        \hline \hline
        Simulation & Reference & $\#$ of Hosts & Mass Selection & Other Selection Criteria \\ \hline
        Aquarius (DMO) & \protect\cite{Springel2008} & 6 MW-like & $M_\rmn{200c} = 10^{11.9 - 12.3}~\rmn{M}_\odot$ & \makecell{no massive neighbour at $z=0$, \\late-type galaxy via SAM} \\
        Via Lactea II (DMO) & \protect\cite{Diemand2008} & 1 MW-like & \makecell{$M_\rmn{200m} = 10^{12.3}~\rmn{M}_\odot$, \\ i.e. $M_\rmn{200c} = 10^{12.1}~\rmn{M}_\odot$} &\makecell{no recent major mergers} \\ 
        Phat ELVIS (DMO)  & \protect\cite{Kelley2019} & 12 MW-like & \makecell{$M_{\Delta_\rmn{c}} = 10^{11.8 - 12.3}~\rmn{M}_\odot$, \\ i.e. $M_\rmn{200c} = 10^{11.6 - 12}~\rmn{M}_\odot$} & \makecell{isolated within $3~\rmn{Mpc}$} \\ 
        \hline
        APOSTLE (bary.) & \protect\cite{Fattahi2016a} & 12 LG-like & $M_\rmn{200c, LG} = 10^{12.2 - 12.6}~\rmn{M}_\odot$& \makecell{pairs isolated within $2.5~\rmn{Mpc}$, \\separation of $600 - 1000~\rmn{kpc}$} \\
        Latte (bary.) & \protect\cite{Wetzel2016} & 1 MW-like & \makecell{$M_\rmn{200m} = 10^{12.1}~\rmn{M}_\odot$, \\ i.e. $M_\rmn{200c} = 10^{12}~\rmn{M}_\odot$} & \makecell{slightly quiescent merger history} \\
        Auriga (bary.) & \protect\cite{Grand2017} & 30 MW-like & $M_\rmn{200c} = 10^{12.0 - 12.3}~\rmn{M}_\odot$& \makecell{isolated, outside of $9 \times R_\rmn{200c}$ \\of any other halo}\\
        FIRE-2 (bary.) & \protect\cite{Garrison-Kimmel2014, Garrison-Kimmel2017, Garrison-Kimmel2019} & 2 LG-like & \makecell{$M_{\Delta_\rmn{c}} = 10^{12.3 - 12.7}~\rmn{M}_\odot$, \\ i.e. $M_\rmn{200c} = 10^{12.2 - 12.6}~\rmn{M}_\odot$} & \makecell{pairs isolated within $2.8~\rmn{Mpc}$, \\no cluster within $7~\rmn{Mpc}$, \\separation of $\sim 800~\rmn{kpc}$} \\
        FIRE-2 (bary.) & \protect\cite{Garrison-Kimmel2014, Garrison-Kimmel2017, Garrison-Kimmel2019} & 6  MW-like & \makecell{$M_\rmn{200m} = 10^{12.0-12.3}~\rmn{M}_\odot$, \\ i.e. $M_\rmn{200c} = 10^{11.9 - 12.2}~\rmn{M}_\odot$} & \makecell{isolated within $2.8~\rmn{Mpc}$} \\
        ARTEMIS (bary.) & \protect\cite{Font2021} & 45 MW-like & $M_\rmn{200c} = 10^{11.9 - 12.3}~\rmn{M}_\odot$ & -- \\
        DC Justice League (bary.) & \protect\cite{Applebaum2021} & 2 MW-like & \makecell{$M_{\Delta_\rmn{c}} = 10^{11.9 - 12.4}~\rmn{M}_\odot$, \\ i.e. $M_\rmn{200c} = 10^{11.8 - 12.3}~\rmn{M}_\odot$} & \makecell{different merger histories} \\
        TNG50 (bary.) & Engler et al. (2021b, this paper) & 198 MW/M31-like & \makecell{$M_* = 10^{10.5} - 10^{11.2}~\rmn{M}_\odot$,\\i.e. $M_\rmn{200c} = 10^{11.9} - 10^{12.5}~\rmn{M}_\odot$} & \makecell{disky shape ($s < 0.45$), \\no massive galaxy within $500~\rmn{kpc}$,\\ host mass $M_\rmn{200c} < 10^{13}~\rmn{M}_\odot$} \\
        \hline
    \end{tabular}
    \caption{{\bf List of simulations of MW/M31- and Local Group-like (LG) hosts referenced throughout this paper, featuring studies of satellite abundances.} This includes both dark matter-only (DMO, upper part) and baryonic simulations (lower section of the Table). We provide references and summarise the sample size, as well as selection criteria of host galaxies and haloes. The latter include constraints on halo and stellar mass range, galaxy morphology, merger histories, and environment. Barring the mass constraints of the APOSTLE simulations, which apply to the total pair mass of the LG-like systems, all other mass selection criteria refer to individual MW/M31-like hosts. For TNG50, these can still include pairs of MW/M31-like hosts: in such cases, the quoted halo mass values refer to the host mass evaluated in a spherical-overdensity fashion starting from the halo center of the two galaxies. See Section~\ref{sec:sample_mw} for further details on the host selection in TNG50. Different simulations adopt different measures of halo mass: $M_\rmn{200c}$ ($M_\rmn{200m}$) corresponds to the total mass of a sphere around the FoF halo centre with a mean density of 200 times the critical (mean) density of the Universe, while $M_{\Delta_\rmn{c}}$ is the total mass of a sphere with a mean density of $\Delta_\rmn{c}$ times the critical density of the Universe. $\Delta_\rmn{c}$ is derived from the collapse of a spherical top-hat perturbation. We convert the host masses of simulations that use either $M_\rmn{200m}$ or $M_{\Delta_\rmn{c}}$ into $M_\rmn{200c}$ masses in order to put them directly into context with our sample of TNG50 MW/M31-like hosts. Note that the conversion between different cosmological mass measurements is based on the TNG50 relations of $M_\rmn{200c}$ vs. $M_\rmn{200m}$ \eqref{eq:M200c_vs_M200m} and $M_\rmn{200c}$ vs. $M_{\Delta_\rmn{c}}$ \eqref{eq:M200c_vs_Mtophat}. The actual $M_\rmn{200c}$ values of other simulations might be slightly different depending on their adopted cosmology. The numbers in this table are relevant for the interpretation of e.g. Figure~\ref{fig:satMF_sims}.}
    \label{tab:hostSelec}
\end{table*}
\endgroup

\subsection{Selecting satellite galaxies}
\label{sec:sample_sats}

Throughout this study, we employ different definitions for satellite galaxies around MW/M31-like hosts. While satellites in IllustrisTNG are generally identified as subhaloes or local overdensities within larger FoF haloes using the \textsc{subfind} algorithm \citep{Springel2001, Dolag2009}, we avoid to restrict our selection of satellites by FoF membership for the most part of this analysis. However, not all luminous subhaloes represent actual galaxies in TNG. Some correspond to fragmentations and clumps within other galaxies due to e.g. disk instabilities that \textsc{subfind} identified as independent objects. We exclude these objects from our sample according to \cite{Nelson2019a}, as these do not represent galaxies that formed by gas collapse at the centre of their DM haloes, which then fall into their $z=0$ hosts.\\

{\it Fiducial satellite selection.} Our fiducial satellite selection is based on the three-dimensional distance to their host MW/M31-like galaxy. We define satellites as galaxies within $300~\rmn{physical~kpc}$ of their host -- corresponding approximately to the virial radii of the MW and M31 -- and require them to have a stellar mass of at least $5 \times 10^6~\rmn{M}_\odot$ (measured within twice the stellar half-mass radius $R_{1/2}^*$). This mass limit ensures an appropriately resolved identification with at least 63 stellar particles per galaxy and corresponds to the minimum stellar mass below which the TNG50 SHMR becomes incomplete due to its stellar mass resolution (see Appendix~\ref{sec:resolution} for details on resolution effects). Furthermore, it corresponds to the mass of the MW's satellite galaxy Leo I. This selection leaves us with a total of 1237 satellite galaxies around 198 MW/M31-like hosts.\\

{\it Observational satellite selections.} We vary this selection in parts of Section~\ref{sec:satMF_LFobs} in order to match the selection criteria of observational surveys by employing no minimum stellar mass and varying the distance limit. For comparisons to the SAGA survey \citep{Geha2017, Mao2021}, we require satellite galaxies to lie within a two-dimensional, randomly projected aperture of $300~\rmn{kpc}$ from their host and to have a line-of-sight velocity within $\pm 250~\rmn{km~s^{-1}}$. For comparisons to Local Volume hosts from \cite{Carlsten2021}, satellites are constrained to a two-dimensional, projected aperture of $150~\rmn{kpc}$ instead. Along the line of sight, satellites are restricted to distances of $\pm 500~\rmn{kpc}$. While the line-of-sight distances of observed satellites in \cite{Carlsten2021} are estimated using either surface brightness fluctuations or the tip of their red giant branch, they also include a comparison to galaxies from TNG100, for which they adopt this physical distance requirement along the line of sight.\\

{\it Alternative satellite selections.} In Section~\ref{sec:res_satMF_acc}, in order to compare the subhalo mass functions across numerical models, including DM-only simulations, we consider satellites as those within 300 physical kpc from their host centers and with a minimum dynamical mass -- the sum of all gravitationally bound particles -- of $5 \times 10^7~\rmn{M}_\odot$. According to the effective SHMR of TNG50, this value corresponds to the smallest total subhalo mass below which finite mass resolution makes the SHMR incomplete and artificially bends it -- see bottom left panel of Figure~\ref{fig:resTest_lumSats}.
Finally, in cases where we compare all accreted satellites to their present-day population of survivors (Section~\ref{sec:res_satMF_acc}), we do require them to be members of their host galaxy's FoF halo since their time of accretion is defined with respect to halo membership \citep{Chua2017}.

\begin{figure*}
    \includegraphics[width=.85\textwidth]{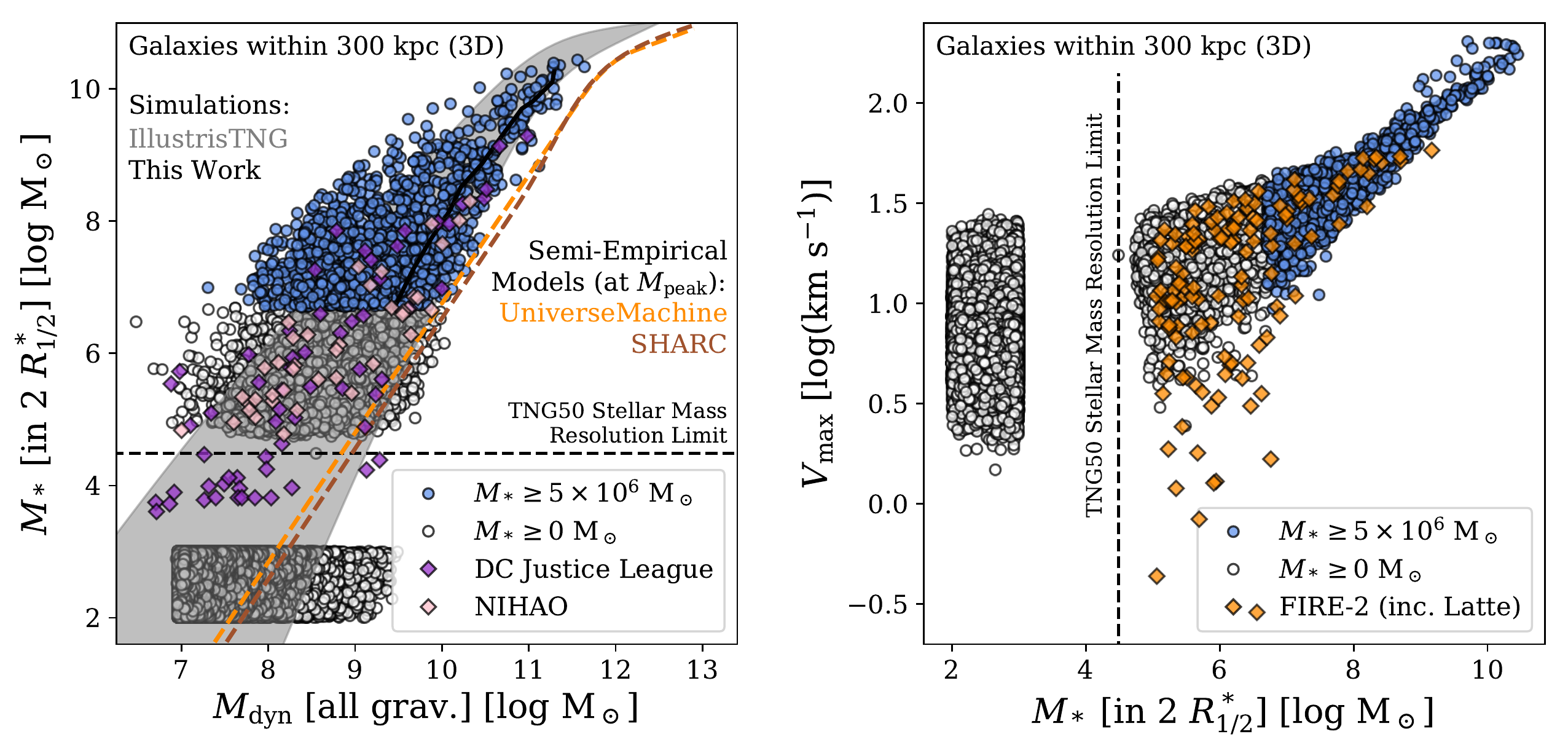}
    \includegraphics[width=.98\textwidth]{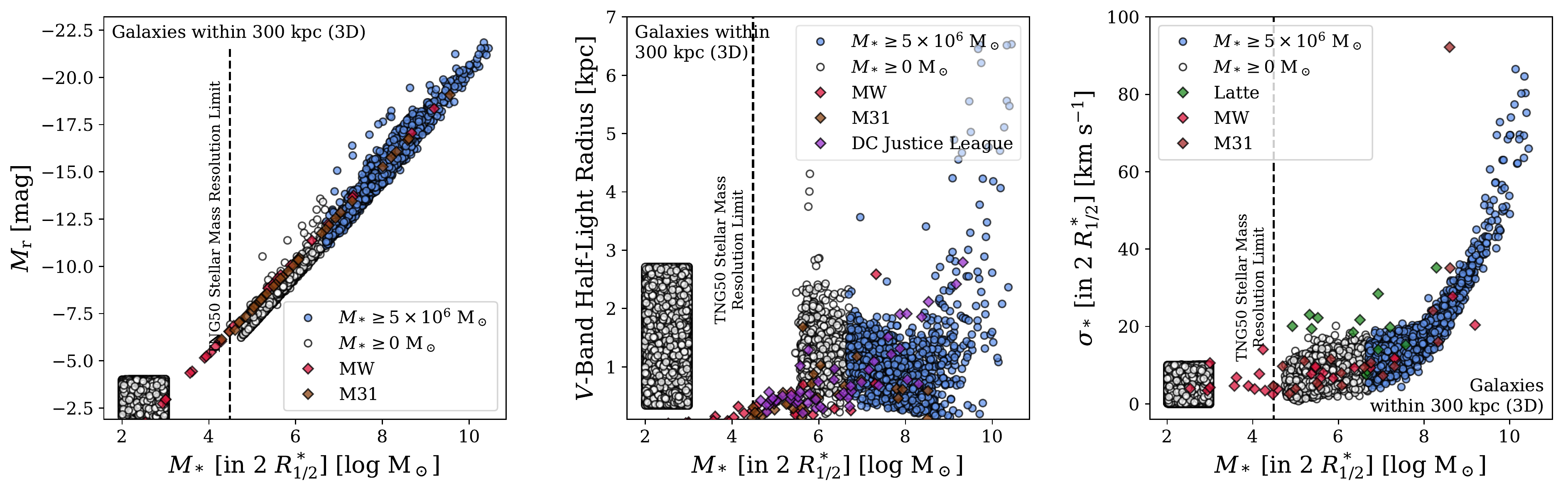}
    \caption{{\bf Properties and scaling relations of satellites and subhaloes around MW/M31-like galaxies in TNG50.} In all panels, we require galaxies and subhaloes to be located within $300~\rmn{kpc}$ of their host. Blue circles correspond to satellite galaxies with a stellar mass of $M_* \geq 5 \times 10^6~\rmn{M}_\odot$, open circles show satellites and subhaloes with $M_* < 5 \times 10^6~\rmn{M}_\odot$, including dark subhaloes that do not host a stellar component. In this case, we assign random values to their stellar properties, disconnected from the main scaling relations. The dashed, black lines denote TNG50's stellar mass resolution limit at $10^{4.5}~\rmn{M}_\odot$, slightly below the mass of the target stellar particle after accounting for mass loss. \textit{Top left panel:} stellar-to-halo mass relation (SHMR), as stellar mass $M_*$ (within twice the stellar half-mass radius) as a function of dynamical mass $M_\rmn{dyn}$ (all gravitationally bound particles). For visualisation purposes, dark subhaloes are assigned a random stellar mass between $10^2$ and $10^3~\rmn{M}_\odot$. We compare the SHMR of our sample of satellite galaxies (solid, black curve) to results from simulations and semi-empirical models. The grey shaded area denotes the scatter for the SHMR of satellites in hosts of $10^{12}-10^{13}~\rmn{M}_\odot$ from \protect\cite{Engler2021} (their lowest bin in host mass); the purple diamonds display satellite galaxies from the DC Justice League simulations of MW analogues \protect\citep{Applebaum2021}, while the pink diamonds correspond to satellites of MW-like galaxies from NIHAO \protect\citep{Buck2019}. The dashed curves display extrapolations for satellite galaxies from semi-empirical models: \textsc{UniverseMachine} \protect\citep[orange curve,][]{Behroozi2019} and SHARC \protect\citep[brown curve,][]{RodPueb2017}. Note that both semi-empirical models define subhalo mass as peak mass, not as a dynamical mass at the present-day. \textit{Top right panel:} maximum circular velocity $V_\rmn{max}$ as a function of stellar mass. Additionally, we compare to satellites of Local Group- and MW-like hosts from the FIRE-2 simulations \protect\citep[orange diamonds,][]{Garrison-Kimmel2019}; this includes satellites of the Latte simulation \protect\citep{Wetzel2016}. \textit{Bottom left panel:} absolute $r$-band magnitude $M_\rmn{r}$ as a function of stellar mass. Dark subhaloes are assigned absolute magnitudes ranging from $-2$ to $-4$. We compare to satellite galaxies from the MW \protect\citep[red diamonds,][]{McConnachie2012} and M31 \protect\citep[brown diamonds,][]{McConnachie2018}. \textit{Bottom central panel:} size-mass relation using 2D stellar half-light radii ($V$-band) for TNG50 satellites and subhaloes. We compare to observed satellites of the MW and M31 (red and brown diamonds, respectively), as well as the DC Justice League simulations (purple diamonds). As TNG50, these works employ 2D stellar half-light radii ($V$-band). \textit{Bottom right panel:} stellar 3D velocity dispersion $\sigma_*$ as a function of stellar mass. Dark subhaloes are assigned velocity dispersions of $\sigma_* = 0 - 10~\rmn{km~s}^{-1}$. We compare to simulated satellite dwarf galaxies from the Latte simulation (green diamonds) and observed dwarfs from the MW and M31 \protect\citep[red and brown diamonds,][respectively]{McConnachie2012, McConnachie2018}.}
    \label{fig:satProps}
\end{figure*}

\section{Results}
\label{sec:results}

\subsection{Properties of satellite populations in TNG50}
\label{sec:satProps}

Before analysing the abundance of satellite galaxies around individual MW/M31-like hosts in TNG50, we investigate some of the properties and scaling relations of both satellites and subhaloes in Figure~\ref{fig:satProps}: their stellar-to-halo mass relation (SHMR) (top left panel), the maximum of their circular velocity profile $V_\rmn{max}$ as a function of stellar mass $M_*$ (top right panel), absolute $r$-band magnitude $M_\rmn{r}$ as a function of stellar mass (bottom left panel), and stellar 3D velocity dispersion $\sigma_*$ as a function of stellar mass (bottom right panel). Satellite galaxies with a stellar mass of $M_* \geq 5 \times 10^6~\rmn{M}_\odot$ are denoted as blue circles, satellites and subhaloes with smaller or no stellar mass whatsoever are marked as open circles. In order to illustrate the abundance of dark subhaloes, we assign random values to their stellar properties, detached from the main scaling relation: stellar masses of $10^2 - 10^3~\rmn{M}_\odot$, absolute $r$-band magnitudes of $-2$ to $-4$, and stellar 3D velocity dispersions of $0 - 10~\rmn{km~s}^{-1}$. We mark the transition between luminous and dark regimes -- the stellar mass resolution limit of TNG50 -- as a dashed line in all panels. This corresponds to the minimum stellar mass that subhaloes contain, slightly lower than the target mass of a single stellar particle due to mass loss.

\subsubsection{Satellite SHMR}
Throughout the paper, the satellites' stellar mass $M_*$ is measured within twice the stellar half-mass radius, while their dynamical mass $M_\rmn{dyn}$ corresponds to the sum of all gravitationally bound particles as defined by the \textsc{subfind} algorithm (i.e. dark matter, stars, gas, and black holes).

The SHMR of satellites of MW/M31-like galaxies in TNG50 in the top left panel of Figure~\ref{fig:satProps} exhibits significant scatter. A large number of satellites show substantial offsets towards lower dynamical masses from their median SHMR (black curve) -- predominantly due to tidal stripping of their dark matter haloes by their host's gravitational potential \citep[e.g.][]{Joshi2019, Engler2021}. Furthermore, we compare our sample to several SHMRs from simulations and semi-empirical models. The shaded grey region denotes the scatter of the SHMR for satellite galaxies in hosts of $10^{12}-10^{13}~\rmn{M}_\odot$ in IllustrisTNG from \cite{Engler2021}. Although this range of host mass does not exactly match the selection of MW/M31-like hosts, the SHMRs are well in agreement, as to be expected modulo resolution effects. The purple and pink diamonds show the SHMR of satellites of MW analogues from other simulations -- the DC Justice League \citep[][]{Applebaum2021} and NIHAO \citep[][]{Buck2019}, respectively. The satellites of both simulations agree well with the overall SHMR in TNG50. The DC Justice League extends to particularly small stellar masses due to their high level of resolution. 

Finally, we compare to {\it extrapolations} for satellite galaxies of the semi-empirical models \textsc{UniverseMachine} \citep[dashed orange curve,][]{Behroozi2019} and SHARC \citep[dashed, brown curve][]{RodPueb2017}. Since both models consider satellite dynamical masses at their peak -- as opposed to the present-day -- the shift of the \textsc{UniverseMachine} and SHARC SHMRs towards larger dynamical masses is expected (see \citealp{Engler2021} for a discussion). These differences between TNG50 and semi-empirical SHMRs hold regardless of whether we consider satellite populations of our full sample of TNG50 MW/M31-like hosts or simply the satellites of hosts that are centrals. Furthermore, it should be noted that both semi-empirical models were calibrated for more massive galaxies: \cite{RodPueb2017} even state that their SHMR should only be trusted down to (sub)halo masses of $10^{10.3}~\rmn{M}_\odot$. We merely report these extrapolations as references. 

\subsubsection{Relation of maximum circular velocity and stellar mass}
The upper right panel depicts the relation of the circular velocity profile's maximum $V_\rmn{max}$ and stellar mass. $V_\rmn{max}$ is defined as the maximum value of a subhalo's spherically-averaged circular velocity curve including all matter components: dark matter, stars, and gas. The satellites form a continuous relation in which their scatter increases substantially towards lower masses. Furthermore, we compare to satellite galaxies from the FIRE-2 simulations \citep[orange diamonds,][]{Garrison-Kimmel2019}, which includes satellites of the Latte simulation \citep{Wetzel2016}. For the most part, their relation of maximum circular velocity and stellar mass is well in agreement with the satellite population of TNG50. At stellar masses below $10^7~\rmn{M}_\odot$, their scatter surpasses TNG50's relation significantly, with FIRE-2 galaxies reaching smaller $V_\rmn{max}$ values than any TNG50 galaxy at fixed satellite stellar mass. 

\begin{figure*}
    \centering
    \includegraphics[width=.48\textwidth]{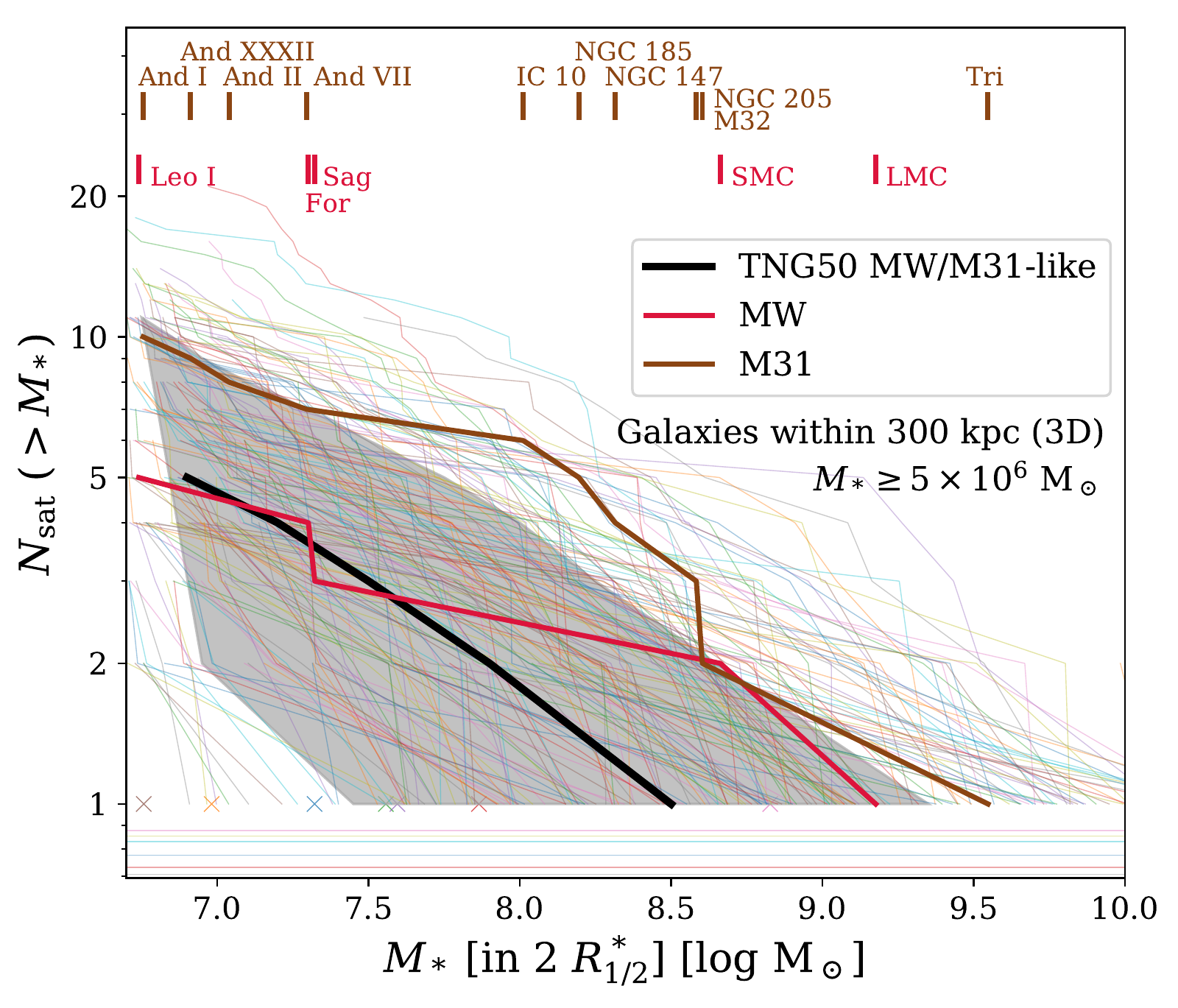}
    \includegraphics[width=.48\textwidth]{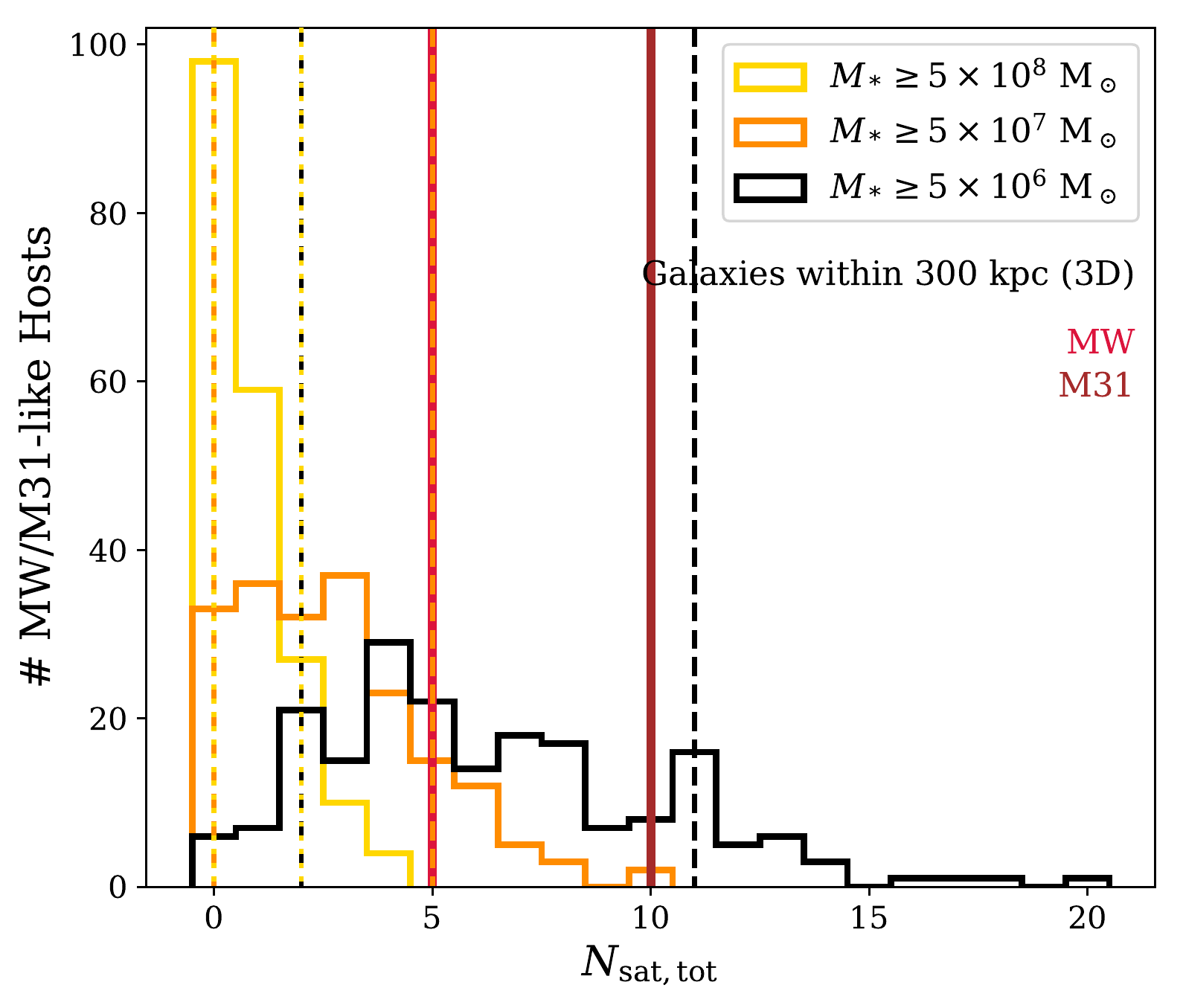}
    \caption{{\bf Satellite demographics around MW/M31-like galaxies in the TNG50 simulation at $\mathbf{z=0}$.} \textit{Left panel:} satellite abundance as cumulative stellar mass function for MW/M31-like hosts in TNG50. We define satellites as galaxies within $300~\rmn{physical~kpc}$~(3D) of their host and with stellar masses  of at least $5 \times 10^6~\rmn{M}_\odot$ (within twice the stellar half-mass radius $R_{1/2}^*$). The thin, coloured curves in the background illustrate the satellite systems of individual TNG50 hosts with crosses corresponding to systems with only a single satellite and horizontal lines with $N_\rmn{sat} < 1$ denoting systems with no satellites meeting the selection: these are 6 among 198 systems. The thick, black curve and grey shaded area depict their median and scatter as $16^\rmn{th}$ and $84^\rmn{th}$ percentiles, computed in bins of satellite stellar mass and including galaxies with zero satellites in the calculations. Furthermore, we compare our findings in TNG50 to satellite abundances of the MW \protect\citep[red curve,][]{McConnachie2012} and M31 \protect\citep[brown curve,][]{McConnachie2018}. \textit{Right panel:} distribution of total satellite abundance $N_\rmn{sat, tot}$ around MW/M31-like hosts in TNG50 and its dependence on the imposed minimum stellar mass. We compare satellite systems with $M_* \geq 5 \times 10^8~\rmn{M}_\odot$ (yellow histogram), $M_* \geq 5 \times 10^7~\rmn{M}_\odot$ (orange histogram), and $M_* \geq 5 \times 10^6~\rmn{M}_\odot$ (black histogram; our fiducial selection). Dashed and dotted lines denote their scatter as $16^\rmn{th}$ and $84^\rmn{th}$ percentiles: 0 and 2 for satellites with $M_* \geq 5 \times 10^8~\rmn{M}_\odot$, 0 and 5 for $M_* \geq 5 \times 10^7~\rmn{M}_\odot$, as well as 2 and 11 for $M_* \geq 5 \times 10^6~\rmn{M}_\odot$. Furthermore, we include the total satellite abundances of the MW and M31 as solid, vertical lines (red and brown lines, respectively). Alternative versions of this figure, where satellite stellar masses are normalised to either host halo or host stellar masses, are given in Appendix~\ref{sec:normSatAb}.}
    \label{fig:MFTNG50_Nsat_hist}
\end{figure*}

\subsubsection{Other observable dwarf properties}
We illustrate the relation of absolute $r$-band magnitudes $M_\rmn{r}$ and stellar mass in the bottom left panel of Figure~\ref{fig:satProps}. The TNG50 satellites exhibit a tight correlation, in agreement with satellite galaxies of the MW and M31. It should be noted that \cite{McConnachie2012} and \cite{McConnachie2018} assume a $V$-band mass-to-light ratio of 1 for simplicity, so the MW and M31 satellites should in reality have slightly different stellar masses. However, these differences should not be significant. In this case, we convert absolute $V$-band luminosities from \cite{McConnachie2012} and \cite{McConnachie2018} to the $r$-band using the luminosities' correlation in TNG50. This relation was determined using least squares minimisation:

\begin{equation}
    M_\rmn{r} = M_\rmn{V} - 0.23~\rmn{mag}.
    \label{eq:Mr_vs_MV}
\end{equation}

The bottom central panel of Figure~\ref{fig:satProps} depicts the size-mass relation using 2D stellar half-light radii ($V$-band) for TNG50 satellites with $M_* > 2 \times 10^6~\MSUN$, as well as for satellites of the MW (red diamonds), M31 (brown diamonds), and the DC Justice League simulations (purple diamonds). There is a reasonable level of agreement between TNG50 and the satellites of both the Galaxy and Andromeda, as well as of the DC Justice League simulations, at the masses where a comparison is possible. In fact, here TNG50 sizes are 2D circularised radii from random projections and are not measured for galaxies with fewer than 10 stellar particles. More information on the size-mass relation in TNG50, outside of the context of MW/M31-like galaxies, can be found in \cite{Pillepich2019} and \cite{Zanisi2020}.

Finally, we present the stellar 3D velocity dispersion $\sigma_*$ as a function of stellar mass in the bottom right panel. The stellar 3D velocity dispersion of simulated satellites is measured as the standard deviation of the velocities of all stellar particles within two times the stellar half-mass radius weighted by their respective stellar mass. The satellites form a continuous relation, in which the stellar 3D velocity dispersion increases significantly for more massive satellites. We include satellites from the Latte simulation \citep[green diamonds,][]{Wetzel2016}, as well as from the MW and M31 as comparison \citep{McConnachie2012, McConnachie2018}. While many of Latte's satellites exhibit slightly larger velocity dispersions than TNG50, most MW and M31 satellites agree reasonably well\footnote{There might be an apparent discrepancy between the stellar velocity dispersion and their maximum circular velocity in the top right panel of Figure~\ref{fig:satProps}, as the Latte satellites are included in the FIRE-2 sample of \cite{Garrison-Kimmel2019}. We speculate that this difference is due to different ways of measuring the velocity properties of satellites between the two studies.}.

\subsection{Satellite abundance of MW/M31-like galaxies in TNG50}
\label{sec:satMF_LFobs}

We present the satellite abundance of all 198 MW/M31-like galaxies in TNG50 as cumulative stellar mass functions in the left panel of Figure~\ref{fig:MFTNG50_Nsat_hist}. Satellites are defined as galaxies within a three-dimensional aperture of $300~\rmn{kpc}$ of their host and are required to have a stellar mass of at least $5 \times 10^6~\rmn{M}_\odot$. This minimum mass allows for an adequate level of resolution (see Appendix~\ref{sec:resolution} for a detailed discussion) and approximately corresponds to the MW's own satellite galaxy Leo I. 

Thin coloured curves in the background of Figure~\ref{fig:MFTNG50_Nsat_hist}, left panel, correspond to the individual satellite stellar mass functions of all MW/M31-like hosts in TNG50, crosses denote systems with only a single satellite, while hosts with no satellites whatsoever are depicted as curves with $N_\rmn{sat} < 1$: these are 6 of 198 hosts. 

MW/M31-like galaxies in TNG50 exhibit a remarkable diversity with significant host-to-host scatter regarding their satellite populations, with total satellite counts ranging from 0 to 20. The thick, black curve displays the median satellite stellar mass function for MW/M31-like galaxies in TNG50, starting with the most massive satellite $M_* \sim 10^{8.5}~\rmn{M}_\odot$ and reaching a total number of five satellites down to $M_* \sim 10^{6.9}~\rmn{M}_\odot$. However, there is a significant amount of scatter, as shown by the grey shaded area denoting the $16^\rmn{th}$ and $84^\rmn{th}$ percentiles. The most massive satellite's stellar mass can vary by $\pm 1~\rmn{dex}$, while the total number of satellites can range from 2 to 11 within the percentile range. This diversity between satellite systems persists independent of host mass: in Appendix~\ref{sec:normSatAb} and Figure~\ref{fig:normSatAb}, we provide the same functions for satellite abundances normalised by host halo mass and host stellar mass.

We compare the findings of TNG50 to the local satellite systems of the MW and M31 (within $300~\rmn{kpc}$), shown as thick, red and brown curves, respectively \citep{McConnachie2012, McConnachie2018}. The MW's satellite stellar mass function falls well within TNG50's $1\sigma$ scatter. While its low-mass end coincides with the TNG50 median, its massive end reaches the upper limits of our scatter due to the presence of both the Small and Large Magellanic Cloud (SMC and LMC, respectively). M31, on the other hand, is slightly more satellite-rich than TNG50's $1\sigma$ scatter. However, it agrees well with many other, individual TNG50 MW/M31-like hosts of our sample. In fact, it should be noted that in our sample of TNG50 MW/M31-like hosts, which is intrinsically volume-limited, the stellar mass distribution is skewed towards masses more similar to the MW rather than the more massive M31 (see Figure~\ref{fig:hostProps}): if instead we were to focus specifically on TNG50 M31 analogues in the left panel of Figure~\ref{fig:MFTNG50_Nsat_hist}, the agreement between the TNG50 median and the observed M31's satellite mass function would be notable. We explore the dependence of satellite abundance on both host selection and host properties further in Sections~\ref{sec:hostSelection} and~\ref{sec:hostProps}.

\begingroup
\setlength{\tabcolsep}{6pt} 
\renewcommand{\arraystretch}{1.15} 
\begin{table}
    \centering
    \begin{tabular}{c c c c}
        \hline \hline
        \# MW/M31-like hosts with & $\pm 0.1~\rmn{dex}$ & $\pm 0.15~\rmn{dex}$ & $\pm 0.2~\rmn{dex}$  \\ \hline 
         SMC & 42 & 62 & 77 \\
        LMC & 12 (21) & 20 (32) & 27 (42) \\
        LMC \& SMC & 6 (7) & 10 (16) & 18 (25) \\ \hline
        M32/NGC205 & 48 & 63 & 75 \\
        Tri & 7 (11) & 14 (20) & 21 (25) \\
        Tri \& M32/NGC205 & 2 (4) & 4 (7) & 11 (13) \\
        Tri \& M32 \& NGC205 & 0 (0) & 1 (1) & 2 (3) \\\hline \hline
        Med. $M_\rmn{host}$ [$\log~\rmn{M}_\odot$] & & &   \\ \hline 
        SMC & 12.2 & 12.3 & 12.3 \\
        LMC & 12.3 & 12.3 & 12.2 (12.3)\\
        LMC \& SMC & 12.2 & 12.3 & 12.3 \\ \hline
        M32/NGC205 & 12.3 & 12.2 & 12.3 \\
        Tri & 12.3 (12.2) & 12.3 & 12.3 \\
        Tri \& M32/NGC205 & 12.4 & 12.4 & 12.4 \\ 
        Tri \& M32 \& NGC205 & -- & 12.4 & 12.4 \\\hline
    \end{tabular}
    \caption{Number of TNG50 MW/M31-like hosts in our fiducial sample with massive satellites such as the SMC and LMC, or M32, NGC205 and Triangulum (Tri). We adopt various mass bins for the reference mass of these massive satellites in order to reflect uncertainties in their measurements: $\pm 0.1~\rmn{dex}$, $\pm 0.15~\rmn{dex}$, and $\pm 0.2~\rmn{dex}$. LMC and Triangulum numbers without parentheses assume them to be the most massive satellite of their host, while the numbers inside parentheses allow for even more massive satellites in the same system. We adopt stellar mass estimates from \protect\cite{McConnachie2012} and \protect\cite{McConnachie2018}: $4.5 \times 10^8~\rmn{M}_\odot$ for the SMC, $1.5 \times 10^9~\rmn{M}_\odot$ for the LMC, $3.8 \times 10^8~\rmn{M}_\odot$ for M32 and NGC205, as well as $3.5 \times 10^9~\rmn{M}_\odot$ for Triangulum. Furthermore, we give the median host mass for all subsamples in the bottom part of the table. We add values in parentheses for systems with even more massive satellites than the LMC or Triangulum if their median host mass is different from the samples in which the LMC or Triangulum is the most massive satellite in the system.}
    \label{tab:SMC_LMC}
\end{table}
\endgroup

We quantify the scatter in total satellite abundance, as well as the effects of our satellite selection and the imposed minimum stellar mass for satellite galaxies in the right panel of Figure~\ref{fig:MFTNG50_Nsat_hist}. 
Here, we show the distribution of the total number of satellites $N_\rmn{sat, tot}$ within $300~\rmn{kpc}$ (3D) of MW/M31-like hosts for three selections in stellar mass: $M_* \geq 5 \times 10^8~\rmn{M}_\odot$ (yellow histogram), $M_* \geq 5 \times 10^7~\rmn{M}_\odot$ (orange histogram), and $M_* \geq 5 \times 10^6~\rmn{M}_\odot$ (black histogram; our fiducial selection). The dotted and dashed vertical lines illustrate the scatter of the distribution of total satellite abundance as $16^\rmn{th}$ and $84^\rmn{th}$ percentiles. As it can be seen, the abundance of satellites with $M_* \geq 5 \times 10^6~\rmn{M}_\odot$ ranges from 6~MW/M31-like hosts with no satellites to one host with 20 satellites and peaks at a total of 4 satellites for 29 MW/M31-like hosts. For more massive satellites, e.g. $M_* \geq 5 \times 10^8~\rmn{M}_\odot$, the number of hosts with no satellites whatsoever increases to 98. In the right panels of Figure~\ref{fig:normSatAb} in Appendix~\ref{sec:normSatAb}, we provide a similar quantification of the scatter in satellite abundances where the satellites are counted based on their mass normalised to their host's total or stellar mass: when the mass of satellites is normalised to their host's, the degree of diversity in total satellite abundance remains the same regardless of the adopted minimum satellite stellar mass, consistently with Poisson statistics \citep[see also][for a discussion on how the scatter can in fact become super-Poissonian at very low (subhalo) mass ratios]{Chua2017}.

\subsubsection{Massive satellites around MW/M31-like hosts}
The presence of massive satellites such as the LMC and SMC around the Galaxy, or Triangulum, M32, and NGC205 around Andromeda is an interesting feature that has been studied in previous theoretical works regarding their general abundance, orbital evolution, or correlations with the mass of their host \citep{Tollerud2011, Busha2011a, Boylan-Kolchin2011, Patel2017, Shao2018}. However, as previous works thus far have been based mostly on N-body only or DM-only calculations, it is useful to query the TNG50 systems as to the presence of massive satellites.

\begin{figure}
    \centering
	\includegraphics[width=8cm]{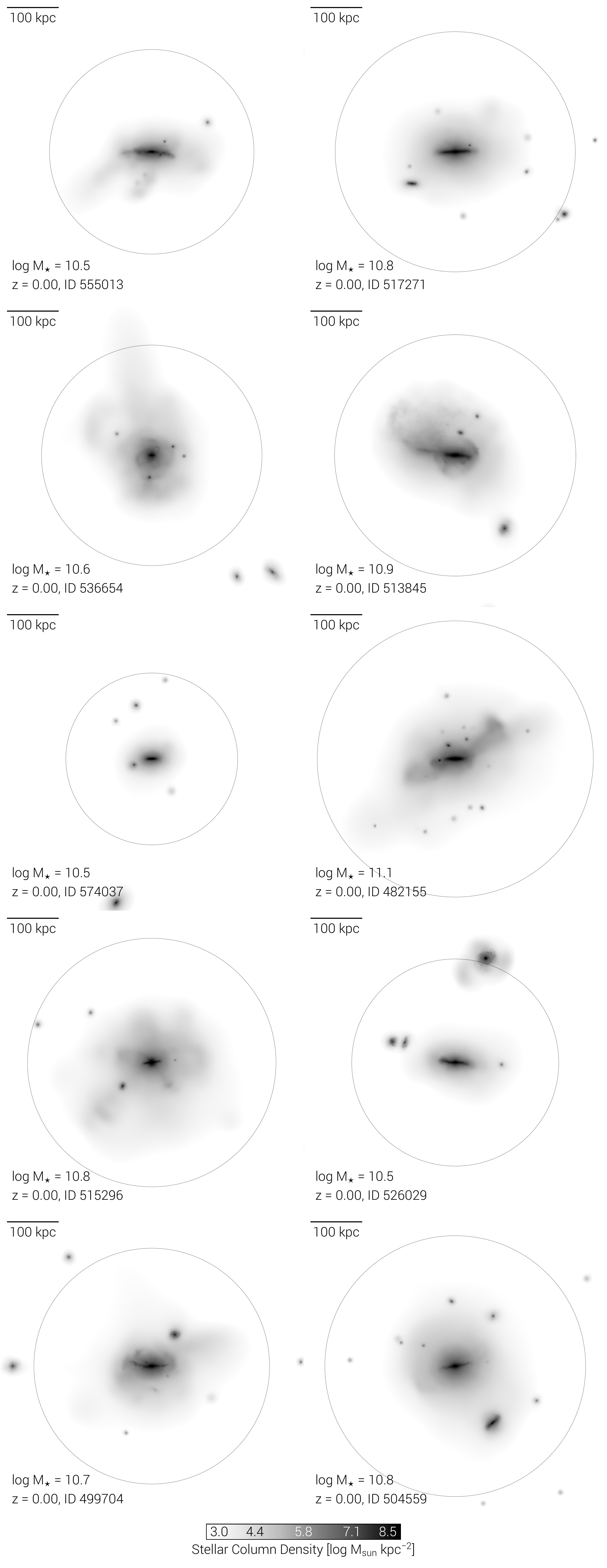}
    \caption{Stellar column density on $600~\rmn{kpc}$ per side of the ten TNG50 galaxies at $z=0$ (edge-on projection) whose satellite stellar mass function is the most similar to that of the Galaxy. Circles denote the virial radius ($\RHOST$) of the underlying DM host. }
    \label{fig:vis1}
\end{figure}
\begin{figure}
    \centering
	\includegraphics[width=8cm]{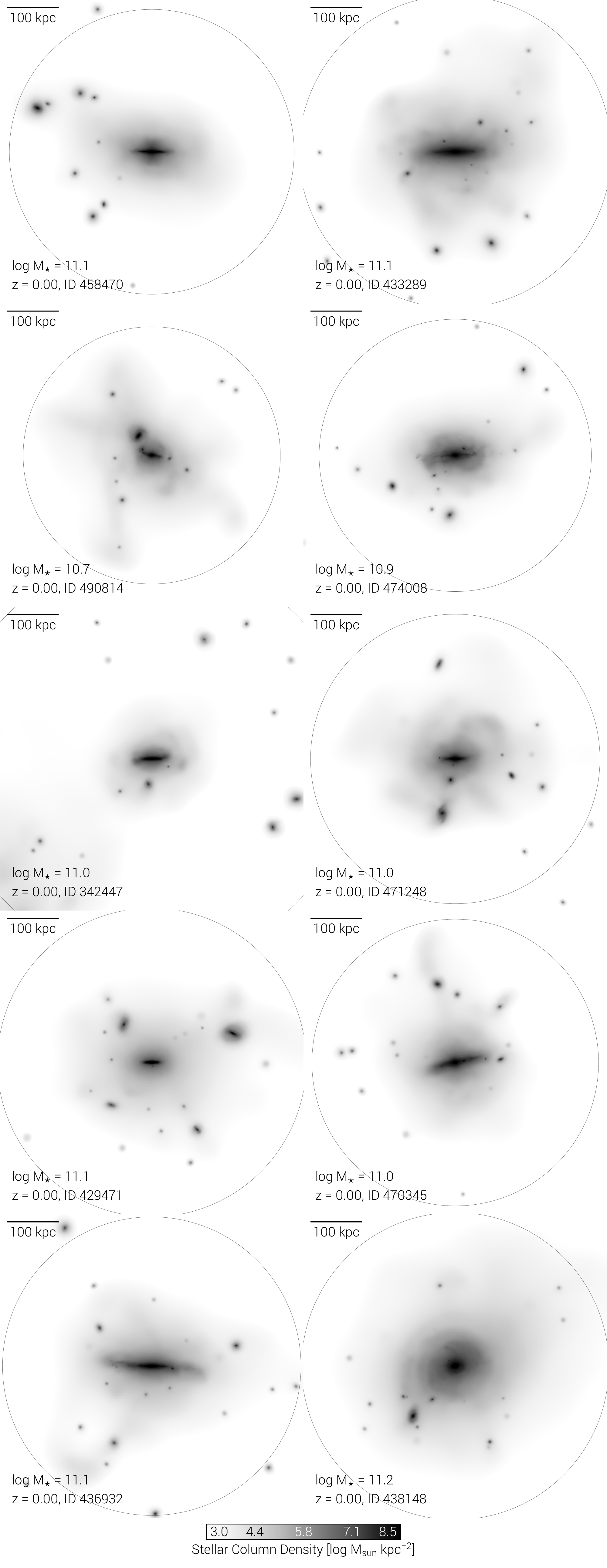}
    \caption{As in Fig.~\ref{fig:vis1} but for the ten TNG50 galaxies at $z=0$ whose satellite stellar mass function is the most similar to that of Andromeda. In TNG50, the latter are usually more massive and reside in more massive haloes (larger $\RHOST$) than the satellite systems that are more similar to the Galaxy's.}
    \label{fig:vis2}
\end{figure}

Interestingly, a significant fraction of TNG50 MW/M31-like galaxies include SMC and/or LMC-like satellites within $300~\rmn{kpc}$. Assuming stellar masses of $10^{8.7 \pm 0.1}~\rmn{M}_\odot$ for SMC-like and $10^{9.2 \pm 0.1}~\rmn{M}_\odot$ for LMC-like satellites \citep{McConnachie2012}, we find that 42~MW/M31-like galaxies host an SMC-like satellite (i.e. 21 per cent of hosts), 12~host an LMC-like galaxy as their most massive satellite (6 per cent of hosts), and 6~MW/M31-like hosts include both an SMC- and an LMC-like galaxy in their satellite population (with the LMC-like galaxy as their most massive satellite). This corresponds to 3 per cent of our MW/M31-like hosting both an SMC- and an LMC-like galaxy, and it is remarkably consistent with the results of \cite{Liu2011}, which were based on SDSS data. However, it should be noted that these observations adopt different, luminosity-based selection criteria for both host and satellite galaxies, as well as spectroscopic distance cuts. As \cite{Busha2011a} show, such differences in sample selections can cause a difference of up to 10~per cent.

We summarise the abundance of SMC- and LMC-like satellites in Table~\ref{tab:SMC_LMC}, including adaptations of different mass bins for their selection. The subhalo IDs of the 6~MW/M31-like hosts with both an SMC- and an LMC-like satellite read: 416713, 430864, 497557, 503437, 511303 and 514829 with host halo masses of $\rmn{log_{10}} ~M_\rmn{200c}/\rmn{M}_\odot = 12.6, 12.3, 12.0, 12.2, 12.1, 12.1$, respectively. These systems, as well as the infall history, spatial distribution, and star formation activity of their massive satellites will be addressed in future studies. However, as the data of TNG50 is publicly available, we provide these IDs for anyone who is interested in studying specifically these systems. In practice, the MW/M31-like hosts containing both an SMC- and an LMC-like satellite (assuming a mass bin of $\pm 0.1~\rmn{dex}$) cover a mass range of $M_\rmn{200c} = 10^{12} - 10^{12.6}~\rmn{M}_\odot$ with a median host mass of $10^{12.2}~\rmn{M}_\odot$, while the TNG50 MW/M31 analogues with an LMC-like galaxy as their most massive satellite have a median host mass of $10^{12.3}~\rmn{M}_\odot$. Whereas both of these mass ranges include either a single or two hosts outside of the $10^\rmn{th}$ and $90^\rmn{th}$ percentiles of the host mass range of all TNG50 MW/M31-like galaxies, they fall well within the range of total halo mass estimates of either the Galaxy or Andromeda (see Figure~\ref{fig:hostProps}). While both the median and the range of host masses for TNG50 MW/M31-like galaxies hosting massive, Magellanic Cloud-like satellites agree with previous host mass estimates from \cite{Busha2011b}, \cite{Cautun2014} predict less massive MW-like haloes. However, these are still consistent with some of the individual TNG50 MW/M31-like hosts with massive satellites. Furthermore, it should be noted that both of these works employ DM-only simulations as opposed to TNG50, which includes baryons. Their host mass estimates originate from measurements of maximum circular velocities of Magellanic Cloud-like subhaloes. We find the same host mass estimates when we increase the uncertainties for the stellar masses of SMC- and LMC-like satellites. With median host masses of $10^{12.3}~\rmn{M}_\odot$, these MW/M31-like systems lie between the median and the $90^\rmn{th}$ percentile of our complete sample of TNG50 MW/M31-like analogues.

Table~\ref{tab:SMC_LMC} shows also the number of TNG50 hosts around which satellites like M32 or NGC205 (with a stellar mass of $3.8\times10^8~\MSUN$), and/or like Triangulum (with $3.5\times10^9~\MSUN$ in stars) orbit \citep{McConnachie2018}. However, the stellar mass functions of some of our TNG50 systems in Figure~\ref{fig:MFTNG50_Nsat_hist} extend even further beyond massive satellites such as the LMC or Triangulum. With stellar masses of more than $10^{10}~\MSUN$, they are almost as massive as the host galaxies themselves. These systems do not represent actual satellites but correspond to galaxies that are about to merge with the MW/M31-like host, representing imminent major mergers (yet, by selection, with $M_* < 10^{10.5}~\MSUN$ at $z=0$).

\subsubsection{TNG50 satellite systems most similar to the MW and M31}
We conclude this overview of TNG50 MW/M31-like galaxies by visually contrasting MW- and M31-like satellite systems in TNG50 in Figures~\ref{fig:vis1} and~\ref{fig:vis2}. Here, we show the projected stellar mass density of the 10 TNG50 galaxies that are most similar to either the MW or Andromeda, respectively. These hosts were selected by computing the residual sum of squares between the simulated and observed satellite mass functions above $5\times10^6~\MSUN$ over all satellite stellar mass bins and by identifying the systems with the lowest values. The subhalo IDs of MW analogues read: 555013, 517271, 536654, 513845, 574037, 482155, 515296, 526029, 499704, 504559. Those of M31 analogues read: 458470, 433289, 490814, 474008, 342447, 471248, 429471, 470345, 436932, 438148.  Evidently, M31-like TNG50 systems are richer in more massive satellites than MW TNG50 analogues and exhibit more luminous, as well as extended stellar haloes. As we will explicitly demonstrate in Section~\ref{sec:hostProps}, such a difference is to a first order related to Andromeda being a more massive galaxy than the MW, possibly also residing in a more massive host halo. As there are apparent qualitative similarities between our TNG50 M31 analogues and observations of M31 and its satellite system from the Pan-Andromeda Archaeological Survey \citep[PAndAS,][]{Martin2013, Ibata2014}, future quantitative comparisons might be warranted.\\

\subsection{Comparisons to observations}

\subsubsection{Comparison to the SAGA survey}
\begin{figure*}
    \centering
    \includegraphics[width=.45\textwidth]{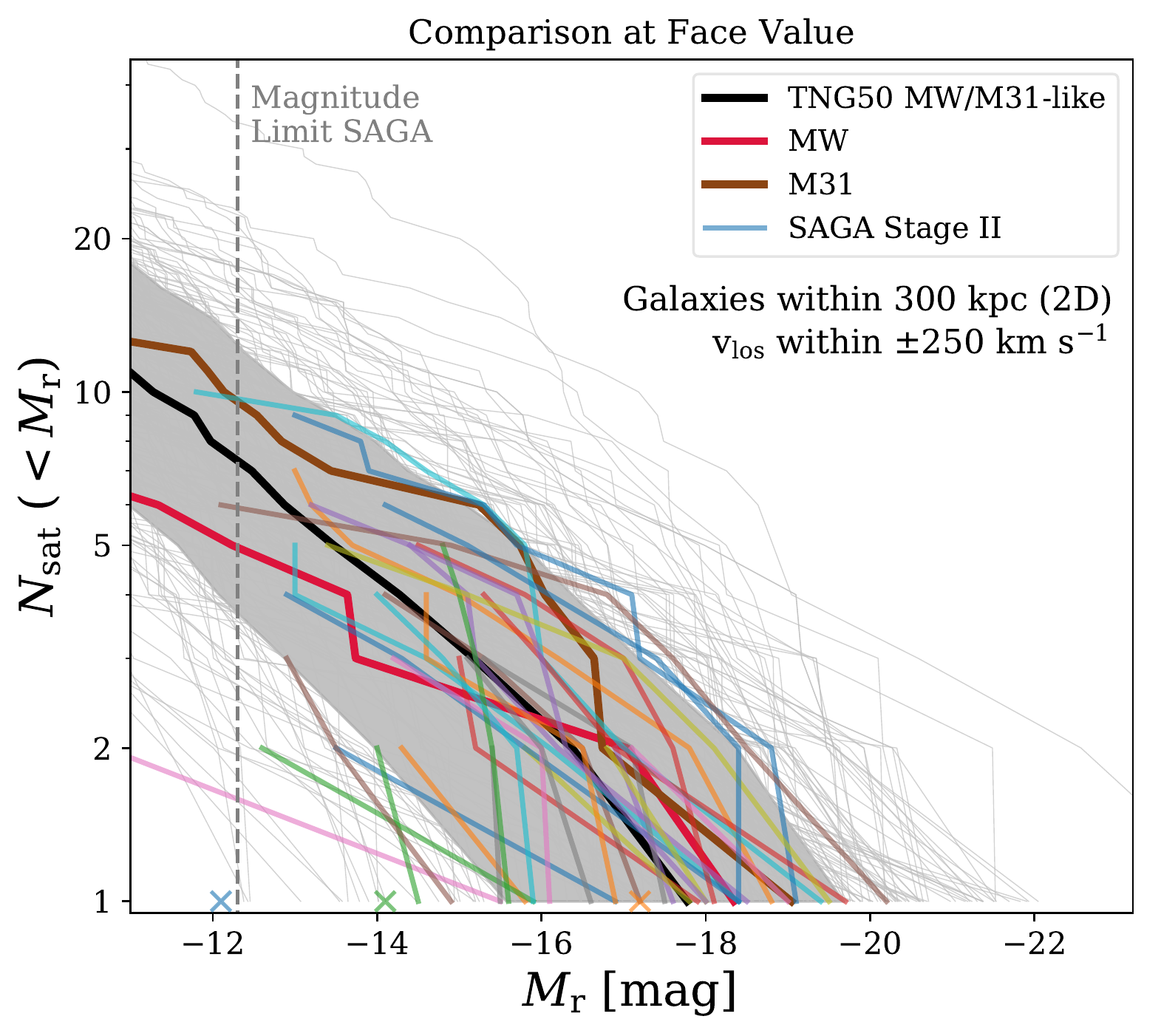} 
    \includegraphics[width=.45\textwidth]{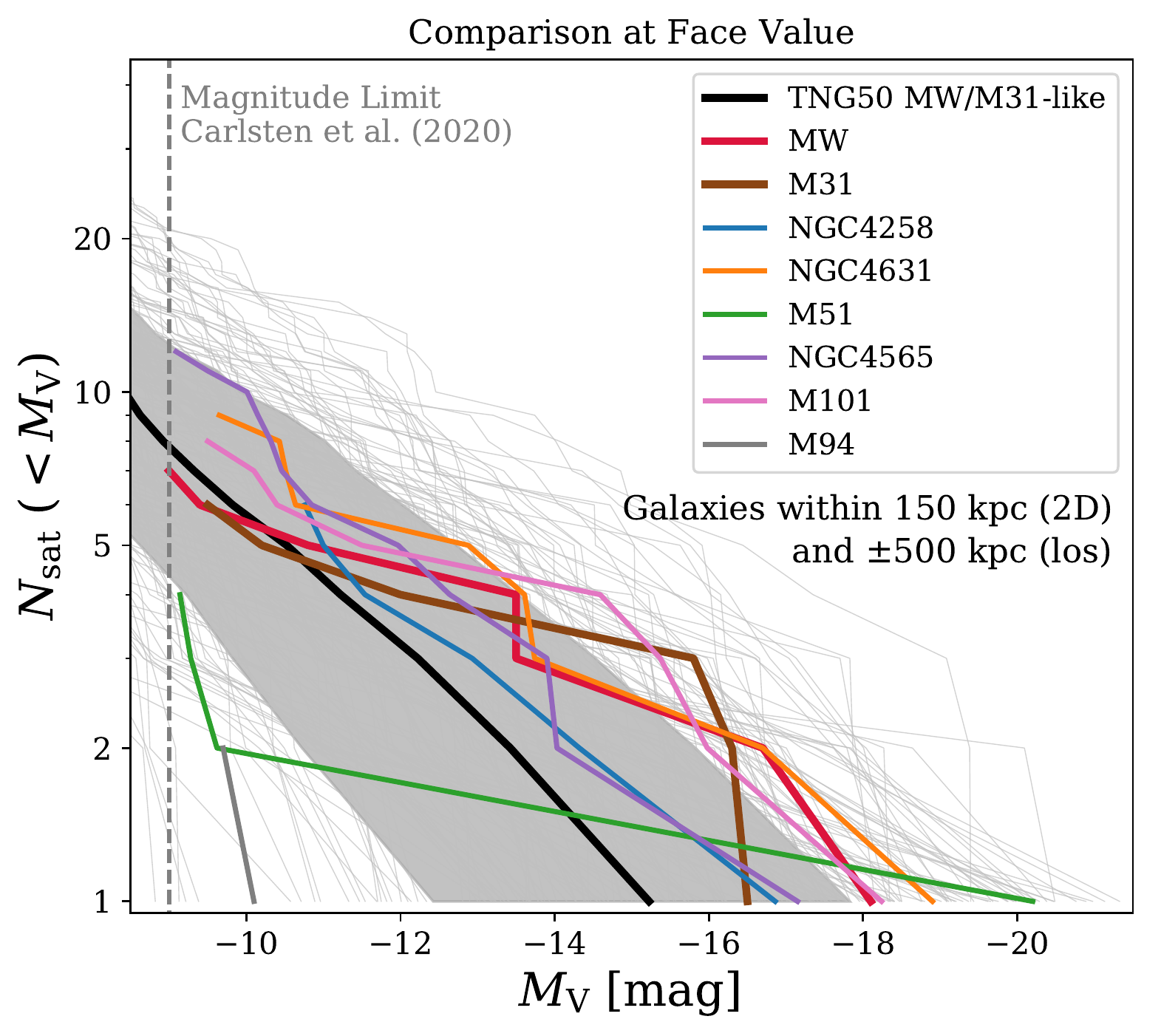}\break
    \includegraphics[width=.45\textwidth]{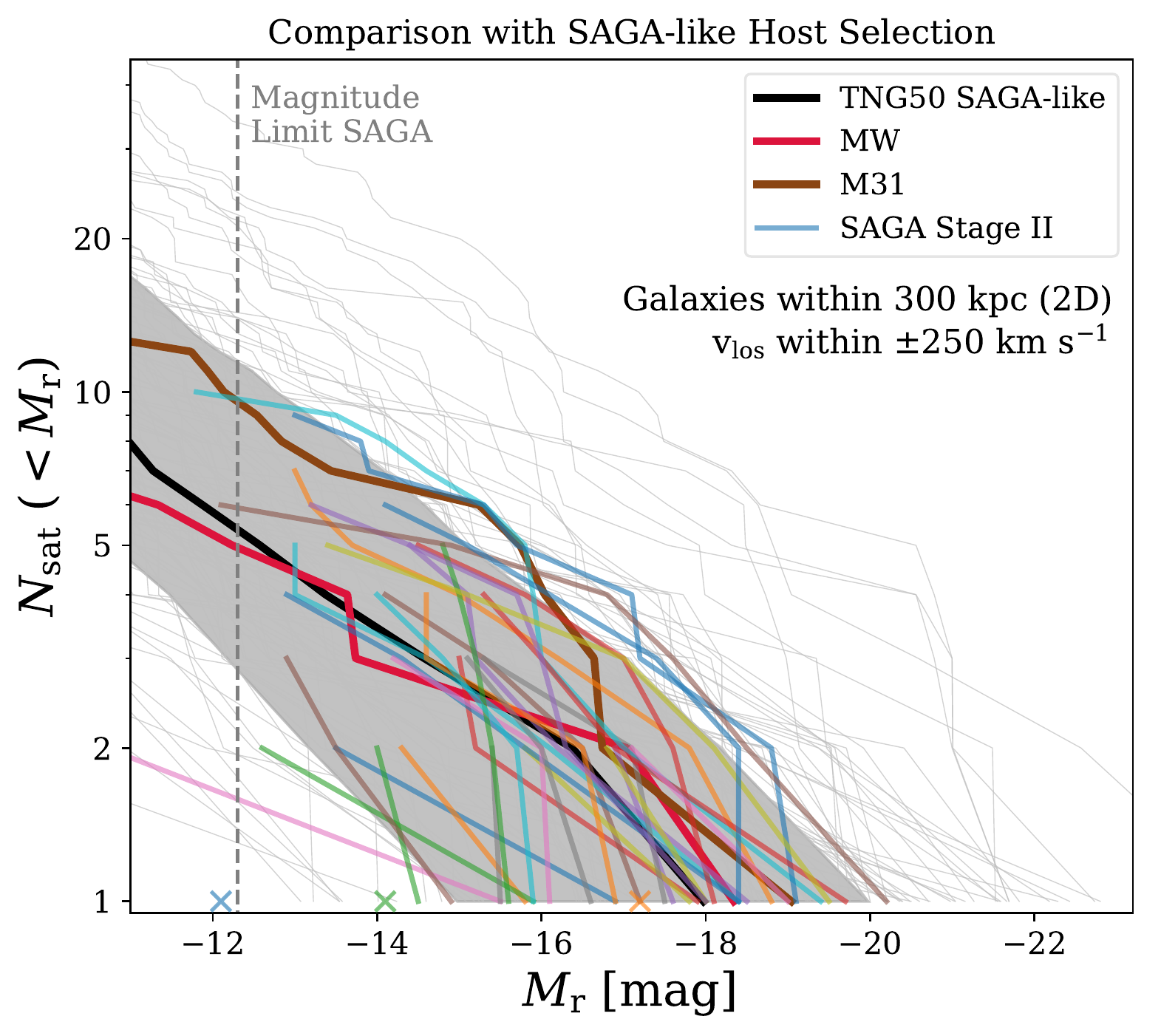} 
    \includegraphics[width=.45\textwidth]{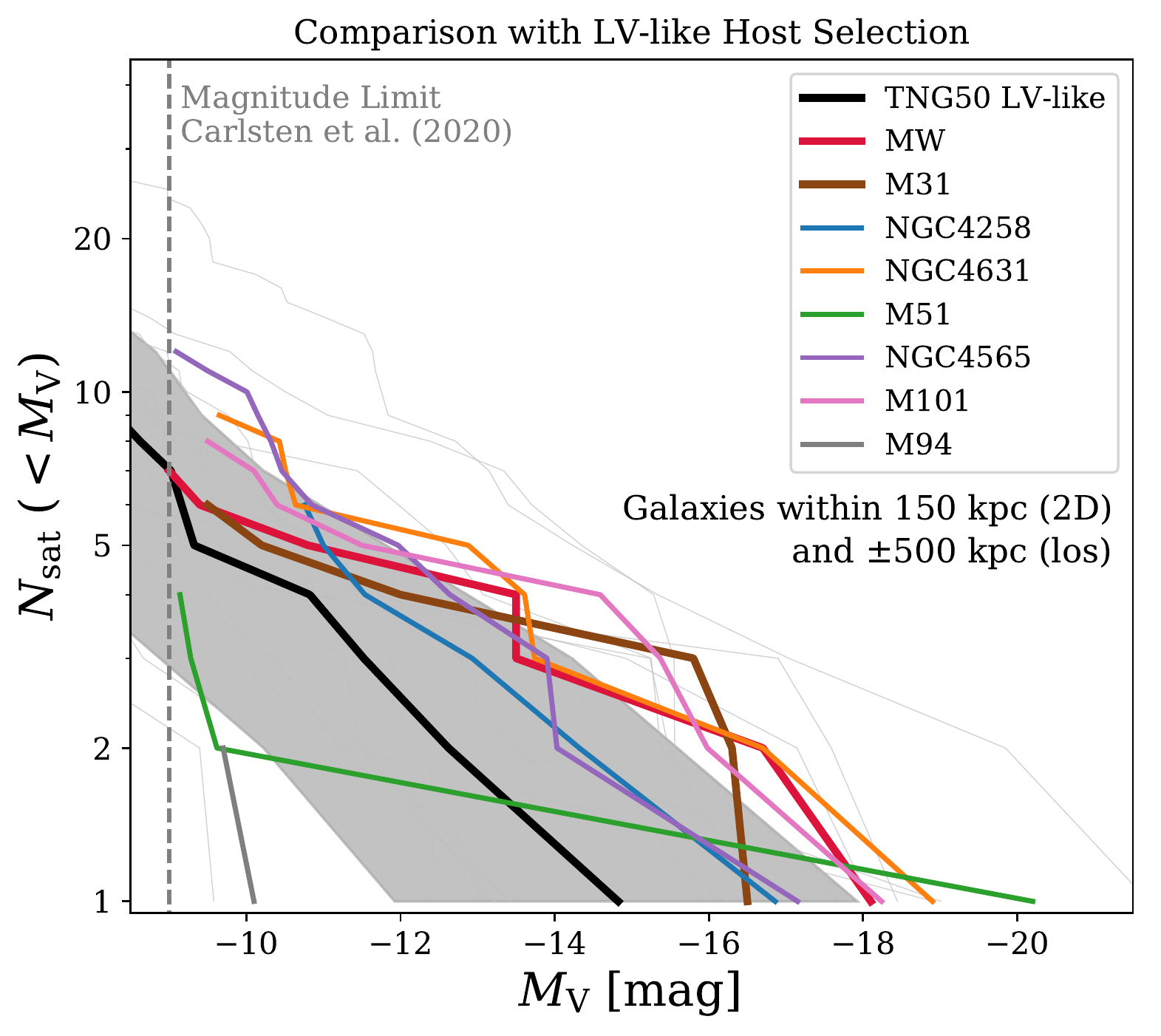}
    \caption{{\bf Satellite abundances in TNG50 and in observations from the SAGA survey and Local Volume galaxies.} In all four panels, the thin, grey curves in the background illustrate the satellite systems of individual TNG50 hosts, while the thick, black curves and grey shaded areas depict their median and scatter as $16^\rmn{th}$ and $84^\rmn{th}$ percentiles, computed in bins of satellite luminosity and including galaxies with zero satellites in the calculations. The top panels depict comparisons at face value, i.e. we compare the satellite systems of observed hosts to those of our fiducial selection of TNG50 MW/M31-like hosts. In the bottom panels, we select TNG50 hosts according to the selection criteria of the observations; furthermore, for each observed host, we choose and use only the three TNG50 hosts with closest $K$-band luminosity (see Section~\ref{sec:sample_mw} for details). \textit{Left panels:} satellite abundances in TNG50 and the SAGA survey. We characterise satellite systems using satellite luminosity functions in the $r$-band for galaxies within a projected aperture of $300~\rmn{kpc}$ and line-of-sight velocities of $\pm 250~\rmn{km~s^{-1}}$ of their host galaxy (with no requirement on stellar mass). Thick, coloured curves depict the satellite systems of 34~observed MW-like galaxies from the SAGA survey's second stage \protect\citep[][two more of their MW analogues host no satellites whatsoever and are thus not shown here]{Mao2021}, red and brown curves show the observed satellite systems of the MW and M31 \protect\citep{McConnachie2012, McConnachie2018}. \textit{Right panels:} satellite luminosity function in the $V$-band comparing TNG50 to 8 hosts in the Local Volume from \protect\cite{Carlsten2021} (including the MW and M31) as thick, coloured curves. Satellites are restricted to the inner (projected) $150~\rmn{kpc}$ of their host systems and to $\pm 500~\rmn{kpc}$ along the line of sight (los).}
    \label{fig:satLF_obs}
\end{figure*}

We compare the TNG50 results to findings from the SAGA survey \citep{Geha2017, Mao2021} in the left panels of Figure~\ref{fig:satLF_obs}.

In the top left panel, we compare TNG50 and SAGA at face value, i.e. using our fiducial selection of TNG50 MW/M31-like hosts (as detailed in Section~\ref{sec:sample_fiducial}). In the bottom panel of Figure~\ref{fig:satLF_obs}, instead, we compare the results of the SAGA survey to the satellite abundances of TNG50 hosts matched to replicate the SAGA host selection (as described in Section~\ref{sec:sample_saga}). In both cases, we characterise satellite abundances in terms of their $r$-band luminosity and include 34 MW-like hosts from the second stage of the SAGA survey \citep{Mao2021}. Their two remaining MW analogues host no satellites whatsoever and are not shown here. The SAGA galaxies are located at a distance of $20-40~\rmn{Mpc}$ and were selected by assuming a MW-like halo mass of $0.6 - 2.7 \times 10^{12}~\rmn{M}_\odot$ and by using abundance matching to infer a $K$-band luminosity range of $-23 > M_\rmn{K} > -24.6$ as a proxy for stellar mass. Furthermore, SAGA hosts are required to be isolated in order to match the MW's large-scale environment. For these hosts, satellites are considered within a projected aperture of $300~\rmn{kpc}$ and a line-of-sight velocity of $\pm 250~\rmn{km~s^{-1}}$ with an absolute $r$-band magnitude of $M_\rmn{r} < -12.3$, which are depicted as thick, coloured curves in the top left panel of Figure~\ref{fig:satLF_obs}. The redshifts of all SAGA satellites have been spectroscopically confirmed.

Similar to Figure~\ref{fig:MFTNG50_Nsat_hist}, the satellite systems of individual TNG50 MW/M31-like hosts are shown in the background as thin, grey curves, while the thick, black curve corresponds to the median TNG50 satellite luminosity function and the grey shaded region denotes its scatter as $16^\rmn{th}$ and $84^\rmn{th}$ percentiles. For this comparison, we do not impose a minimum stellar mass to TNG50 satellites and rather match the satellite selection criteria of the SAGA survey (see Section~\ref{sec:sample_sats}). Furthermore, we include the satellite luminosity functions of both MW and M31 (within $300~\rmn{kpc}$) as an additional comparison (red and brown curves, respectively), with $M_\rmn{r}$ for MW/M31 satellites obtained from the $M_\rmn{V}$ values of \cite{McConnachie2012} and \citep{McConnachie2018} using the TNG50 $M_\rmn{r}$ vs. $M_\rmn{V}$ relation from Equation~\eqref{eq:Mr_vs_MV}. 

Overall, the top left panel of Figure~\ref{fig:satLF_obs} displays a remarkable agreement between the r-band satellite abundances of TNG50, the SAGA galaxies, as well as the MW and M31. Whereas this level of agreement may be coincidental -- as there the comparison is made at face value, i.e. without ensuring the compatibility of the hosts in the simulation and observed samples --, this is in fact confirmed by the results in the bottom panel of Figure~\ref{fig:satLF_obs}, where we employ a SAGA-like selection of TNG50 hosts (see Section~\ref{sec:sample_saga} for details). The satellite systems of the SAGA-like TNG50 hosts are well in agreement with the MW and M31, as well as the actual MW-like galaxies of the SAGA survey. All satellite luminosity functions of observed galaxies lie either within the $1\sigma$ scatter of TNG50 or are consistent with the satellite abundances of individual TNG50 hosts. Compared to our fiducial selection of TNG50 MW/M31-like hosts, the SAGA-like selection is slightly less satellite rich. Furthermore, they exhibit a slightly larger scatter both in terms of their brightest satellite and their total satellite abundance.

\subsubsection{Comparison to Local Volume hosts}
In the right panels of Figure~\ref{fig:satLF_obs}, we quantify the comparison between TNG50 and the satellite systems of hosts in the Local Volume (i.e. within $12~\rmn{Mpc}$), examined by \cite{Carlsten2021} using CFHT/MegaCam data. In the top panel, the comparison is ``at face value'', i.e. made in comparison to the TNG50 fiducial sample of MW/M31-like galaxies, whereas in the bottom panel we attempt to match the host selection of the \cite{Carlsten2021} sample, as detailed in Section~\ref{sec:sample_mw}. Namely, for each observed galaxy, we select three TNG50 central and disky  galaxies with the closest $K$-band luminosity to the observed ones (see Section~\ref{sec:sample_lv}).

The hosts of \cite{Carlsten2021} span a halo mass range of $0.8 - 3 \times 10^{12}~\rmn{M}_\odot$, similar to the $10^\rmn{th}$ and $90^\rmn{th}$ percentiles of our TNG50 host halo mass range. In order to ensure completeness of the observed satellite systems down to $M_\rmn{V} \sim -9$, they exclusively consider satellites within the inner $150~\rmn{kpc}$ (3D) of their host galaxies. The satellite galaxies' line-of-sight distances are estimated using either surface brightness fluctuations (SBF) or the tip of the red giant branch (TRGB). As in the observations, we only count TNG50 galaxies within a projected aperture of $150~\rmn{kpc}$ of their host as satellites and do not require a minimum stellar mass. Furthermore, we limit satellite galaxies to line-of-sight distances of $500~\rmn{kpc}$. \cite{Carlsten2021} include a comparison with satellite systems from TNG100 in their study and adopt this line-of-sight criterion as a compromise between SBF and TRGB distance estimates. 

In the right panels of Figure~\ref{fig:satLF_obs}, the six satellite systems from \cite{Carlsten2021}, as well as those of the MW and M31 are depicted as $V$-band luminosity functions by thick, coloured curves. The median and scatter of the TNG50 hosts (thin, grey curves) are shown as thick, black curve and grey shaded area. Both the top and bottom right panels of Figure~\ref{fig:satLF_obs} show that, whereas most observed Local Volume systems overall fall within the satellite abundances from TNG50 and lie largely within the TNG50 $1\sigma$ scatter, they are somewhat more concentrated on the satellite-richer side. Imposing an LV-like selection makes the TNG50 median and scatter to shift to slightly lower satellite abundances than those of the fiducial MW/M31-like selection (top vs. bottom panels).

Interestingly, the galaxies observed by \cite{Carlsten2021} exhibit a smaller host-to-host scatter for satellites in their host's inner regions, i.e. within 150 projected kpc, in comparison to satellite abundances measured across larger apertures: NGC4258, NGC4631, NGC4565, and M101, as well as the MW and M31, exhibit remarkably similar satellite abundances compared to their satellites counted across larger apertures (left vs. right top panels of Figure~\ref{fig:satLF_obs}). While the scatter between individual TNG50 MW/M31-like hosts seems to be slightly smaller compared to our fiducial satellite selection in the left panel of Figure~\ref{fig:MFTNG50_Nsat_hist}, further analysis of the spatial distribution of satellite galaxies -- especially comparing the inner and outer regions of TNG50 hosts -- will be addressed in future studies (\textcolor{blue}{Bose et al. in prep.}).

\begin{figure*}
    \centering
    \includegraphics[width=.45\textwidth]{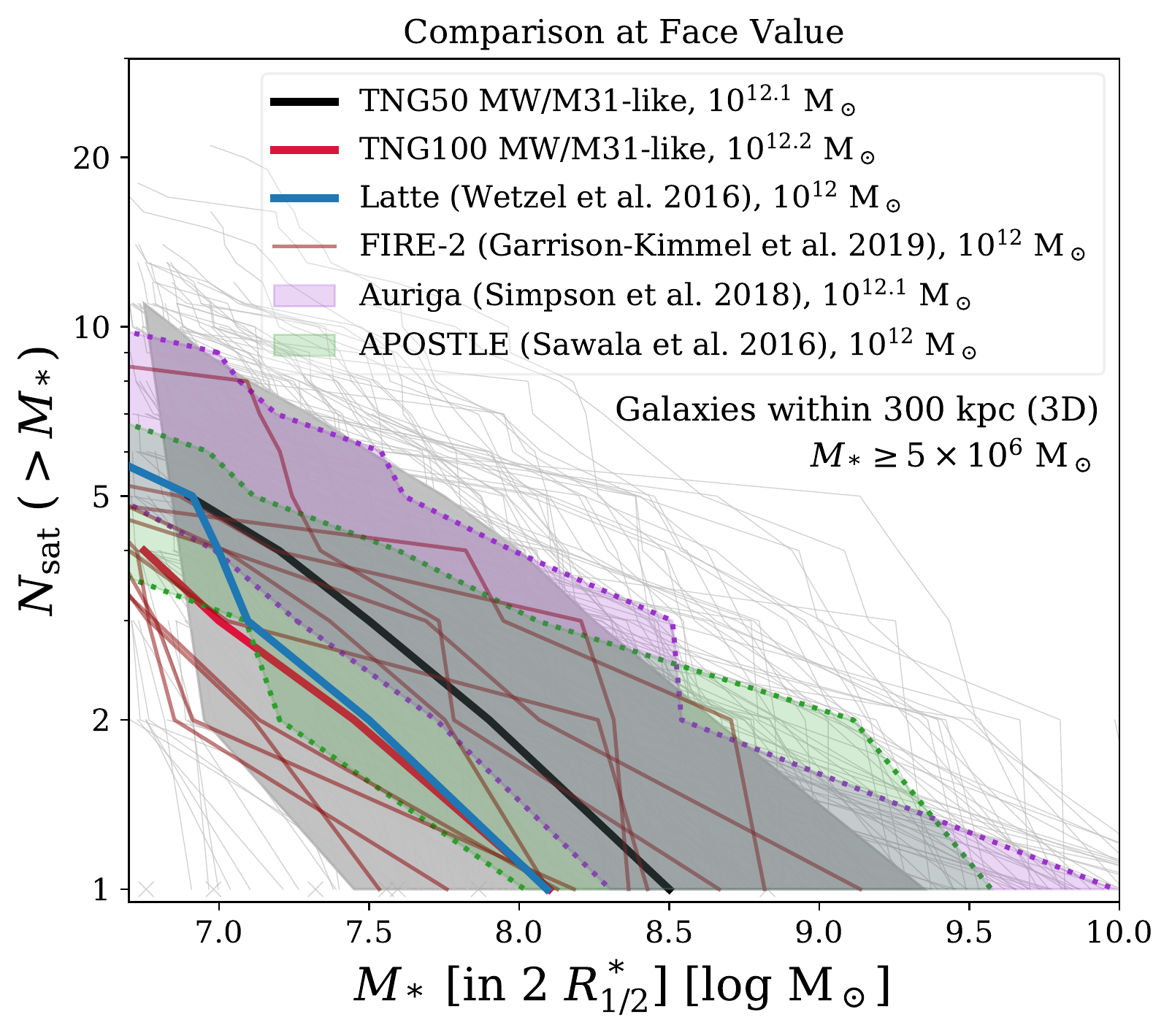}
    \includegraphics[width=.45\textwidth]{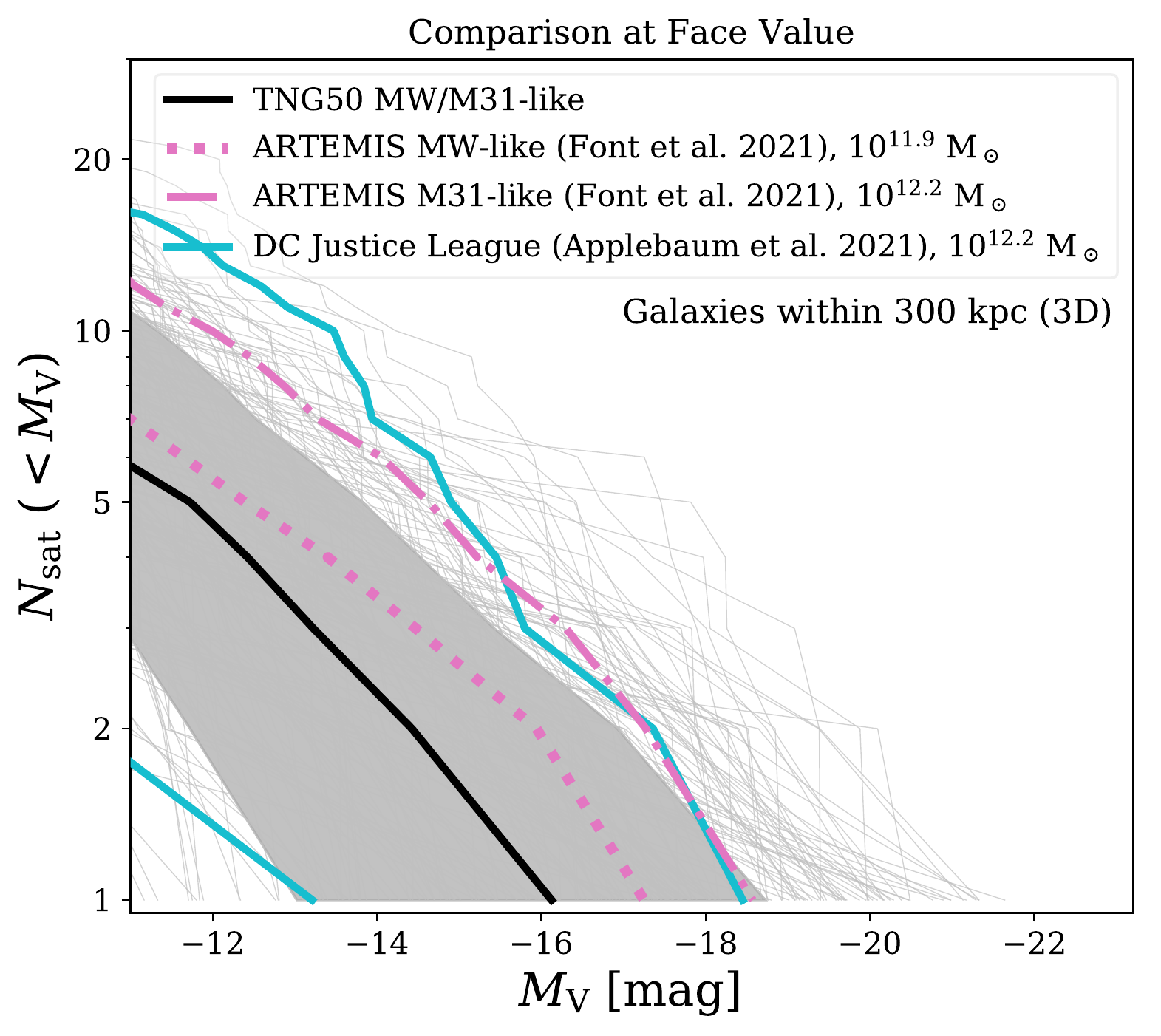}\break
    \includegraphics[width=.45\textwidth]{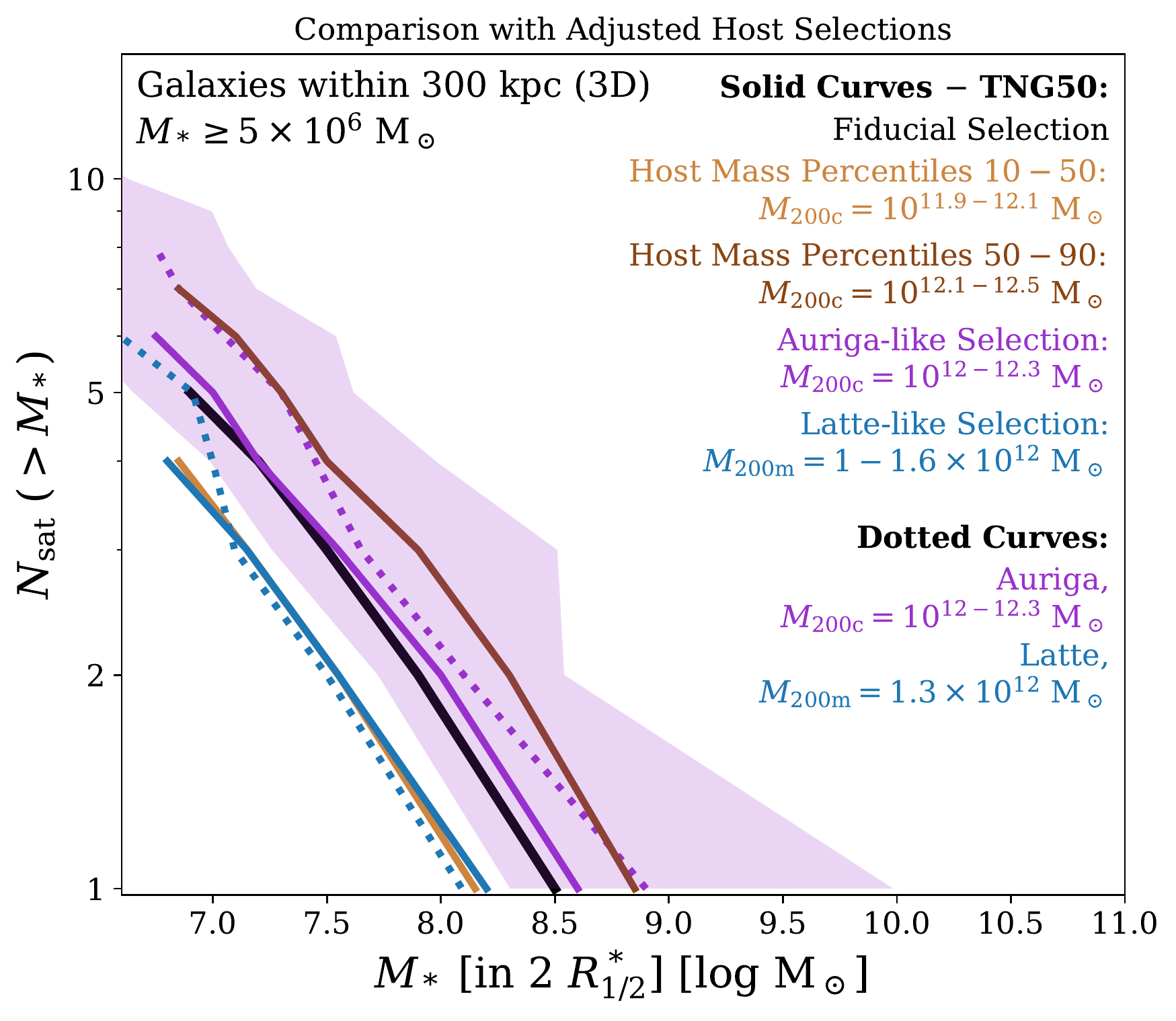}
    \caption{{\bf Satellite abundance in TNG50 and recent cosmological hydrodynamical simulations.} The median host halo mass or host mass range ($M_\rmn{200c}$) of each simulation, i.e. curve, is given in the legend. The top panels depict comparisons at face value, i.e. we compare the satellite systems of previous simulations to those of our fiducial selection of TNG50 MW/M31-like hosts. In both panels, the thin, grey curves in the background illustrate the satellite systems of individual TNG50 hosts, while the thick, black curves and grey shaded areas depict their median and scatter as $16^\rmn{th}$ and $84^\rmn{th}$ percentiles, computed in bins of satellite stellar mass and including galaxies with zero satellites in the calculations. \textit{Top left panel:} satellite stellar mass function comparing the satellite abundance of MW/M31-like galaxies in TNG50 to TNG100 (red curve), Latte \protect\citep[blue curve,][]{Wetzel2016}, FIRE-2 \protect\citep[brown curves,][]{Garrison-Kimmel2019}, Auriga \protect\citep[purple shaded area,][]{Simpson2018}, and APOSTLE \protect\citep[green shaded area,][]{Sawala2016b}. \textit{Top right panel:} satellite luminosity function in the $V$-band comparing TNG50 to the ARTEMIS \protect\citep[pink curves,][]{Font2021} and DC Justice League simulations \protect\citep[blue curves,][]{Applebaum2021}. The pink, dotted curve depicts the MW-like subsample of ARTEMIS, consisting of hosts with $M_\rmn{200c} < 10^{12}~\rmn{M}_\odot$, while the pink, dash-dotted curve shows their M31-like subsample with $M_\rmn{200c} > 1.2 \times 10^{12}~\rmn{M}_\odot$ (see Table~\ref{tab:hostSelec} for their general selection criteria). We include the median host halo masses of all simulations in terms of $M_\rmn{200c}$ in the legend. \textit{Bottom panel:} satellite stellar mass functions for various definitions of MW/M31-like hosts. TNG50 satellites are required to be located within a 3D aperture of $300~\rmn{kpc}$ of their host and to have a stellar mass of $M_* \geq 5 \times 10^6~\rmn{M}_\odot$. We compare our fiducial selection of MW/M31-like galaxies (black curve, as in Figure~\ref{fig:MFTNG50_Nsat_hist}) to several halo mass-based selections. The latter are limited to only central galaxies with no further requirements (e.g. on morphology or environment). Light and dark brown curves show satellite systems of lower- and higher-mass host haloes in TNG50. Their mass ranges correspond to the $10^\rmn{th}$ to $50^\rmn{th}$ percentiles, as well as the $50^\rmn{th}$ to $90^\rmn{th}$ percentiles of total halo mass covered by our fiducial host stellar mass range (see Figure~\ref{fig:hostProps}). Furthermore, we match the host mass range of TNG50 hosts to those of the Auriga (purple curve) and Latte simulations (blue curve). For Latte, we assume a range of $\pm 0.1~\rmn{dex}$ in halo mass (see Table~\ref{tab:hostSelec} for host selection criteria). The purple dotted curve and shaded region denote the median and scatter in satellite abundance of the actual Auriga simulations \protect\citep[][and private communication]{Simpson2018}; the blue, dotted curve shows the satellite system of Latte \protect\citep{Wetzel2016}.}
    \label{fig:satMF_sims}
\end{figure*}

\subsection{Comparison to previous cosmological hydrodynamical simulations}
\label{sec:prevSims}

We compare the satellite abundance of MW/M31-like galaxies in TNG50 to previous cosmological hydrodynamical galaxy simulations in the top panels of Figure~\ref{fig:satMF_sims}. Note that all of these simulations employ different definitions of MW- and/or M31-like galaxies with different host mass ranges on different mass types, as well as various other criteria on their morphology, environment, or merger history. Hence, as we find it useful to see all previous and current results in single plots, we consider such comparisons at face value. We summarise the host selection criteria of all simulations addressed in this study in Table~\ref{tab:hostSelec}. This means that although there is overlap between models with respect to their adopted mass range or some of their other criteria, all simulations study somewhat different kinds of host galaxies or samples with different distributions of galaxy or host mass. Therefore, some deviations concerning their individual or average satellite abundances are to be expected. Conversely, agreement across galaxy formation models cannot be over-interpreted as consistency, at least not before galaxy-to-galaxy variations and trends with host properties are accounted for.

The top left panel of Figure~\ref{fig:satMF_sims} compares the satellite stellar mass functions of TNG50 -- individual hosts as thin, grey curves in the background, as well as their median and scatter as thick, black curve and grey shaded area -- to the TNG100 median (red curve), the Latte simulation \citep[blue curve,][]{Wetzel2016}, FIRE-2 \citep[brown curves,][]{Garrison-Kimmel2019}, as well as the scatter of Auriga \citep[purple shaded area,][]{Simpson2018} and APOSTLE \citep[green shaded area,][]{Sawala2016b}. These comparisons are discussed here at face value, i.e. without adjusting for the different nominal selections of the underlying host properties -- we expand on the effects of different host selections in the bottom panel of Figure~\ref{fig:satMF_sims}.

MW/M31-like systems in TNG100 contain systematically both a lower number of satellites and overall less massive satellites -- an effect of decreased numerical resolution. We extensively discuss the impact of resolution on the abundance of both luminous satellite galaxies and dark subhaloes in more detail in Appendix~\ref{sec:resolution}. The Latte simulation mostly exhibits a lower satellite abundance than the TNG50 median -- albeit well within TNG50's $1\sigma$ scatter -- but rises to meet the TNG50 median at the low-mass end. While their most massive satellite is $\sim 0.4~\rmn{dex}$ less massive than the TNG50 average, Latte only simulates a single MW-like galaxy. However, satellites as massive as the LMC tend to occur relatively rarely \citep[e.g.][]{Busha2011a, Liu2011, Tollerud2011, Gonzalez2013} and it should be noted that \cite{Wetzel2016}, in their figure~3, exclude the LMC for its mass and Sagittarius due to its disruptive state from the reported MW satellite mass function in their comparison of Latte with the actual MW. While the two Local Group-like (LG) pairs and six isolated MW-like hosts of the FIRE-2 simulations exhibit significant scatter, their satellite abundance is overall consistent with MW/M31-like hosts in TNG50, well within TNG50's $1\sigma$ scatter. The overall abundances in Auriga extend to both larger numbers and more massive satellite galaxies than the TNG50 median -- we expand on this comparison in Section~\ref{sec:hostSelection}. The mass of their most massive satellites surpasses our $1\sigma$ scatter, however, it is still in agreement compared to individual TNG50 hosts. The LG-like hosts of APOSTLE exhibit similar trends and include a larger number of massive satellites. However, their satellite abundances are overall consistent with either TNG50's median and $1\sigma$ scatter or the individual TNG50 MW/M31-like hosts.

We show another set of $V$-band satellite luminosity functions in the bottom right panel of Figure~\ref{fig:satMF_sims} and compare TNG50 to recent results from the ARTEMIS \citep[pink curves,][]{Font2021} and DC Justice League simulations \citep[blue curves,][]{Applebaum2021}. While ARTEMIS consists of a sample of 45 hosts, we show the average abundances of ARTEMIS subsamples that were specifically matched to either the MW (dotted curve, $M_\rmn{200c} < 10^{12}~\rmn{M}_\odot$) or M31 (dash-dotted curve, $M_\rmn{200c} > 1.2 \times 10^{12}~\rmn{M}_\odot$). Both their selections exhibit consistently larger satellite abundances than the TNG50 median, with their MW-like selection within and their M31-like selection outside of TNG50's scatter. However, M31-like hosts in ARTEMIS are still in agreement with the overall satellite abundance of individual TNG50 analogue hosts. The curves from the DC Justice League simulations illustrate the satellite abundances of two individual MW-like galaxies. Both of them lie outside of TNG50's $1\sigma$ scatter -- one of them on the satellite-richer, the other on the satellite-poorer side -- but are consistent with the satellite abundance of individual TNG50 MW/M31-like hosts. These deviations are expected considering the masses of the DC Justice League host haloes. With $0.75$ and $2.4 \times 10^{12}~\rmn{M}_\odot$, their halo masses are close to the $10^\rmn{th}$ and $90^\rmn{th}$ percentiles of the halo mass range of TNG50 MW/M31-like hosts.
 
Overall, we find an encouraging level of apparent agreement with other cosmological simulations. The number of satellites above a given stellar mass is consistently within a factor of 2 of TNG50's median from $M_* = 10^7 - 10^{8.5}~\rmn{M}_\odot$. The satellite abundance of MW-, M31-, or LG-like hosts holds across different definitions of host galaxies, different physical models, and various levels of numerical resolution -- not just compared to TNG50 but among all simulations in general.
 
\subsubsection{Dependence on host selection}
\label{sec:hostSelection}

While satellite abundances of different simulations are consistent when compared at face value, the effects of specific host galaxy and halo selections on their present-day satellites remain to be explored. We vary the criteria for the selection of MW/M31-like galaxies in the bottom panel of Figure~\ref{fig:satMF_sims} and compare their satellite systems to our fiducial selection (black curve, see Section~\ref{sec:sample_mw} for its specific criteria). For all TNG50 hosts, we define satellites as galaxies with $M_* \geq 5 \times 10^6~\rmn{M}_\odot$ and within $300~\rmn{kpc}$~(3D) of their host galaxy. While we did also study a a subsample of our fiducial selection consisting of MW/M31-like hosts in the vicinity of a Virgo-like galaxy cluster (i.e. within $10~\rmn{Mpc}$), we do not show their satellite abundances here. Their overall distributions lie well within the range of our entire fiducial sample with no distinct concentration on the satellite-richer or -poorer side and their median would coincide with the satellite stellar mass function of our fiducial sample.

For the remaining curves in the bottom panel of Figure~\ref{fig:satMF_sims}, we adopt alternative selections: namely, we consider TNG50 hosts with a certain total host halo mass, we exclusively consider centrals, and waive any further limitations on their morphology, merger history, or environment. We define low- and high-mass samples based on the halo masses that our fiducial stellar mass range covers in Figure~\ref{fig:hostProps}: the low-mass sample covers the $10^\rmn{th}$ to $50^\rmn{th}$ percentile ($M_\rmn{200c} = 10^{11.9} - 10^{12.1}~\rmn{M}_\odot$, light brown curve), the high-mass sample corresponds to the $50^\rmn{th}$ to $90^\rmn{th}$ percentile of this range ($M_\rmn{200c} = 10^{12.1} - 10^{12.5}~\rmn{M}_\odot$, dark brown curve). Both curves display a distinct offset from our fiducial satellite stellar mass function: the low-mass sample hosts a smaller number of satellites, which are less massive, while the high-mass sample extends to both larger satellite masses and higher total abundances. 

Furthermore, we match the selection mass range to two previous cosmological simulations and show the corresponding TNG50 results: an Auriga-like selection (solid, purple curve) with $M_\rmn{200c} = 10^{12} - 10^{12.3}~\rmn{M}_\odot$ \citep{Grand2017} and Latte (solid, blue curve). Since the Latte simulation consists of only a single MW-like galaxy with $M_\rmn{200m} = 1.3 \times 10^{12}~\rmn{M}_\odot$ \citep{Wetzel2016}, we simply assume a mass bin of $\pm 0.1~\rmn{dex}$, resulting in a range of $M_\rmn{200m} = 1 - 1.6 \times 10^{12}~\rmn{M}_\odot$. $M_\rmn{200m}$ corresponds to the total mass of a sphere around the FoF halo's centre with a mean density of 200 times the \textit{mean} density of the Universe (as opposed to the Universe's \textit{critical} density for $M_\rmn{200c}$). The satellite abundances of the actual Latte simulation \citep{Wetzel2016} and the Auriga sample \citep{Simpson2018} are given as a blue, dotted curve and purple, dotted curve and shaded area, respectively. 

The TNG50 median satellite mass functions with the Auriga-like selection are very similar to the TNG50 median of our fiducial selection of MW/M31-like hosts, with a slight offset towards larger satellite abundances. Considering that the median host halo mass of our fiducial selection is at $M_\rmn{200c} = 10^{12.1}~\rmn{M}_\odot$, this result is reasonable. However, the median of the actual Auriga simulations displays slightly larger satellite abundances than our TNG50 Auriga-like selection and agrees more with our high-mass host sample. The Latte-like selection, on the other hand, returns TNG50 hosts whose satellite mass functions are in excellent agreement with that of the Latte simulation (blue solid vs. blue dotted curves), at least for satellite stellar masses above $10^7~\MSUN$. Thus, when the host selection is properly matched, the TNG50 and Latte models (i.e. FIRE models since Latte employs physical models from FIRE) predict essentially identical MW-like satellite mass functions, despite starkly different numerical resolution and galaxy formation model assumptions.

\subsection{Evolution of luminous and dark satellite populations through time}
\label{sec:res_satMF_acc}

\begin{figure*}
    \includegraphics[width=.55\textwidth]{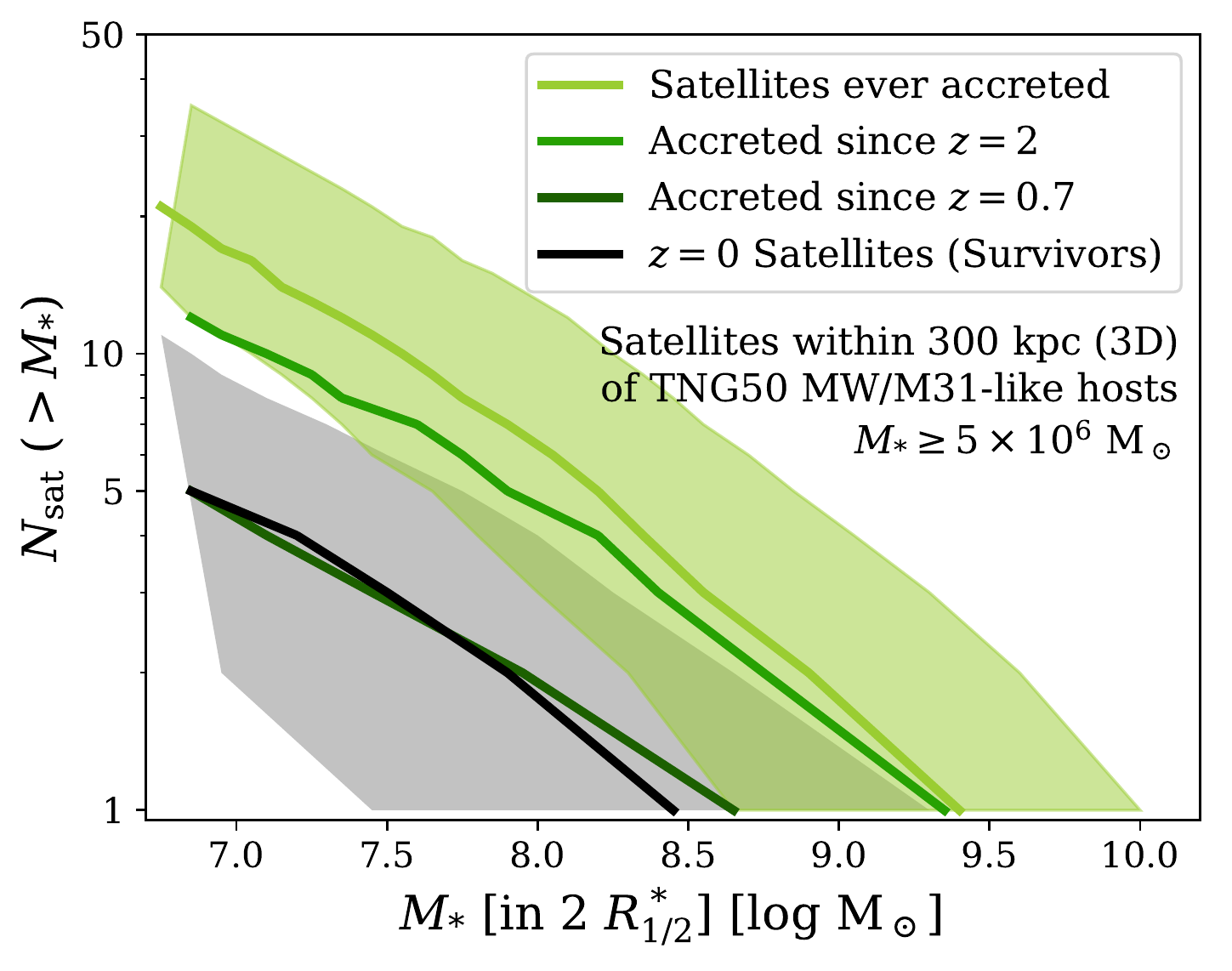}
	\includegraphics[width=.45\textwidth]{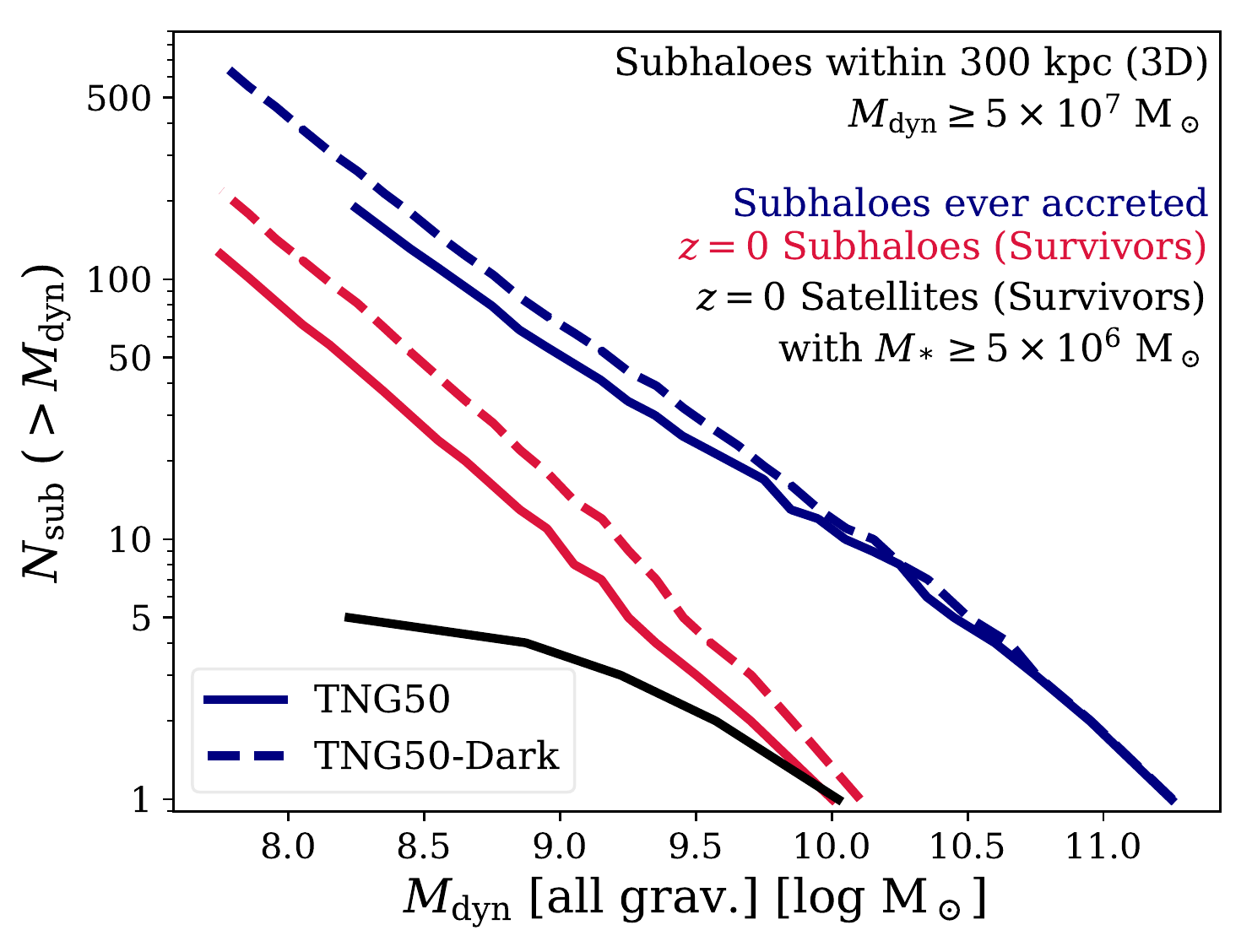} \hspace{.3cm}
	\includegraphics[width=.45\textwidth]{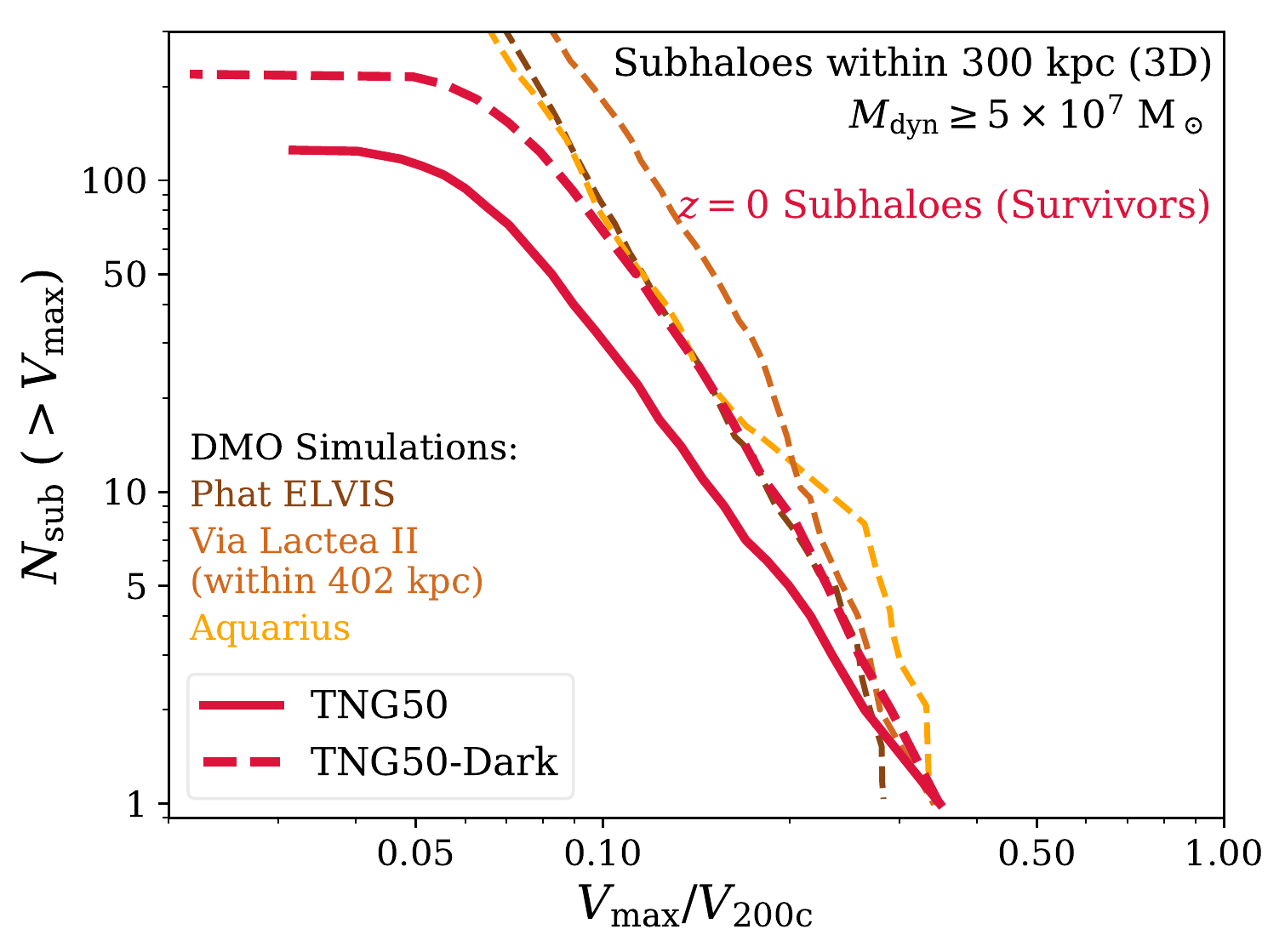}
    \caption{{\bf Satellite and subhalo abundance of MW/M31-like hosts in TNG50 comparing past to present-day satellite populations and baryonic to dark matter-only simulations}. \textit{Top panel:} median satellite stellar mass function (within two stellar half-mass radii $R_{1/2}^*$) for satellite galaxies within $300~\rmn{kpc}$ of their MW/M31-like host (3D) and with a stellar mass of at least $5 \times 10^6~\rmn{M}_\odot$ for present-day satellites at $z=0$ (black curve, as in Fig.~\ref{fig:MFTNG50_Nsat_hist}) and all satellites that have ever been accreted by their host (light green curve). Grey and green shaded areas denote their scatter as $16^\rmn{th}$ and $84^\rmn{th}$ percentiles. Furthermore, we limit satellites by their time of accretion to $z=2$ (medium green curve) and $z=0.7$ (dark green curve) to show the evolution of the median satellite population. \textit{Bottom panels:} median subhalo abundance for subhaloes with dynamical mass $M_\rmn{dyn} \geq 5 \times 10^7~\rmn{M}_\odot$ in terms of either total dynamical mass $M_\rmn{dyn}$ (left panel) or normalised maximum circular velocity $V_\rmn{max}/V_\rmn{200c}$ (right panel) using all gravitationally bound particles in both TNG50 (solid curves) and its dark matter-only analogue TNG50-Dark (dashed curves). Note that subhaloes are not required to include a luminous component. We compare all subhaloes that have ever been accreted by MW/M31-like hosts (blue curves), the surviving subhaloes at $z=0$ (red curves), as well as subhaloes that host luminous satellite galaxies with stellar masses of at least $5 \times 10^6~\rmn{M}_\odot$. Furthermore, we include the subhalo abundance of several previous DM-only simulations in terms of their maximum circular velocity normalised by their host virial velocity $V_\rmn{max}/V_\rmn{200c}$ in the right panel: Phat ELVIS \protect\citep[][]{Kelley2019}, Via Lactea II \protect\citep{Diemand2008}, and Aquarius \protect\citep{Springel2008} (brown to orange, dashed curves). As the Phat ELVIS and Via Lactea II simulations employ different definitions of host virial properties, we convert their host velocities to $V_\rmn{200c}$ using the underlying relationships within TNG50 and neglect deviations due to the different adopted cosmological parameters. While Phat ELVIS and Aquarius require their subhaloes to be located within $300~\rmn{kpc}$ of their host as well, this distance limit is extended to $402~\rmn{kpc}$ for Via Lactea II subhaloes. We summarise their host selection criteria in Table~\ref{tab:hostSelec}.}
    \label{fig:satMF_acc}
\end{figure*}

The satellites we observe today orbiting around galaxies like the MW and M31 do not represent the whole sample of galaxies that have ever been accreted. Most of the galaxies that enter the gravitational field of more massive haloes are destined to be destroyed and to ultimately form the diffuse stellar halo of their host galaxy \citep{Purcell2007, Sales2007, Fattahi2020}.

In this section, we compare present-day populations of satellite galaxies around MW/M31-like hosts in TNG50 to their abundance throughout cosmic time by taking all satellites ever accreted into account. This includes disrupted and merged satellites, backsplash galaxies, as well as present-day survivors.  
We compare $z=0$ and ever accreted median satellite stellar mass functions in the top panel of Figure~\ref{fig:satMF_acc}. TNG50 satellites are required to be located within $300~\rmn{kpc}$ (3D) of their host and to have a minimum stellar mass of $5 \times 10^6~\rmn{M}_\odot$. Note that, contrary to the rest of this paper, the definition of satellite galaxies in this section is not solely based on the distance from their host but additionally requires them to be members of the same FoF halo, since the identification of accreted satellites is based on their infall into a more massive host halo. Therefore, we limit our sample of MW/M31-like hosts to centrals, which leaves us with a sample of 190 hosts.

The black curve and grey scatter in the top panel of Figure~\ref{fig:satMF_acc} show the median satellite stellar mass function and its scatter as $16^\rmn{th}$ and $84^\rmn{th}$ percentiles of present-day, surviving satellites -- practically the median in the left panel of Figure~\ref{fig:MFTNG50_Nsat_hist} -- while the light green curve and the corresponding shaded area display median and scatter of the stellar mass function of all satellites that have ever been accreted by these MW/M31-like hosts using the satellites' stellar mass at infall. Including all satellites ever accreted extends the stellar mass function both towards more massive satellites of $10^{9.4 \pm 0.5}~\rmn{M}_\odot$, as well as towards larger total satellite abundances of $21^{+13}_{-7}$. At fixed minimum satellite stellar mass, the median abundance of all satellites ever accreted is larger than the median abundance of present-day survivors by a factor of $4-5$. This offset between surviving and accreted satellite populations is in qualitative agreement with previous DM-only and hydrodynamic simulations \citep{Purcell2007, Sales2007}. \cite{Fattahi2020} find a similar quantitative difference for satellites from the Auriga simulations.

Furthermore, we analyse the evolution of the satellite stellar mass function through time by limiting satellite galaxies according to their time of accretion either to satellites that have been accreted since $z=2$ (medium green curve) or those that were accreted onto their MW/M31-like host by $z=0.7$ (dark green curve). Overall, the number of satellites that have been accreted since $z=2$ decreases compared to all satellites ever accreted. While the massive end exhibits barely a difference, less massive satellites become increasingly affected since such galaxies are less resistant to environmental effects and are thus more prone to be disrupted. Shifting this limit on accretion time further towards the present day continues to decrease the number of satellites. In fact, the mass function of the present-day population of surviving satellite galaxies is similar to the one of satellites accreted since $z \sim 0.7$. So on average, present-day satellites of the MW and Andromeda fell into the gravitational potential of their host not earlier than $z\sim 0.7-1$.

\subsection{Baryonic vs. DM-only simulation expectations}
\label{sec:bary_vs_dmo}
We examine differences between luminous satellite and dark subhalo populations in the bottom panels of Figure~\ref{fig:satMF_acc}. The lower left panel shows a subhalo mass function in terms of their total dynamical mass $M_\rmn{dyn}$, i.e. all particles that are gravitationally bound to satellites, including dark matter, stars, gas, and black holes. Subhaloes are required to share the same FoF halo as their host, to be located within $300~\rmn{kpc}$ (3D) of their host galaxy, and to have a minimum dynamical mass of $5 \times 10^7~\rmn{M}_\odot$. This value corresponds to the smallest total subhalo mass below which the SHMR becomes incomplete and is artificially bent due to finite mass resolution (see Figure~\ref{fig:resTest_lumSats}, bottom left panel). Note that these subhaloes do not necessarily need to include a stellar component, meaning they can be either luminous or dark subhaloes. We compare the abundance of all subhaloes that have ever been accreted (blue curves) to the present-day population of surviving subhaloes at $z=0$ (red curves), as well as surviving satellite galaxies at $z=0$ with a stellar mass of $M_* \geq 5 \times 10^6~\rmn{M}_\odot$ (black curve). Furthermore, we illustrate differences between baryonic and DM-only simulations in the bottom panels of Figure~\ref{fig:satMF_acc},  by contrasting the subhalo samples from both TNG50 (solid curves), as well as its DM-only analogue simulation TNG50-Dark (dashed curves). 

While the abundance of subhaloes at $z=0$ is significantly larger than for luminous satellite galaxies, reaching 120 (200) in TNG50 (TNG50-Dark), surviving subhaloes and all subhaloes ever accreted exhibit a similar -- albeit slightly smaller -- difference as for satellite galaxies. At fixed dynamical mass, the number of surviving subhaloes at $z=0$ is smaller than those that have ever been accreted onto MW/M31-like hosts by a factor of $3-5$. The abundance of subhaloes at $z=0$ is always larger in TNG50-Dark than in the baryonic run by a factor of up to 2. However, this trend varies slightly when considering all subhaloes that have ever been accreted. Here, the abundance of massive subhaloes is the same in both TNG50 and TNG50-Dark. However, below dynamical masses of $10^{10}~\rmn{M}_\odot$, the number of subhaloes in TNG50-Dark becomes larger than the number of subhaloes in TNG50. The inclusion of baryonic processes affects the evolution of subhalo populations significantly both before and after accretion into their present-day host environment. Low-mass galaxies in baryonic simulations can experience substantial gas outflows due to galactic winds, leading to a redistribution of dark matter and lower masses compared to DM-only simulations. Intermediate-mass galaxies, on the other hand, have deeper potential wells and are therefore able to accrete more gas, resulting in larger masses in baryonic simulations \citep{Chua2017}. Furthermore, mass stripping and the survivability of subhaloes after infall correlates with the underlying structure of both subhalo and host. While a steeper matter density in subhaloes makes them more resistant to environmental effects, mass loss and total disruption, the same feature makes host haloes more efficient at tidal stripping, decreasing the survivability of subhaloes \citep{Jiang2016}. Introducing baryonic processes therefore changes both the overall mass range of subhaloes, as well as their survivability inside their host environment, resulting in different subhalo mass functions -- both at accretion and at $z=0$. Since DM-only simulations are missing complex and non-negligible physical processes, near-field cosmology as well as the analysis of MW/M31-like satellite systems and their evolution must rely on baryonic simulations.

We find similar trends for the abundance of subhaloes with $M_\rmn{dyn} \geq 5 \times 10^7~\rmn{M}_\odot$ in terms of their maximum circular velocity $V_\rmn{max}$ in the bottom right panel of Figure~\ref{fig:satMF_acc}. As the subhalo abundance in terms of $V_\rmn{max}$ is expected to be essentially scale free when normalised to their host (\citealt{Wang2012}, however, see \citealt{Chua2017} for the breaking of self-similarity in full-physics hydrodynamical simulations), we normalise the subhalo abundance by the virial velocity of their host halo $V_\rmn{200c}$. The number of surviving subhaloes at $z=0$ (red curves) reaches 120 in TNG50 (solid curve) and 200 in TNG50-Dark (dashed curves) for $V_\rmn{max}/V_\rmn{200c} \geq 0.02-0.07$, while their overall abundance is always larger in TNG50-Dark. Subhalo populations extend to normalised maximum circular velocities of $\sim 0.35$ in both TNG50 and TNG50-Dark. However, their distributions begin to flatten considerably towards the low-velocity end for subhaloes with $V_\rmn{max}/V_\rmn{200c} < 0.07$. 

Furthermore, we compare our findings from TNG50-Dark (red dashed curve) to other DM-only simulations: Phat ELVIS \citep{Kelley2019}, Via Lactea II \citep{Diemand2008}, and Aquarius \citep{Springel2008} (brown to orange, dashed curves). All of the simulations employ different definitions of MW/M31-like haloes. We summarise their host selection criteria in Table~\ref{tab:hostSelec}. As both Phat ELVIS and Via Lactea II employ different measurements of their host haloes' virial properties -- $V_{\Delta_\rmn{c}}$ (i.e. the total velocity of a sphere with a mean density of $\Delta_\rmn{c}$ times the critical density of the Universe, where $\Delta_\rmn{c}$ is derived from the collapse of a spherical top-hat perturbation), and $V_\rmn{200m}$ (i.e. the velocity of a sphere around the FoF halo centre with a mean density of 200 times the mean density of the Universe), respectively --, we convert their host velocities into $V_\rmn{200c}$ using the TNG50 relations of these different velocity measurements. Note that the actual $V_\rmn{200c}$ velocities of these simulations might be slightly different depending on their adopted cosmology.
%

%
Both Phat ELVIS and Aquarius require satellites to be located within $300~\rmn{kpc}$ of their host galaxies, however, this aperture is extended in Via Lactea II to $402~\rmn{kpc}$. After normalising the subhaloes' maximum rotational velocities to their respective host velocity, all simulations exhibit overall consistent subhalo abundances. While the abundance of Via Lactea II is slightly larger than in TNG50-Dark, the abundances of Phat ELVIS and Aquarius coincide with TNG50-Dark at almost all velocities. Despite different definitions of MW/M31-like haloes -- including limitations on mass ranges, morphologies, isolation criteria, and merger histories -- we find a reasonable agreement and consistent subhalo abundances between TNG50-Dark and other, previous DM-only simulations.

\subsection{Dependence on host properties}
\label{sec:hostProps}

\begin{figure*}
    \centering
	\includegraphics[width=.81\textwidth]{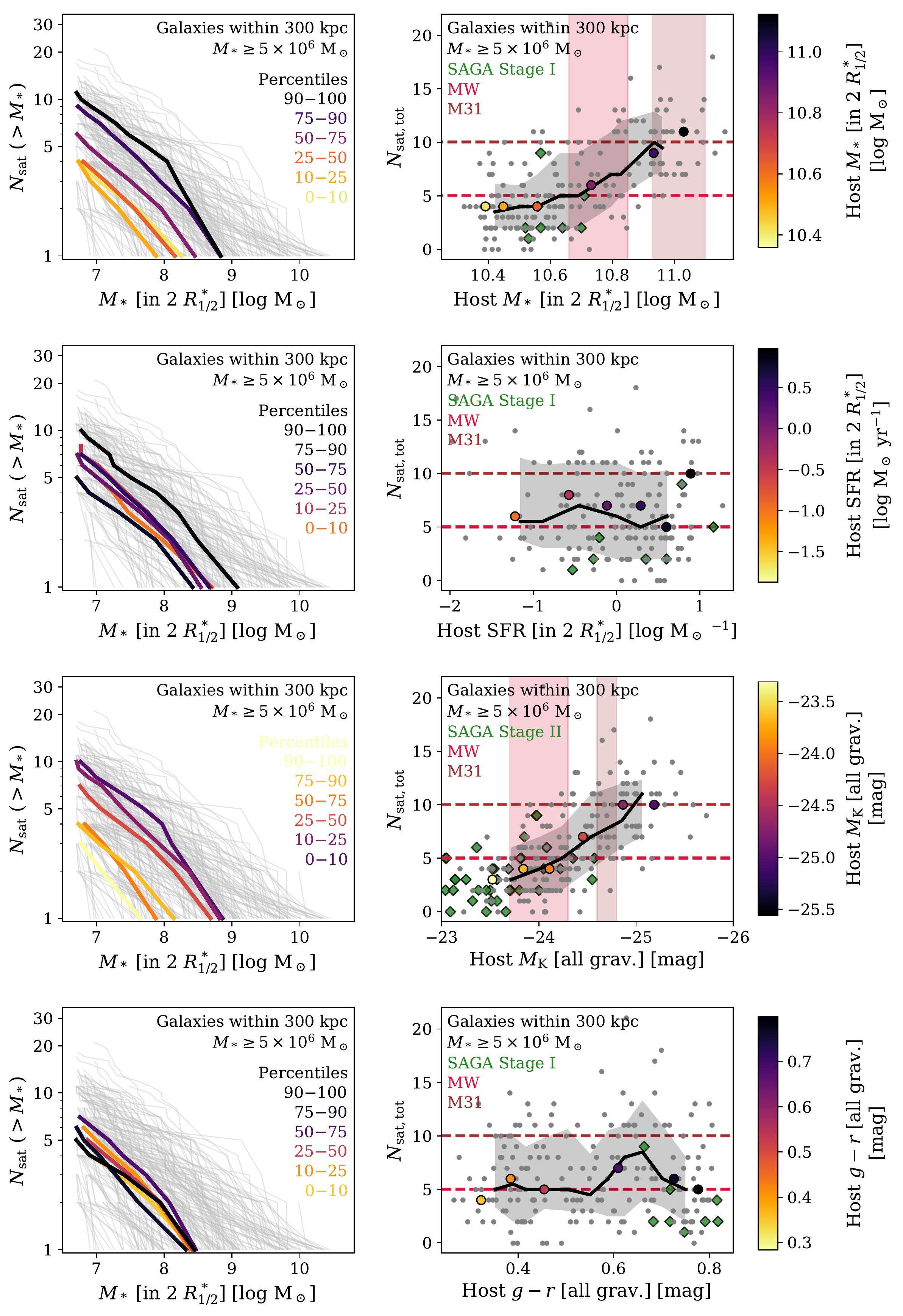}
    \caption{{\bf Dependence of satellite abundances on host galaxy properties} for satellites within $300~\rmn{kpc}$ (3D) of their MW/M31-like host and with a stellar mass of at least $5 \times 10^6~\rmn{M}_\odot$. Each row investigates a different host property (from top to bottom): stellar mass $M_*$, star formation rate SFR (both within twice the stellar half-mass radius), $K$-band luminosity $M_\rmn{K}$, and $g-r$ colour. \textit{Left panels:} median satellite stellar mass functions in various percentiles of the host property in question (thick, yellow to black curves). The thin, grey curves in the background denote satellite stellar mass functions of individual TNG50 MW/M31-like hosts as a reference. \textit{Right panels:} total number of satellites as a function host properties for the percentiles (yellow to black circles), all TNG50 MW/M31-like galaxies (grey circles), as well as their running median (black curves) and scatter (grey shaded area, $16^\rmn{th}$ and $84^\rmn{th}$ percentiles). Furthermore, we include MW-like hosts from the SAGA survey as a comparison (green diamonds): its first stage \protect\citep{Geha2017} consisting of 8 MW-like hosts for stellar mass, SFR and $g-r$ colour, as well as 36 hosts from its second stage \protect\citep{Mao2021} for $K$-band luminosity. The horizontal, dashed lines mark the total satellite abundance of the MW (red line) and M31 (brown line), while the vertical, shaded areas denote estimates of stellar mass \citep{Licquia2015, Sick2015, Boardman2020} and $K$-band luminosity \citep{Drimmel2001, Hammer2007}.}
    \label{fig:satMF_galProps}
\end{figure*}

\begin{figure*}
    \centering
	\includegraphics[width=.81\textwidth]{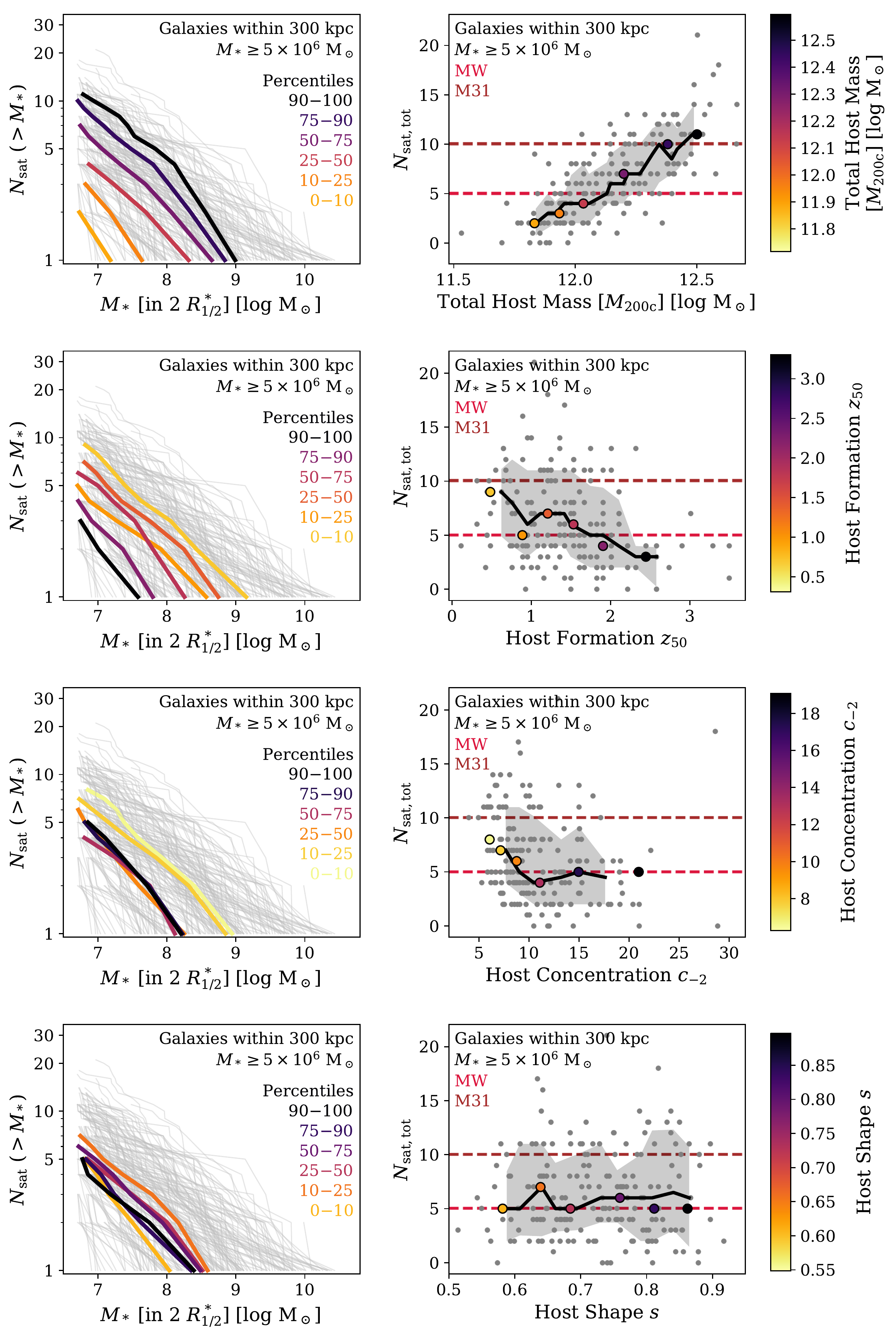}
    \caption{{\bf Dependence of satellite abundances on host halo properties (instead of host galaxy properties)} for satellites within $300~\rmn{kpc}$ (3D) of their MW/M31-like host and with a stellar mass of at least $5 \times 10^6~\rmn{M}_\odot$. Each row investigates a different host property (from top to bottom): total mass $M_\rmn{200c}$, halo assembly time $z_\rmn{50}$, i.e. the redshift at which the MW/M31-like host had assembled 50 per cent of its mass, as well as halo concentration $c_{-2}$ and halo shape as minor-to-major axis ratio $s$. \textit{Left panels:} median satellite stellar mass functions in various percentiles of the host property in question (thick, yellow to black curves). The thin, grey curves in the background denote satellite stellar mass functions of individual TNG50 MW/M31-like hosts as a reference. \textit{Right panels:} total number of satellites as a function host properties for the percentiles (yellow to black circles), all TNG50 MW/M31-like galaxies (grey circles), as well as their running median (black curves) and scatter (grey shaded area, $16^\rmn{th}$ and $84^\rmn{th}$ percentiles). The horizontal, dashed lines mark the total satellite abundance of the MW (red line) and M31 (brown line).}
    \label{fig:satMF_haloProps}
\end{figure*}

In this section, we investigate the correlations of satellite abundance with properties of their host and look for possible physical origins of the scatter in satellite abundance. The large number of MW/M31-like galaxies in TNG50 allows us to search for trends in a statistically significant manner and hence to provide the baryonic physics counterpart to previous results on subhaloes that were based on DM-only simulations \citep[e.g.][and subsequent similar analyses]{Gao2004} or semi-analytic models \citep[e.g.][]{Sales2013, Starkenburg2013, WangWhite2012}. We characterise satellite systems using their stellar mass functions (limited to satellites within $300~\rmn{kpc}$ of their host and with $M_* \geq 5 \times 10^6~\rmn{M}_\odot$) in general, as well as in terms of the total number of satellites within $300~\rmn{kpc}$.

\subsubsection{Dependence on host galaxy properties}
We illustrate dependencies of satellite abundances on host galaxy properties in Figure~\ref{fig:satMF_galProps}. Panels in the left column depict satellite stellar mass functions for subpopulations in various percentiles of the host property in question, with yellow to black curves. The specific colours of percentiles vary slightly depending on the distribution of the respective host property under consideration. The grey curves in the background show the individual stellar mass functions as reference. Right panels display the total number of satellites as a function of the respective host property for individual hosts and percentiles, as well as their running median and scatter (black curve and grey shaded area). These curves do not necessarily cover the entire range of the host property in question. Since the population of satellites might not be complete around the boundaries of the respective host property range, the running median and scatter could otherwise display misleading trends. Furthermore, we include either the 8 hosts of the SAGA survey's first stage \citep{Geha2017} or the 36 hosts of its second stage \citep{Mao2021} as comparison (green diamonds) and indicate the total abundance of MW and M31 satellites with $M_* \geq 5 \times 10^6~\rmn{M}_\odot$ (red and brown dashed lines, respectively). We characterise our hosts using stellar mass, star formation rate, $K$-band luminosity, and $g-r$ colour (from top to bottom). For stellar mass and $K$-band luminosity, we include estimates for the MW and M31 (red and brown shaded areas, respectively).

In Figure~\ref{fig:satMF_galProps}, while there is overlap among the stellar mass functions of different percentiles and significant scatter in the total number of satellites as a function of host properties, we find the clearest trends with stellar mass and its observational counterpart, the $K$-band luminosity. Namely, more massive and brighter MW/M31-like galaxies do host a larger number of satellite galaxies. Furthermore, the strength of these correlations becomes more significant at higher masses and brighter luminosities. At fixed host stellar mass, there is a \textit{normalised} $1\sigma$ scatter in total satellite abundance (calculated as $|N_\rmn{sat,~tot} - \langle N_\rmn{sat,~tot} \rangle| / \langle N_\rmn{sat,~tot} \rangle$) of up to $\pm 0.5$ at $10^{10.5}~\MSUN$ and $\pm 0.6$ at $10^{10.8}~\MSUN$; at fixed host $K$-band luminosity, there is a normalised scatter of up to $\pm 0.8$ at $-24~\rmn{mag}$ and $\pm 0.5$ at $-24.8~\rmn{mag}$. While we did inspect correlations with the stellar mass function slope as well, we find either no or only very minor dependencies, e.g. the satellite stellar mass functions of more massive hosts tend to be slightly steeper. Considering the comparison to observations, we see that the TNG50 trends are well in agreement with those from the SAGA hosts -- both exhibit a significant amount of scatter. Furthermore, \cite{Mao2021} recover the same significant correlation of satellite abundance and host $K$-band luminosity. Considering the MW and M31, we find their total satellite abundances to be consistent with the distributions of TNG50 MW/M31-like hosts given their estimates of stellar mass and $K$-band luminosity and amid a significant degree of scatter in both directions. Clearly, the observed satellite system of Andromeda is more compatible to those of TNG50 hosts with larger stellar masses and smaller K-band magnitudes, and it is in fact well reproduced by TNG50.

\subsubsection{Dependence on host halo properties}
We extend this investigation to host halo properties in Figure~\ref{fig:satMF_haloProps}. As in Figure~\ref{fig:satMF_galProps}, we illustrate trends with percentile stellar mass functions (left panels) and the total number of satellites as a function of host halo properties (right panels). Once more, it should be noted that the specific colours of percentiles vary slightly depending on the distribution of the respective host property in question. Since we consider properties of the FoF halo, we limit our sample of MW/M31-like hosts to centrals, which leaves us with a sample of 190 hosts. Furthermore, we indicate the total satellite abundances of the MW and M31 with $M_* \geq 5 \times 10^6~\rmn{M}_\odot$ (red and brown dashed lines, respectively).

We characterise the host halo by its total mass $M_\rmn{200c}$, its assembly time, for which we employ $z_\rmn{50}$ -- the redshift at which 50 per cent of its present-day total mass had been assembled --, its concentration $c_{-2}$, and its shape $s$ as its minor-to-major axis ratio (from top to bottom). We compute halo concentration by fitting an Einasto profile \citep{Einasto1965, Navarro2004} to the radial distribution of dark matter density $\rho_\rmn{DM}(r)$, following \textcolor{blue}{Pillepich et al. (in prep.)}. The concentration parameter $c_{-2}$ corresponds to the ratio of the virial radius and the radius at which the DM density profile's slope takes on an isothermal value. We measure DM halo shapes following \cite{Chua2019}.

Overall, we find much stronger trends with halo mass and assembly than with host galaxy properties: more massive MW/M31-like haloes and those with a later assembly $z_{50}$ host a larger number of surviving satellite galaxies by $z=0$. At fixed total host mass, there is a \textit{normalised} $1\sigma$ scatter in total satellite abundance (calculated as $|N_\rmn{sat,~tot} - \langle N_\rmn{sat,~tot} \rangle| / \langle N_\rmn{sat,~tot} \rangle$) of up to $\pm  0.8$ at $10^{12}~\MSUN$ and $\pm 0.3$ at $10^{12.4}~\MSUN$. However, the trend with host assembly evolves throughout time with different stages in halo formation: for earlier characterisations of the assembly times ($z_\rmn{10}$), the correlation with total satellite abundance is stronger than with later characterisations of halo assembly ($z_\rmn{90}$, see Appendix~\ref{sec:haloAss} and Figure~\ref{fig:haloAssembly} for details). Host halo concentration has only a slight impact on the total abundance of present-day satellite galaxies: whereas less concentrated haloes host a larger number of $z=0$ satellites, this trend flattens and disappears towards higher concentrations. Finally, there are no discernible trends with host halo shape. 

As for Figure~\ref{fig:satMF_galProps}, we also inspected correlations with the stellar mass function slope, however, they are not shown since we find either no or only very minor dependencies, e.g. the satellite stellar mass functions of host haloes with a later assembly tend to be slightly less steep. We confirm the correlations described so far using Spearman correlation coefficients. While host $K$-band luminosity yields the best correlation coefficient of the host galaxy properties with $-0.65$, the trend with total host halo mass is even stronger with a Spearman correlation coefficient of $0.75$. Therefore, total satellite abundances are more closely related to host halo than host galaxy properties. This is consistent with results from the APOSTLE and ARTEMIS simulations \citep{Fattahi2016a, Font2021}.

The TNG-based findings uncovered in this paper are qualitatively consistent with the trends between subhalo abundances and host halo properties in the Illustris simulation \citep{Chua2017}. But some differences do emerge: we expand on the host-dependent trends of subhalo, rather than satellite, abundances in TNG50 in Appendix~\ref{sec:subMF_haloProps}. 
As shown in Appendix~\ref{sec:subMF_haloProps} and Figure~\ref{fig:haloProps_subhaloes}, when considering subhalo abundance in our DM-only analogue simulation TNG50-Dark, the total abundance of subhaloes displays a strong dependence on host concentration. However, host haloes in TNG50-Dark are generally less concentrated than in TNG50: baryonic effects in TNG50 cause the host haloes to contract (see also \citealp{Chua2017, Lovell2018} for Illustris and IllustrisTNG, as well as \citealp{Duffy2010, Marinacci2014, Zhu2016, Fiacconi2016} for other previous cosmological and zoom-in simulations). This washes out the correlations between total satellite/subhalo abundance and host concentration that have been previously quantified with N-body only models.

To our knowledge, this is the first time that it has been possible to quantify the scatter and the dependence on host properties of the satellite abundance of MW/M31-like hosts with a full-physics, hydrodynamical galaxy-formation simulation that samples many tens, in fact a couple of hundred, MW/M31-like hosts. Both Figures~\ref{fig:satMF_galProps} and~\ref{fig:satMF_haloProps} demonstrate that the observed satellite abundances of the MW and M31 are well reproduced by TNG50 as they are in agreement with the distribution of TNG50 MW/M31-like hosts, even at fixed observed host properties. Whereas a significant degree of host-to-host variation remains also at fixed total satellite abundance, we find that, also according to TNG50, the larger number of satellites around Andromeda compared to those of the Galaxy suggest a higher total halo mass, more recent halo formation time, and lower halo concentration for the host halo of M31 in comparison to the MW's.

\section{Summary \& conclusions}
\label{sec:conc}

We have analysed the abundance of satellite galaxies at $z=0$ around 198 MW- and M31-like galaxies in TNG50, the final instalment in the IllustrisTNG suite of cosmological magnetohydrodynamical simulations. Thanks to the available volume and the zoom-in-like resolution of TNG50, we have obtained a statistically-significant sample of both MW/M31-like galaxies as well as their satellite populations within $300~\rmn{kpc}$ (3D) and were able to reliably resolve satellites down to stellar masses of $5 \times 10^6~\rmn{M}_\odot$. From TNG50, we have selected MW/M31 analogues as disky galaxies with a stellar mass of $M_* = 10^{10.5 - 11.2}~\rmn{M}_\odot$ in relative isolation at $z=0$ (Figure~\ref{fig:hostProps}). We have compared our findings to both recent observational surveys and previous cosmological zoom-in and large-scale simulations by carefully matching both selection criteria and mass or magnitude distributions of the hosts. We have put the population of present-day survivors in contrast to the abundance of all satellites that have ever been accreted by MW- and M31-like hosts for luminous satellite galaxies in TNG50 as well as subhaloes in both TNG50 and its dark matter-only analogue TNG50-Dark. Furthermore, we have compared the results for TNG50-Dark to subhalo abundances of other, previous DM-only simulations. Finally, we quantified the correlations of satellite abundance with various host galaxy and host halo properties. In an upcoming paper, we will focus on specific properties of MW/M31-like satellites such as their star formation activity and gas fractions, as well as their dependence on host properties and infall times (\textcolor{blue}{Engler et al. in prep.}).

The results of this paper are summarised as follows.

\begin{itemize}
    \item Our sample of TNG50 satellite galaxies around MW/M31-like hosts follows basic scaling relations that are in reasonable agreement with satellite populations from previous zoom-in simulations, semi-empirical models, and observed Local Volume dwarfs (Figure~\ref{fig:satProps}).\\
    \item The abundance of satellite galaxies around MW- and M31-like hosts in TNG50 is remarkably diverse and exhibits a significant host-to-host scatter (Figure~\ref{fig:MFTNG50_Nsat_hist}, left panel). The total number of satellites with $M_* \geq 5 \times 10^6~\rmn{M}_\odot$ around TNG50 hosts ranges from 0 to 20 (i.e. between 2 and 11 within $16^\rmn{th}-84^\rmn{th}$ percentiles). This degree of scatter persists even at fixed host halo mass: the total number of satellites in $10^{12}~\rmn{M}_\odot$ hosts range from 0 to 11 (Figure~\ref{fig:satMF_haloProps}). However, the median TNG50 MW/M31-like galaxy has a total of 5 satellites down to $M_* \sim 5\times 10^{6}~\rmn{M}_\odot$, the most massive of which reaches a stellar mass of $M_* \sim 10^{8.5}~\rmn{M}_\odot$.\\
    \item While the distribution of total satellite abundance appears to be skewed to lower numbers when increasing the minimum satellite stellar mass, this is merely an effect of Poisson statistics (Figure~\ref{fig:MFTNG50_Nsat_hist}, right panel). In fact, their normalised distributions show that the diversity, i.e. scatter, of satellite systems remains the same regardless of the employed minimum stellar mass (Figure~\ref{fig:normSatAb}, right panels).\\
    \item Considering not only the present-day, surviving satellite population at $z=0$ but all satellites that have ever been accreted by MW/M31-like hosts, we show that at a fixed minimum stellar mass, the number of ever accreted satellites is larger by a factor of $4-5$ than those that survive through $z=0$ (Figure~\ref{fig:satMF_acc}, top panel). According to TNG50, on average, present-day satellites of the MW and Andromeda have been accreted more recently than $z\sim 0.7-1$.\\
    \item While there can be up to 120 surviving subhaloes in TNG50 MW/M31-like galaxies with $M_\rmn{dyn} \geq 5 \times 10^7~\rmn{M}_\odot$ and $V_\rmn{max}/V_\rmn{200c} \sim 0.02$, their number is vastly reduced (by more than a factor of 10) when we additionally require them to host a luminous galaxy, e.g. with $M_* \geq 5 \times 10^6~\rmn{M}_\odot$ (Figure~\ref{fig:satMF_acc}, bottom panels).  Moreover, the TNG model returns a suppressed cumulative subhalo mass function in comparison to DM-only predictions.\\
    \item  Using our baryonic simulation, we show that the abundance of satellites depends on host properties. More massive and $K$-band brighter galaxies host more satellites at $z=0$ (Figure~\ref{fig:satMF_galProps}). Furthermore, more massive haloes, haloes that assembled later in time, and those that are less concentrated host a larger number of satellites at present-day times (Figure~\ref{fig:satMF_haloProps}), with the latter correlation being weaker than those with mass and assembly time. Overall, the abundance of satellite galaxies around MW/M31-like galaxies in TNG50 correlates more strongly with host {\it halo} than host {\it galaxy} properties: total satellite abundance exhibits the most significant correlation with host halo mass.\\
    \item Whereas Andromeda holds a richer system of satellites than the $1\sigma$ scatter of the TNG50 MW/M31-like galaxies, this is reasonable since the mass of Andromeda lies at the high-mass end of the TNG50 galaxies selected for the comparison (see Figures~\ref{fig:satMF_galProps} and \ref{fig:satMF_haloProps}). Moreover, while both the Galaxy and Andromeda host a few more massive satellites than the TNG50's average, hosts similar to these do exist in TNG50 (Figures~\ref{fig:vis1} and \ref{fig:vis2}). In fact, there are 6 MW/M31-like galaxies (i.e. 3 per cent of MW/M31-like hosts) in TNG50 that host both a Large and a Small Magellanic Cloud-like satellite (with the LMC-like galaxy as their most massive satellite).\\
    \item Comparing the satellite abundances in TNG50 with observed hosts -- e.g. MW-like galaxies from the SAGA survey \citep{Geha2017, Mao2021} and hosts in the Local Volume \citep{Carlsten2021} -- yields consistent results (Figure~\ref{fig:satLF_obs}). In both the comparison at face value with our fiducial TNG50 sample of MW/M31-like hosts and the comparison with the matched selections of our SAGA- and LV-like host samples, TNG50's median and scatter agree well with the observational results from SAGA, but  less so from \cite{Carlsten2021}.\\
    \item TNG50 MW/M31-like hosts exhibit satellite mass functions that are in good overall agreement compared to previous cosmological hydrodynamical simulations of MW-like hosts, even those with better numerical resolution (Figure~\ref{fig:satMF_sims}, top panels). When compared at face value, there is a significant scatter not only between simulations but also between hosts of the same models. However, these deviations are expected given the large intrinsic galaxy-to-galaxy variations and the host-dependent trends of the satellite abundances, especially with host mass. In fact, when we compare TNG50 results by replicating the host mass selection of the Auriga and Latte simulations, we obtain remarkably consistent results (Figure~\ref{fig:satMF_sims}, bottom panel).\\
\end{itemize}

In conclusion, thanks to the TNG50 simulation we have highlighted and quantified the diversity of satellite populations around MW- and M31-like galaxies, utilising a statistical sample of 198 hosts. The reasons for this diversity in present-day, surviving satellites depend on properties of the host itself and on environmental effects. More massive hosts and hosts with a later halo assembly are richer in surviving satellite galaxies at $z=0$, as accreted satellites can be disrupted or merge inside their host halo. Overall, however, amid such galaxy-to-galaxy diversity and different galaxy formation models, the satellite abundances predicted by TNG50 are consistent with observed galaxies within the Local Volume and beyond, as well as several previously simulated MW- and M31-like galaxies. However, whereas the scientific conclusions from previous comparisons had been de facto impaired by limited host number statistics and by host selections and mass ranges that were not necessarily compatible, we are now able to assess the bounty of the theoretical model while controlling for selection effects. Twenty years after the original formulation of the missing satellites problem, we can confidently put it to rest.

\section*{Acknowledgements}
CE acknowledges support by the Deutsche Forschungsgemeinschaft (DFG, German Research Foundation) through project 394551440 and thanks Yao-Yuan Mao, as well as the participants of the KITP conference ``The Galaxy-Halo Connection Across Cosmic Time: Recent Updates'' for useful discussions and input. We thank Christine Simpson for sharing data on satellite stellar masses from the Auriga simulations.
This work was also funded by the Deutsche Forschungsgemeinschaft (DFG, German Research Foundation) -- Project-ID 138713538 -- SFB 881 (``The Milky Way System'', subprojects A01 and A03). 
FM acknowledges support through the Program "Rita Levi Montalcini" of the Italian MUR.
The primary TNG simulations were realised with compute time granted by the Gauss Centre for Super-computing (GCS): TNG50 under GCS Large-Scale Project GCS-DWAR (2016; PIs Nelson/Pillepich), and TNG100 and TNG300 under GCS-ILLU (2014; PI Springel) on the GCS share of the supercomputer Hazel Hen at the High Performance Computing Center Stuttgart (HLRS). Additional simulations for this paper were carried out on the Draco and Cobra supercomputers at the Max Planck Computing and Data Facility (MPCDF).
 
\section*{Data availability}
As of February 1st, 2021, data of the TNG50 simulation series are publicly available from the IllustrisTNG repository: \url{https://www.tng-project.org}. Data directly referring to content and figures of this publication is available upon request from the corresponding author.



\bibliographystyle{mnras}
\bibliography{TNG50_MW_satPops}



\appendix

\section{Resolution Effects}
\label{sec:resolution}

\begin{figure*}
    \centering
	\includegraphics[width=.44\textwidth]{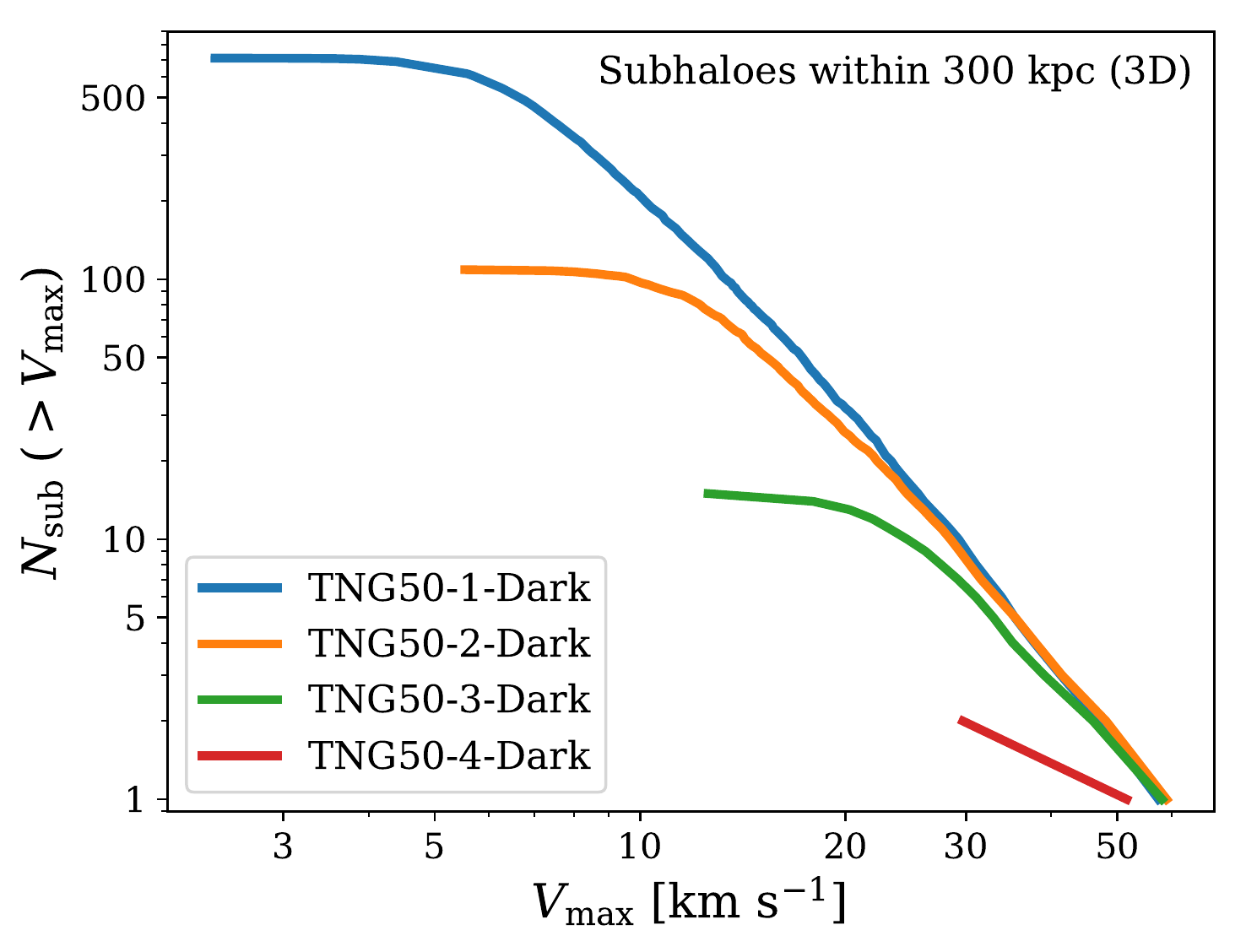}
    \includegraphics[width=.44\textwidth]{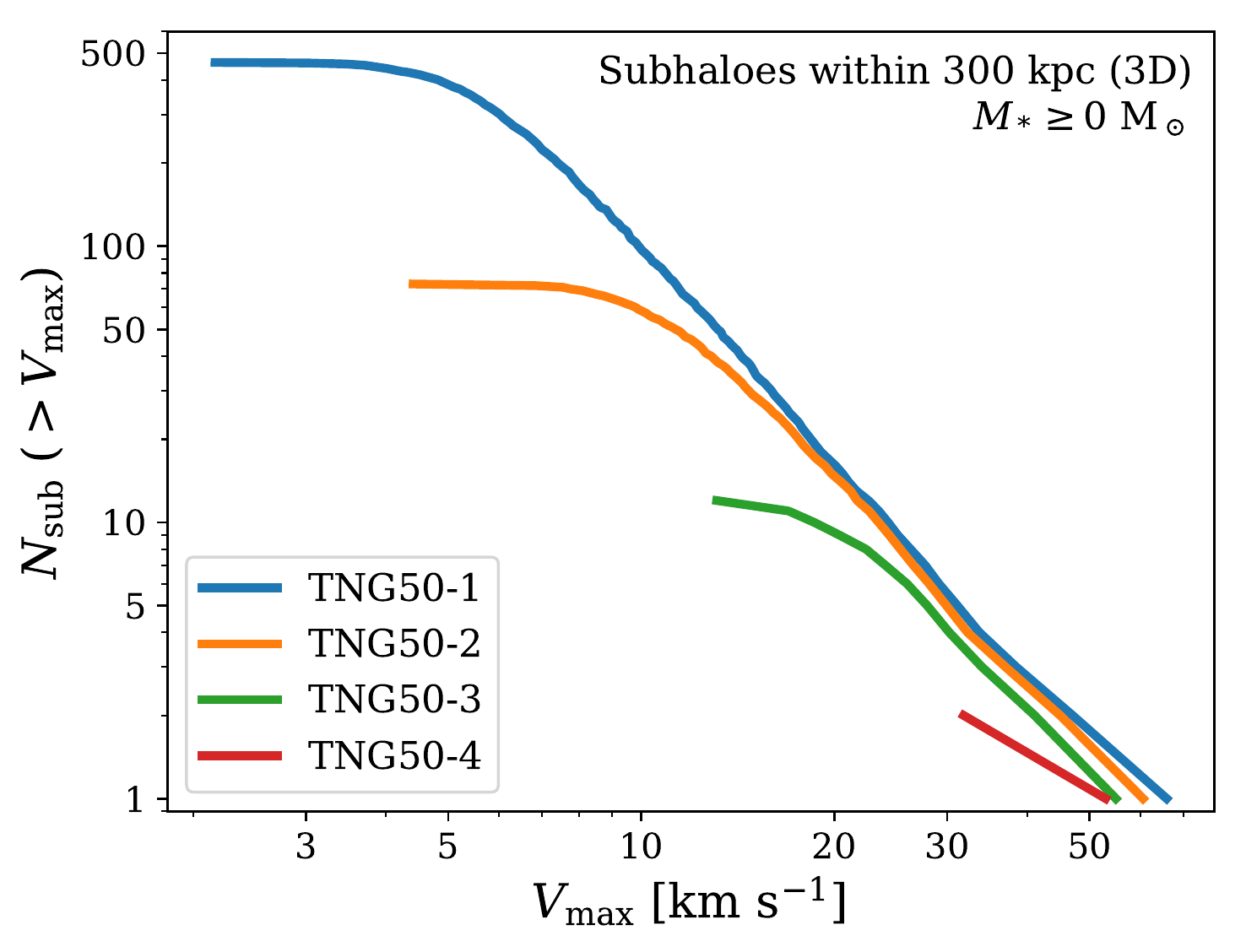}
    \includegraphics[width=.44\textwidth]{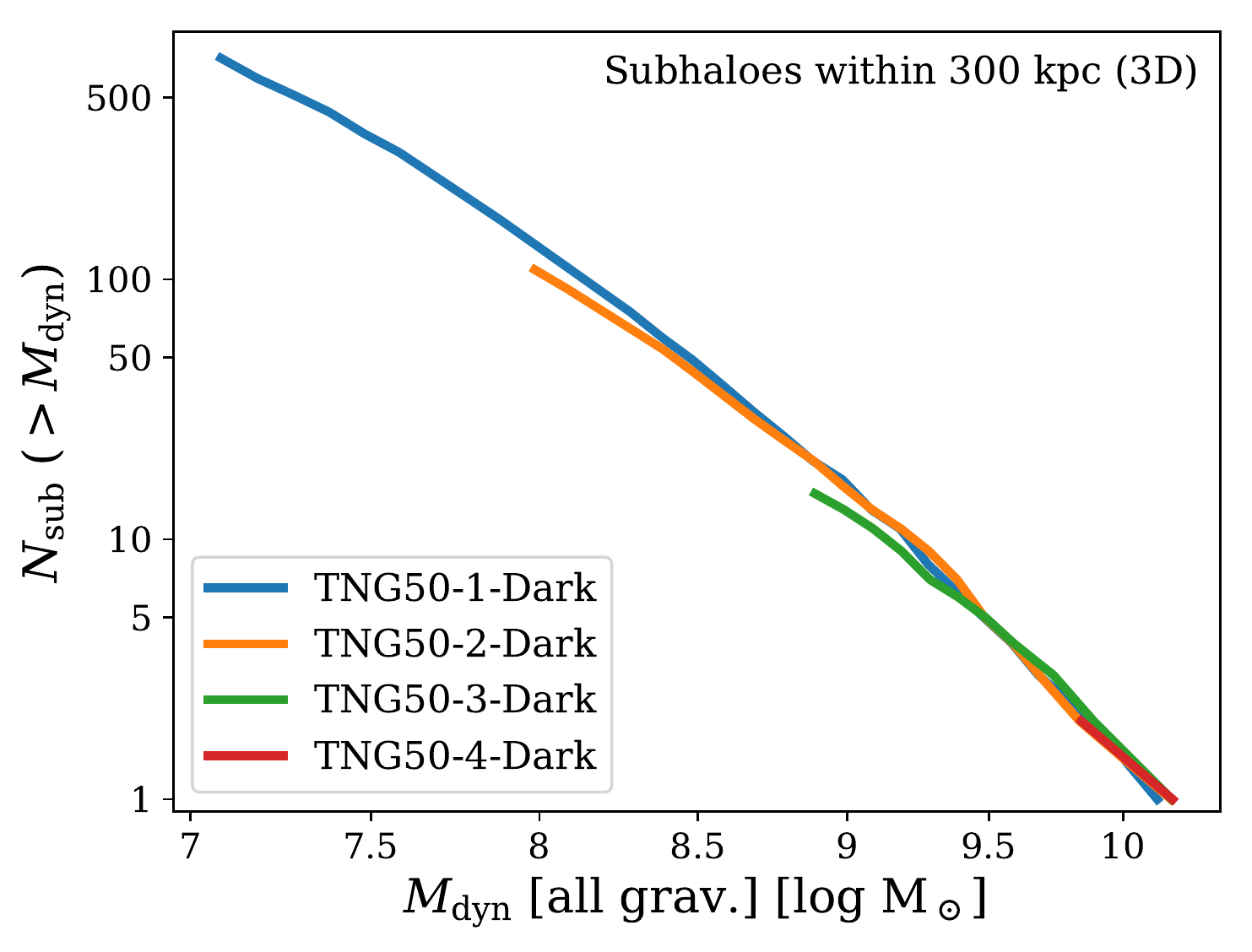}
    \includegraphics[width=.44\textwidth]{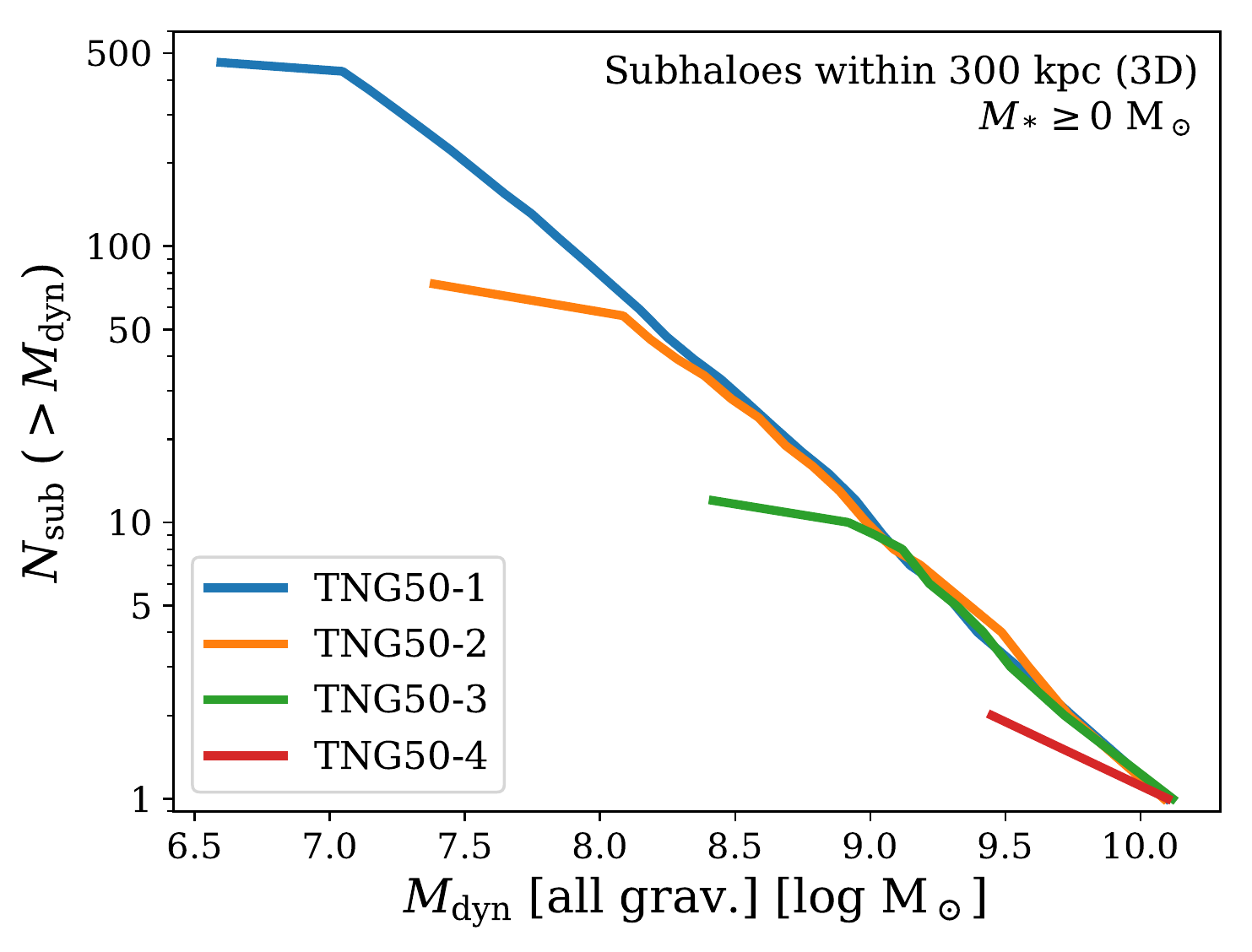}
	\caption{{\bf Impact of numerical resolution on the abundance of dark and luminous subhaloes within $\mathbf{300~\rmn{\textbf{kpc}}}$ of MW/M31-like hosts.} We illustrate trends from DM-only analogues (left panels) and baryonic runs (right panels) at different resolution levels: TNG50-1 and TNG50-1-Dark (blue curves), TNG50-2 and TNG50-2-Dark (orange curves), TNG50-3 and TNG50-3-Dark (green curves), as well as TNG50-4 and TNG50-4-Dark (red curves). \textit{Top panels:} subhalo abundance in terms of their maximum circular velocity $V_\rmn{max}$. \textit{Bottom panels:} subhalo abundance in terms of their dynamical mass $M_\rmn{dyn}$.}
    \label{fig:resTest_allSubs}
\end{figure*}

Throughout this paper, we focus on the highest resolution run of the TNG50 simulation series, TNG50-1 aka simply TNG50, and its DM-only analogue TNG50-1-Dark (aka TNG50-Dark). However, as is evident in the top left panel of Figure~\ref{fig:satMF_sims} when comparing satellite abundances in TNG50 and TNG100, a lower numerical resolution decreases the number of surviving satellite galaxies at $z=0$ significantly -- either through a lower build-up of stellar mass or more effective artificial disruption. 

We examine the impact of different resolution levels on present-day satellite and subhalo populations in this section. We start in Figure~\ref{fig:resTest_allSubs} by presenting  the subhalo abundance around MW- and M31-like hosts for TNG50-Dark (left panels) and TNG50 (right panels), both in terms of the subhaloes' maximum circular velocity $V_\rmn{max}$ (top panels) and their dynamical mass $M_\rmn{dyn}$ (bottom panels). These are global subhalo properties, independent of their stellar mass content. Furthermore, we count all subhaloes, both luminous and dark. In all cases, we show subhalo abundances of different resolution runs: TNG50-1-, -2-, -3- and -4-Dark, as well as TNG50-1, -2, -3, and -4 (blue, orange, green, and red curves, respectively), with progressively poorer spatial and mass resolutions (see \citealp{Pillepich2019, Nelson2019b} for details). In order to compare TNG50 and its DM-only analogue across resolution levels, we simplify our selection of MW/M31-like hosts: we exclusively consider centrals and base our selection on a range in total host halo mass of $M_\rmn{200c} = 10^{11.9} - 10^{12.5}~\rmn{M}_\odot$. This corresponds to the $10^\rmn{th}$ and $90^\rmn{th}$ percentiles of the halo mass range covered by our fiducial sample (see Figure~\ref{fig:hostProps} and Section~\ref{sec:sample_mw}).

Overall, the subhalo abundance is very similar across different resolution runs at larger velocities and dynamical masses, reaching up to $V_\rmn{max} \sim 60-70~\rmn{km~s}^{-1}$ and $M_\rmn{dyn} \sim 10^{10.2}~\rmn{M}_\odot$ in both TNG50-Dark and TNG50. At the low-velocity end, however, the subhalo distribution becomes flat: the numerical resolution is not sufficient to resolve these subhaloes anymore and they are artificially disrupted. This occurs at maximum circular velocities of $5-7~\rmn{km~s}^{-1}$ ($10-12~\rmn{km~s}^{-1}$, $20-22~\rmn{km~s}^{-1}$) in both the DM-only and the baryonic version of TNG50-1~(-2,~-3). 

Although the subhalo distributions flatten not as clearly when viewed in terms of dynamical mass for TNG50-Dark, we can clearly see down to what subhalo masses the subhalo abundances are well converged in the baryonic runs. By comparing across the different resolution levels, we can confidently say that TNG50 satellite abundance results are well converged, i.e. they rise monotonically without being incomplete due to numerical resolution limits, for $V_\rmn{max}\gtrsim 5\rmn{km~s}^{-1}$ and $M_\rmn{dyn} \gtrsim 10^{7}~\rmn{M}_\odot$.

\begin{figure*}
    \centering
    \includegraphics[width=.44\textwidth]{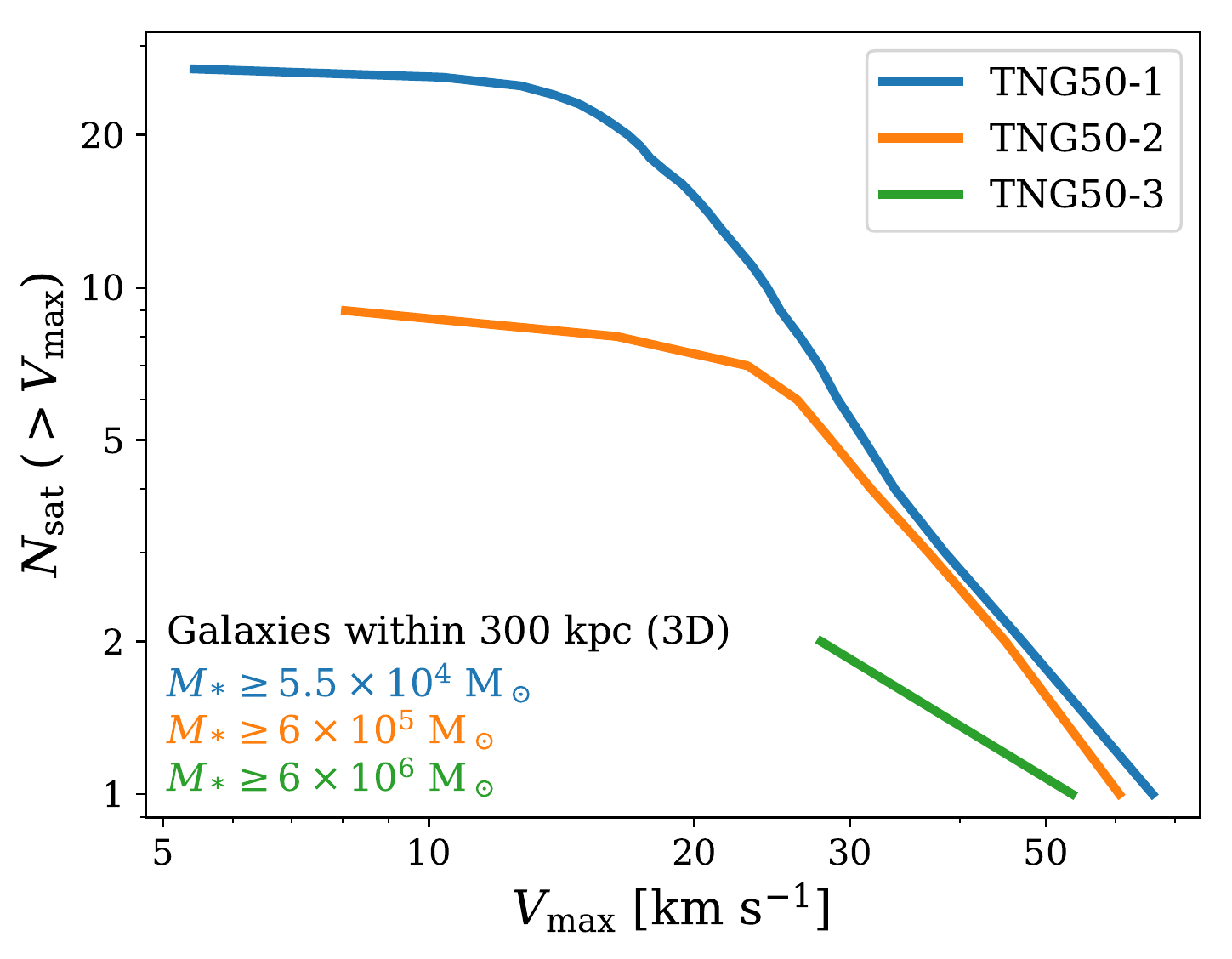}
    \includegraphics[width=.44\textwidth]{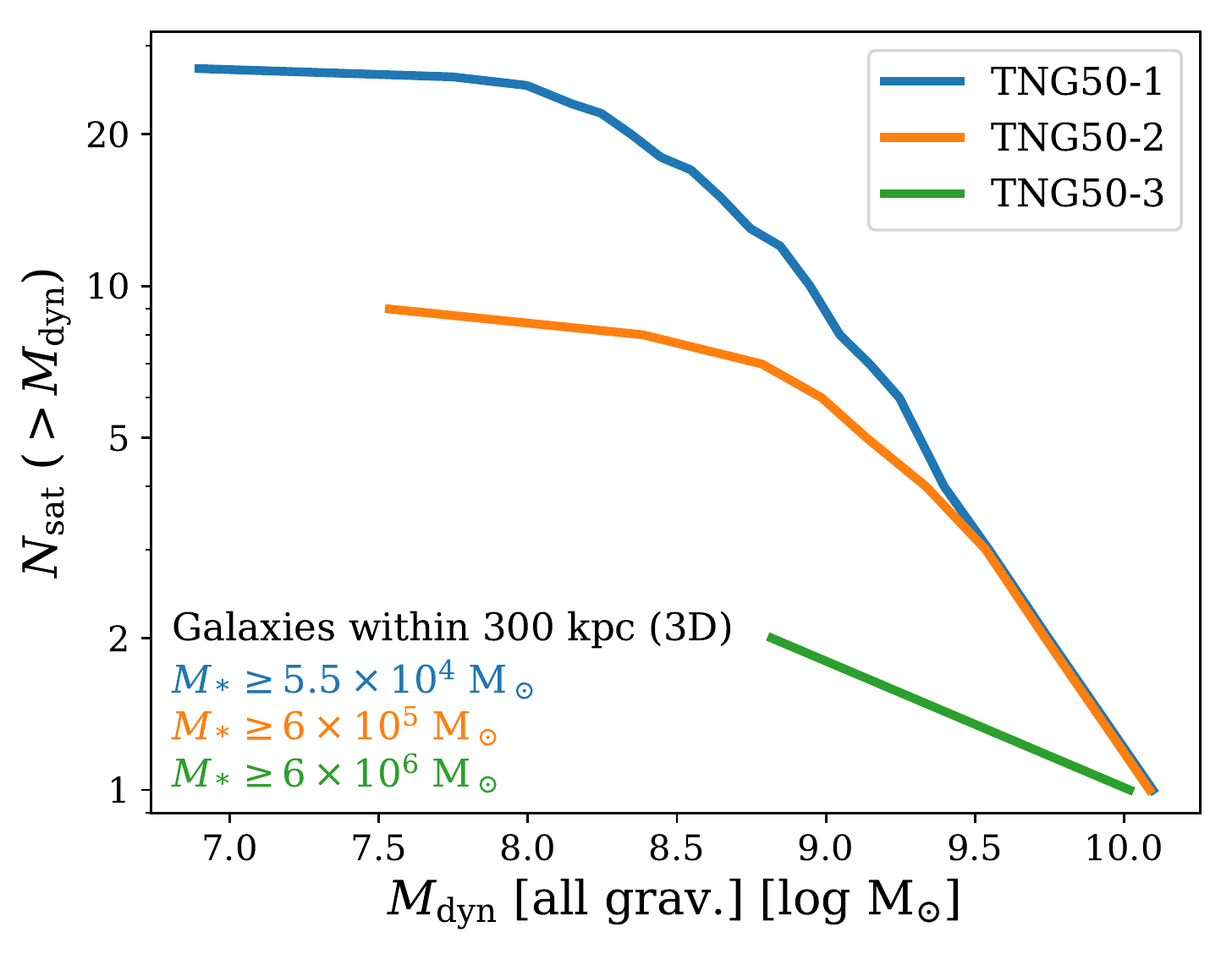}
    \includegraphics[width=.44\textwidth]{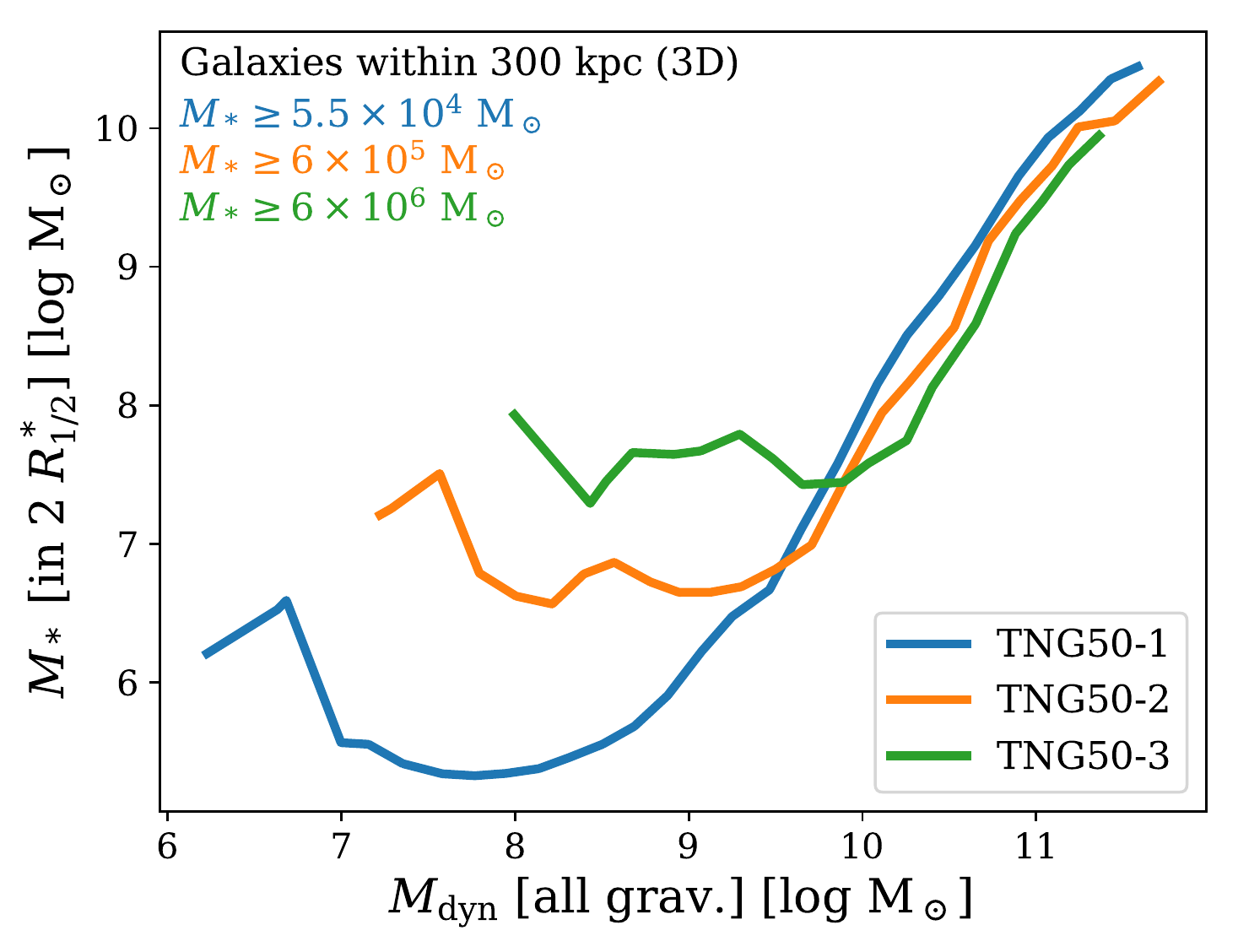}
	\includegraphics[width=.44\textwidth]{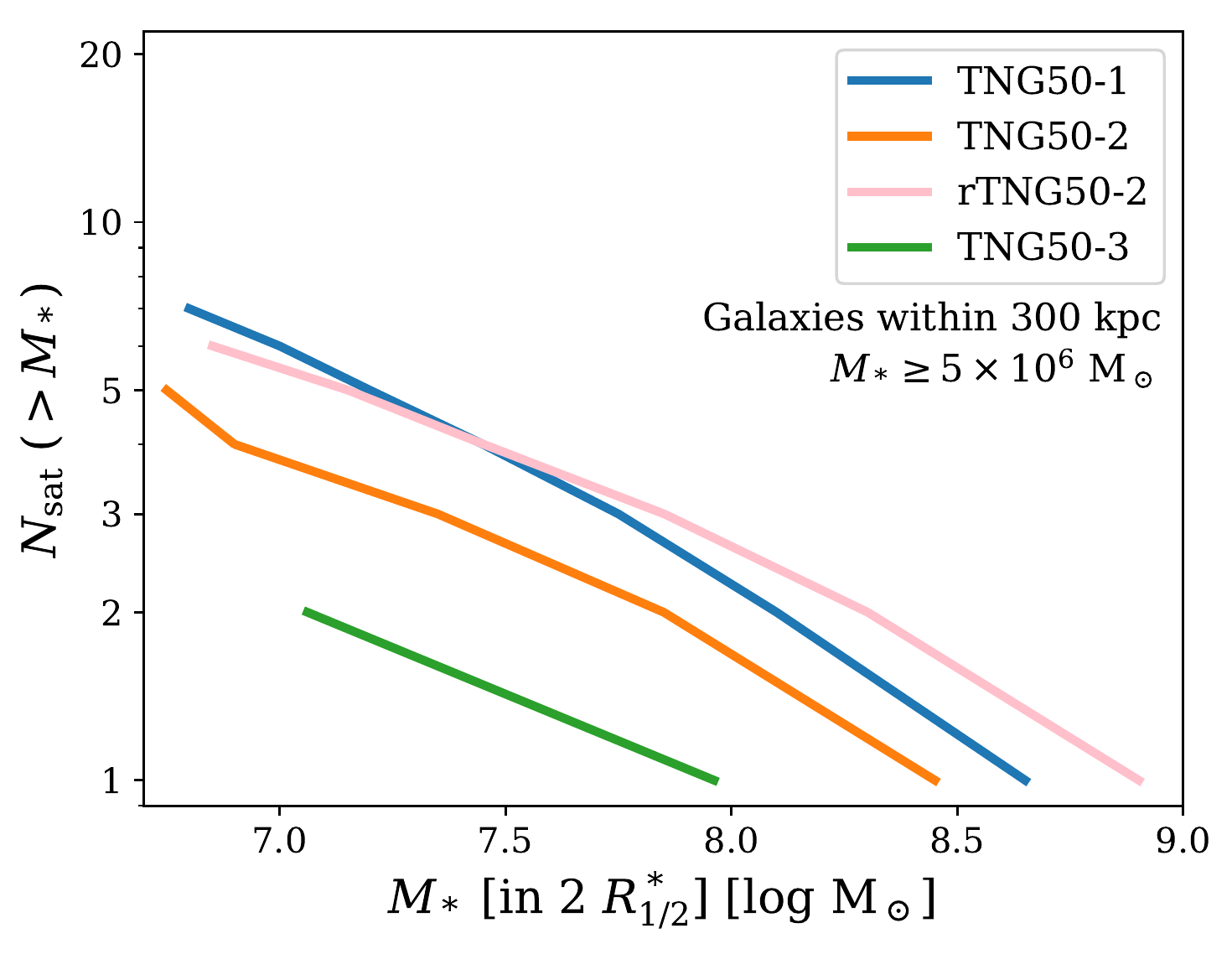}
    \caption{{\bf Resolution effects on luminous satellite galaxies within $\mathbf{300~\rmn{\textbf{kpc}}}$ of MW/M31-like hosts in TNG50.} In all panels, we compare TNG50-1 (blue curve), TNG50-2 (orange curve), and TNG50-3 (green curve). \textit{Top panels:} satellite abundance in terms of maximum circular velocity $V_\rmn{max}$ (top left) and dynamical mass $M_\rmn{dyn}$ (top right), down to the least massive satellite available: $M_* \geq 5.5 \times 10^4~\rmn{M}_\odot$ in TNG50-1, $M_* \geq 6 \times 10^5~\rmn{M}_\odot$ in TNG50-2, and $M_* \geq 6 \times 10^6~\rmn{M}_\odot$ in TNG50-3. \textit{Bottom left panel:}  average stellar-to-halo mass relation for all luminous satellites of MW/M31-like galaxies in TNG50-1, -2, and -3. \textit{Bottom right panel:} satellite stellar mass function using stellar mass within twice the stellar half-mass radius for satellite galaxies according to our fiducial satellite selection of $M_* \geq 5 \times 10^6~\rmn{M}_\odot$. Additionally, we include a sample of TNG50-2 subhaloes that have been repopulated with galaxy stellar masses according to the most similar subhalo in TNG50-1 (rTNG50-2, pink curve).}
    \label{fig:resTest_lumSats}
\end{figure*}

For the abundance of {\it luminous} satellite galaxies, the resolution trends are qualitatively similar as those seen thus far, as is shown in Figure~\ref{fig:resTest_lumSats}. The top panels show the distribution of satellites with a luminous component down to $M_* = 5.5 \times 10^4~\rmn{M}_\odot$ ($6 \times 10^5~\rmn{M}_\odot$, $6 \times 10^6~\rmn{M}_\odot$) in TNG50-1 (-2, -3) (blue, orange, and green curves) in terms of maximum circular velocity $V_\rmn{max}$ (top left panel) and dynamical mass $M_\rmn{dyn}$ (top right panel). Satellite distributions flatten and become incomplete below $V_\rmn{max} \sim 18~\rmn{km~s}^{-1}$ ($25~\rmn{km~s}^{-1}$) and $M_\rmn{dyn} \sim 10^{8.5}~\rmn{M}_\odot$ ($10^{9.3}~\rmn{M}_\odot$) in TNG50-1 (-2). So, when subhaloes and satellites are characterised by properties that relate to their total mass (i.e. $V_\rmn{max}$ or $M_\rmn{dyn}$), the resolution convergence of the subhalo/satellite abundance in baryonic simulations behaves very similar to that in DM-only models, with resolution effects progressively creeping in from the low-mass end.

The resolution convergence becomes more complex when satellite galaxies are counted based on properties related to their stellar mass. In the bottom left panel, stellar mass and dynamical mass are connected as stellar-to-halo mass relation (SHMR). At fixed dynamical mass, higher levels of resolution imply larger stellar masses -- as is evident at the SHMR's massive end (see also \citealp{Engler2021} and \citealp{Pillepich2018b}). At lower dynamical masses, however, the SHMR begins to flatten (at $10^{8.7}~\rmn{M}_\odot$ for TNG50-1, $10^{9.5}~\rmn{M}_\odot$ for TNG50-2, and $10^{10}~\rmn{M}_\odot$ for TNG50-3). This is not a physical part of the relation but a limitation due to the finite stellar mass particle resolution. While the underlying galaxy formation model would entail a certain average SHMR with a related scatter, only satellites with at least one stellar particle can be accounted for: subhaloes that remain dark due the limitations of stellar mass resolution are not included in the average SHMR curves and would otherwise populate the bottom part of the plot. The relations flatten at these dynamical masses since they are ``incomplete''. The shape of the SHMR curves indicate the minimum stellar mass to which we can reliably count satellites in each simulation: for TNG50, this limit emerges at $M_* \gtrsim 10^6~\rmn{M}_\odot$. Therefore, we choose $M_* \geq 5\times 10^6~\rmn{M}_\odot$ as our fiducial minimum satellite stellar mass in TNG50.

Finally, the bottom right panel of Figure~\ref{fig:resTest_lumSats} shows the median satellite stellar mass function for our fiducial selection, i.e. satellites with a stellar mass of $M_* \geq 5 \times 10^6~\rmn{M}_\odot$, across TNG50 resolution levels. While the satellite stellar mass functions are converg{\it ing}, poorer resolution implies artificially suppressed satellite mass functions. However, this mostly seems to relate to the reduced stellar masses in subhaloes of a given dynamical mass instead of the enhanced disruption of subhaloes or satellites at progressively poorer resolution. We confirm this by showing the satellite stellar mass function of an additional sample of repopulated TNG50-2 subhaloes (rTNG50-2, pink curve). Here, each TNG50-2 subhalo that survives through $z=0$ in the selected MW/M31-like hosts has not been assigned its simulated TNG50-2 stellar mass, but the stellar mass of the TNG50-1 subhalo that is most similar in dynamical mass. This not only increases the stellar mass of luminous TNG50-2 satellites in general; it also populates some of its dark subhaloes -- which are otherwise not able to form a galaxy due to the stellar particle mass resolution limitations -- with a luminous component. The stellar mass function of rTNG50-2 satellite galaxies essentially coincides with the satellite abundance of TNG50-1. Therefore, the differences between TNG50 resolution levels (or at least between TNG50-1 and TNG50-2) are almost entirely driven by the lower build-up of stellar mass in the lower-resolution runs. Artificial disruption of subhaloes \citep{vandenbosch2018}, on the other hand, has little to no effect on the differences of satellite abundance between resolution levels in the considered regimes.

\section{Normalised satellite abundances}
\label{sec:normSatAb}

\begin{figure*}
    \centering
    \includegraphics[width=.35\textwidth]{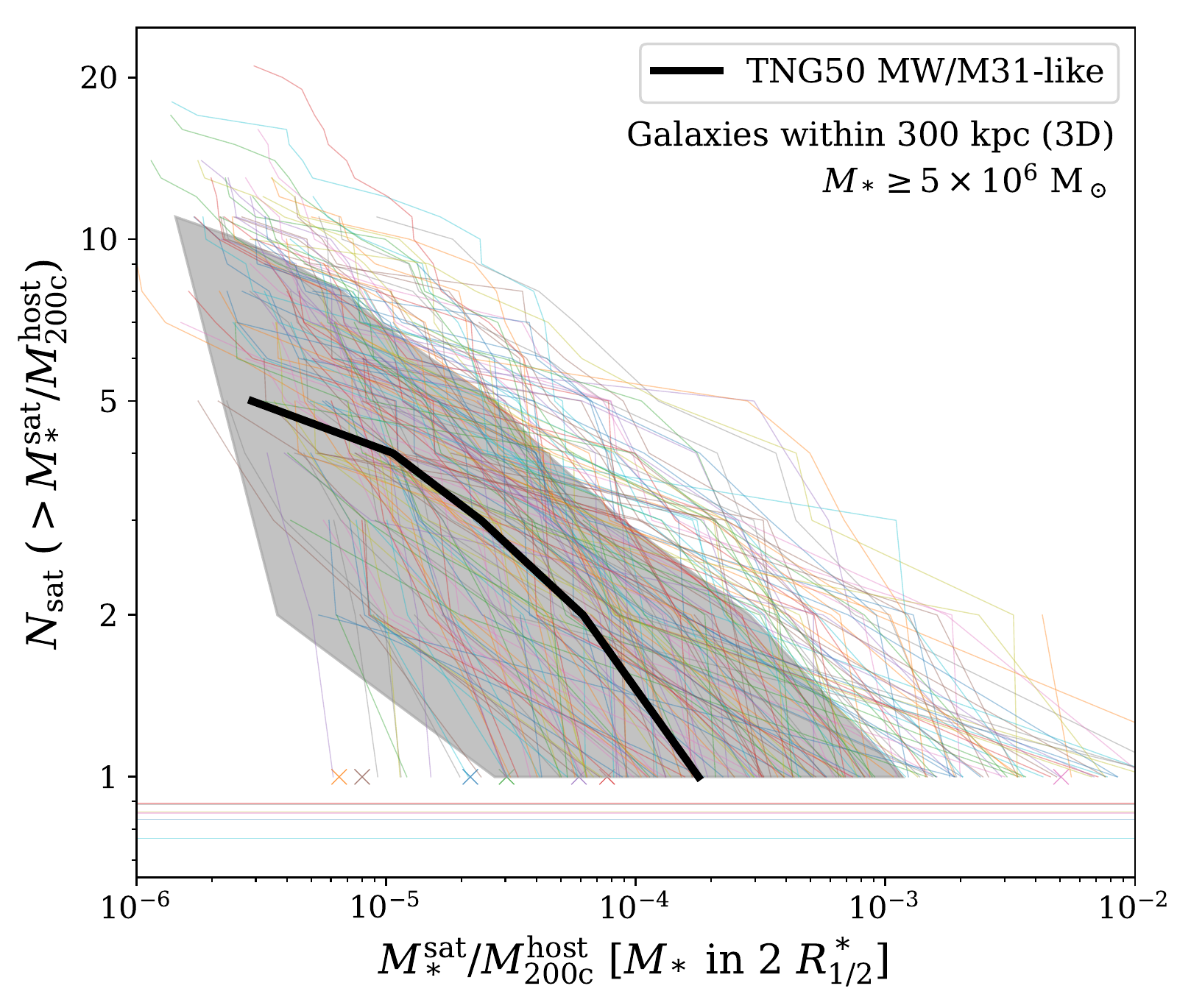}
    \includegraphics[width=.35\textwidth]{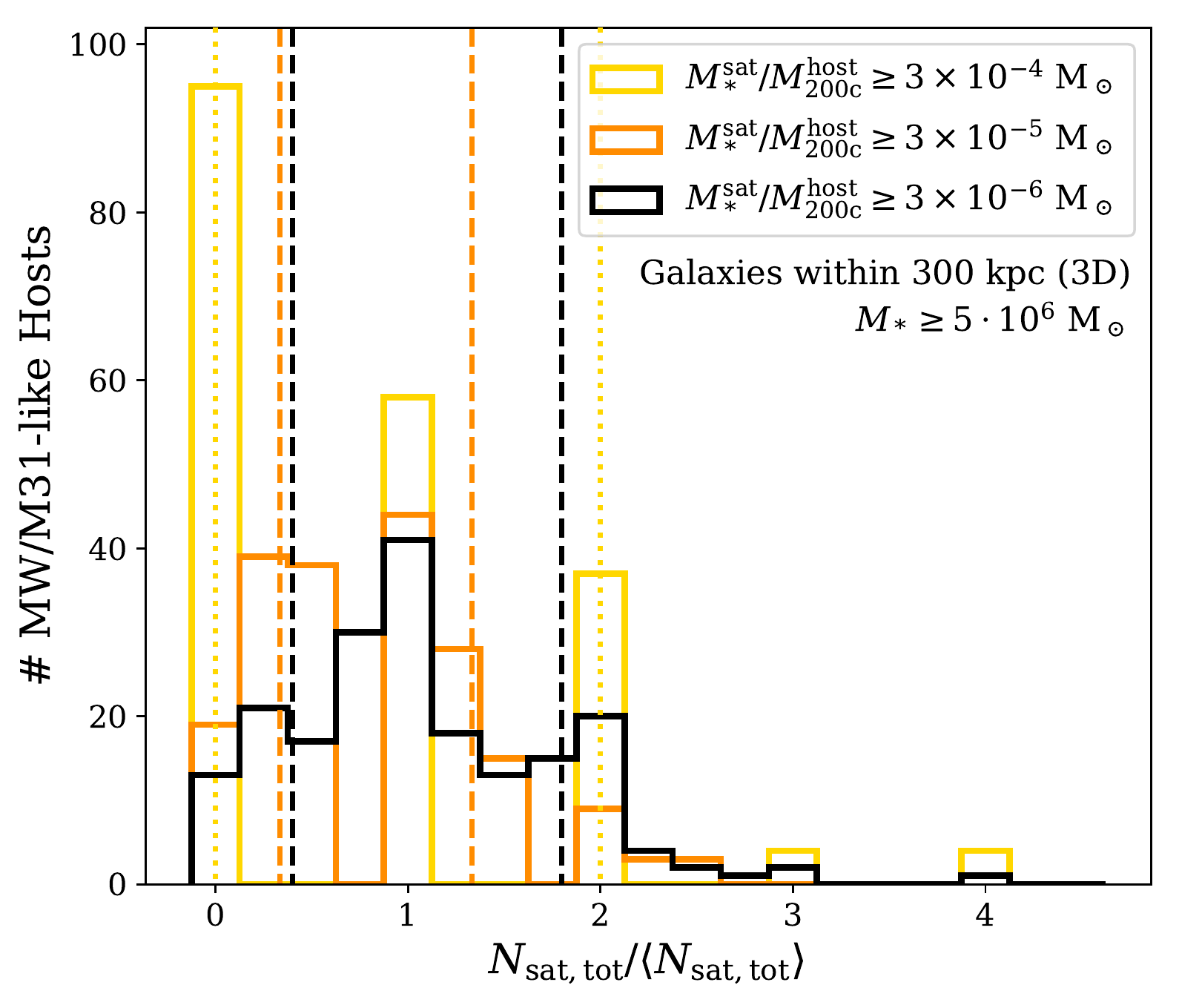}
    \includegraphics[width=.35\textwidth]{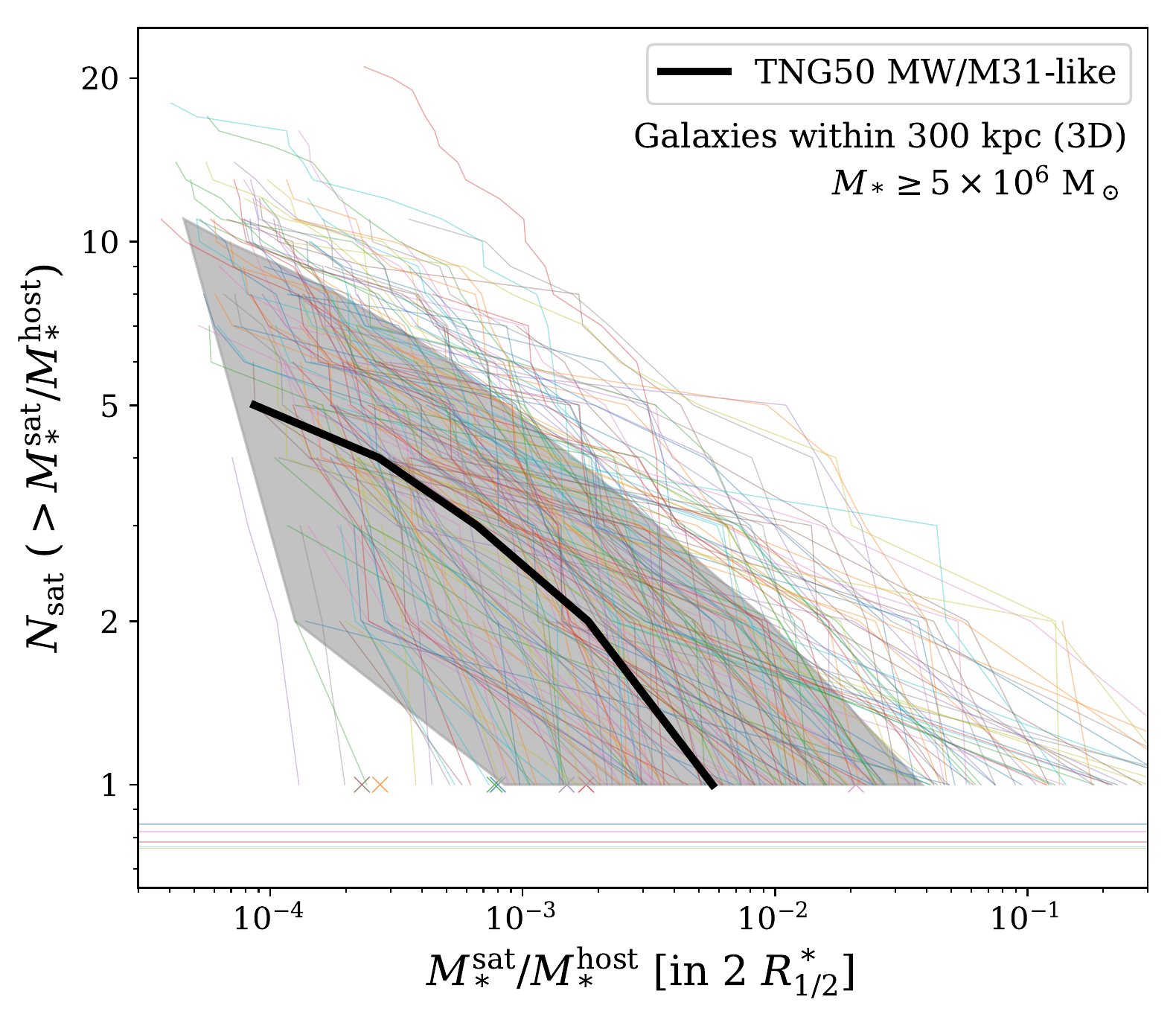}
    \includegraphics[width=.35\textwidth]{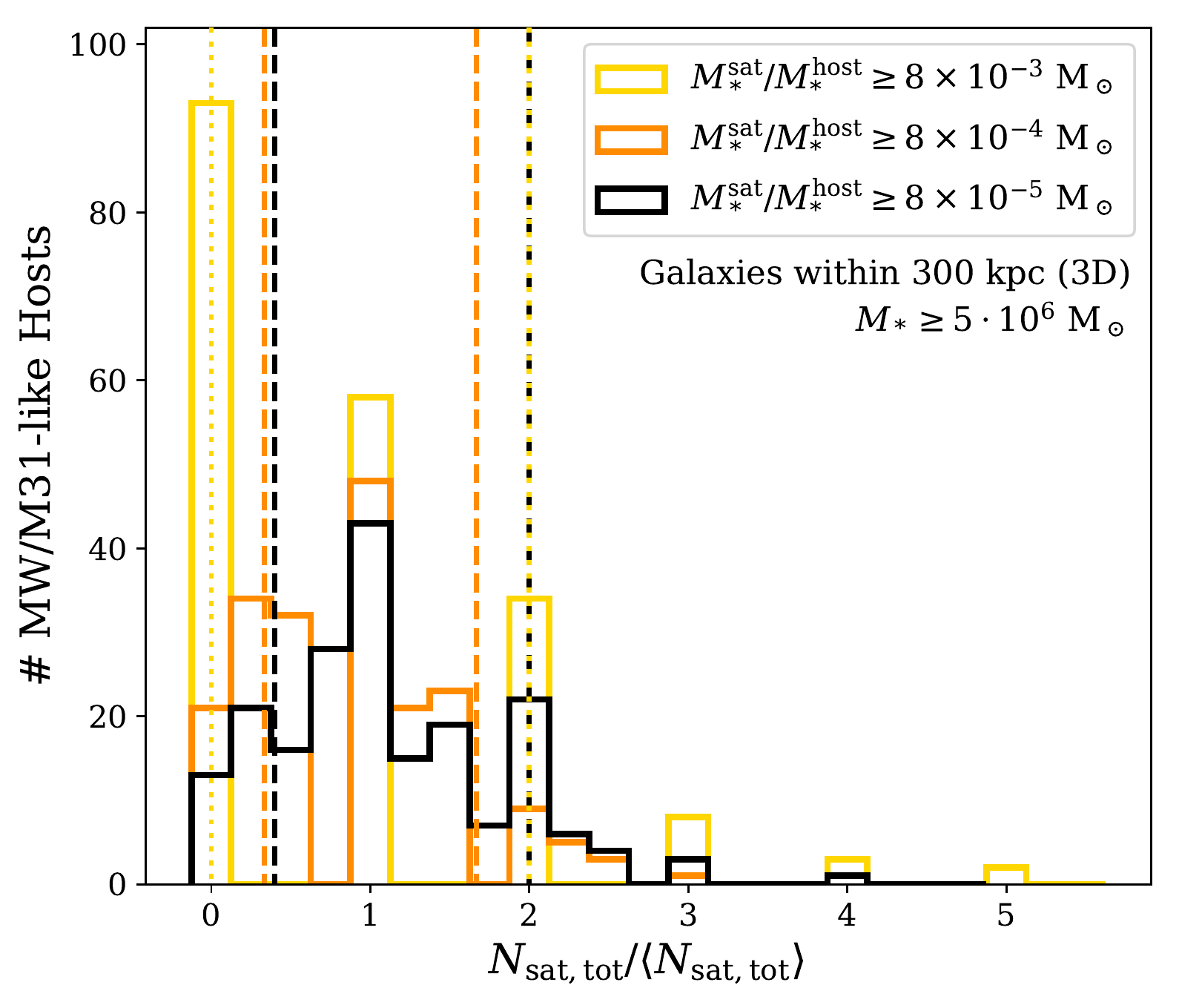}
    \caption{{\bf As in Figure~\ref{fig:MFTNG50_Nsat_hist} but for the {\it normalised} satellite demographics around MW/M31-like galaxies in the TNG50 simulation at $\mathbf{z=0}$.} In all panels, we define satellites as galaxies within $300~\rmn{physical~kpc}$~(3D) of their host and with stellar masses of at least $5 \times 10^6~\rmn{M}_\odot$ (within twice the stellar half-mass radius $R_{1/2}^*$). \textit{Left panels:} cumulative satellite abundance in terms of satellite stellar mass $M_*^\rmn{sat}$ normalised by either the total virial mass of their MW/M31-like host $M_\rmn{200c}^\rmn{host}$ (top left panel) or by host stellar mass $M_*^\rmn{host}$ (bottom left panel). The thin, coloured curves in the background illustrate the satellite systems of individual TNG50 hosts with crosses corresponding to systems with only a single satellite and horizontal lines with $N_\rmn{sat} < 1$ denoting systems with no satellites meeting the selection (6 out of 198 systems). The thick, black curve and grey shaded area depict their median and scatter as $16^\rmn{th}$ and $84^\rmn{th}$ percentiles, computed in bins of normalised satellite stellar mass. \textit{Right panels:} distribution of normalised total satellite abundance $N_\rmn{sat,tot} / \langle N_\rmn{sat,tot} \rangle$ with satellite stellar masses normalised by either the total virial mass of their MW/M31-like host $M_\rmn{200c}^\rmn{host}$ (top right panel) or by host stellar mass $M_*^\rmn{host}$ (bottom right panel), and its dependence on the imposed minimum normalised stellar mass. These bins correspond to the stellar masses in the right panel of Figure~\ref{fig:MFTNG50_Nsat_hist} assuming an average host virial mass of $M_\rmn{200c} = 10^{12.1}~\rmn{M}_\odot$ and an average host stellar mass of $M_* = 10^{10.8}~\rmn{M}_\odot$.}
    \label{fig:normSatAb}
\end{figure*}

While the total number of satellites grows on average with the mass of their host, a significant degree of scatter remains even at fixed host stellar and total masses (top right panels of Figures~\ref{fig:satMF_galProps} and~\ref{fig:satMF_haloProps}, respectively). We verify our results from Section~\ref{sec:satMF_LFobs} regarding the diversity in satellite abundance around MW/M31-like hosts in Figure~\ref{fig:normSatAb} by normalising satellite stellar masses by either total host halo mass (top left panel) or host stellar mass (bottom left panel).

As in the left panel of Figure~\ref{fig:MFTNG50_Nsat_hist}, the thin, coloured curves in the background correspond to the individual satellite stellar mass functions of all MW/M31-like hosts in TNG50, while crosses denote systems with only a single satellite and hosts with no satellites whatsoever are depicted as curves with $N_\rmn{sat}$ < 1 (6 out of 198 hosts). For both the normalisation by total host halo mass and by host stellar mass, the diversity persists: the total satellite counts still range between 0 and 20, while their $16^\rmn{th}$ and $84^\rmn{th}$ percentiles range from 2 to 11 satellites.

Furthermore, we present analogues to the right panel of Figure~\ref{fig:MFTNG50_Nsat_hist} as distributions of total satellite abundance normalised by their median in the right panels of Figure~\ref{fig:normSatAb}. We consider satellites as all galaxies within $300~\rmn{kpc}$ of MW/M31-like hosts with a stellar mass of at least $5 \times 10^6~\rmn{M}_\odot$ and show their distributions for three selections in normalised stellar mass (by host total mass in the top right panel and by host stellar mass in the bottom right panel). Assuming an average total host halo mass of $M_\rmn{200c} = 10^{12.1}~\rmn{M}_\odot$ and an average host stellar mass of $M_* = 10^{10.8}~\rmn{M}_\odot$, the bins of normalised satellite stellar mass in the right panels of Figure~\ref{fig:normSatAb} correspond to the same typical stellar masses as in Figure~\ref{fig:MFTNG50_Nsat_hist}: $5 \times 10^8~\rmn{M}_\odot$ (yellow histogram), $5 \times 10^7~\rmn{M}_\odot$ (orange histogram), and $5 \times 10^6~\rmn{M}_\odot$ (black histogram). For both the normalisation by host total and host stellar mass, the distributions of different normalised stellar mass bins exhibit similar extents and shapes. Their scatter as $16^\rmn{th}$ and $84^\rmn{th}$ percentiles all range between normalised total satellite abundances of 0 and 2. Therefore, the degree of diversity in total satellite abundance remains the same regardless of the employed minimum satellite stellar mass.

\section{Halo Assembly}
\label{sec:haloAss}

\begin{figure*}
    \centering
    \includegraphics[width=.75\textwidth]{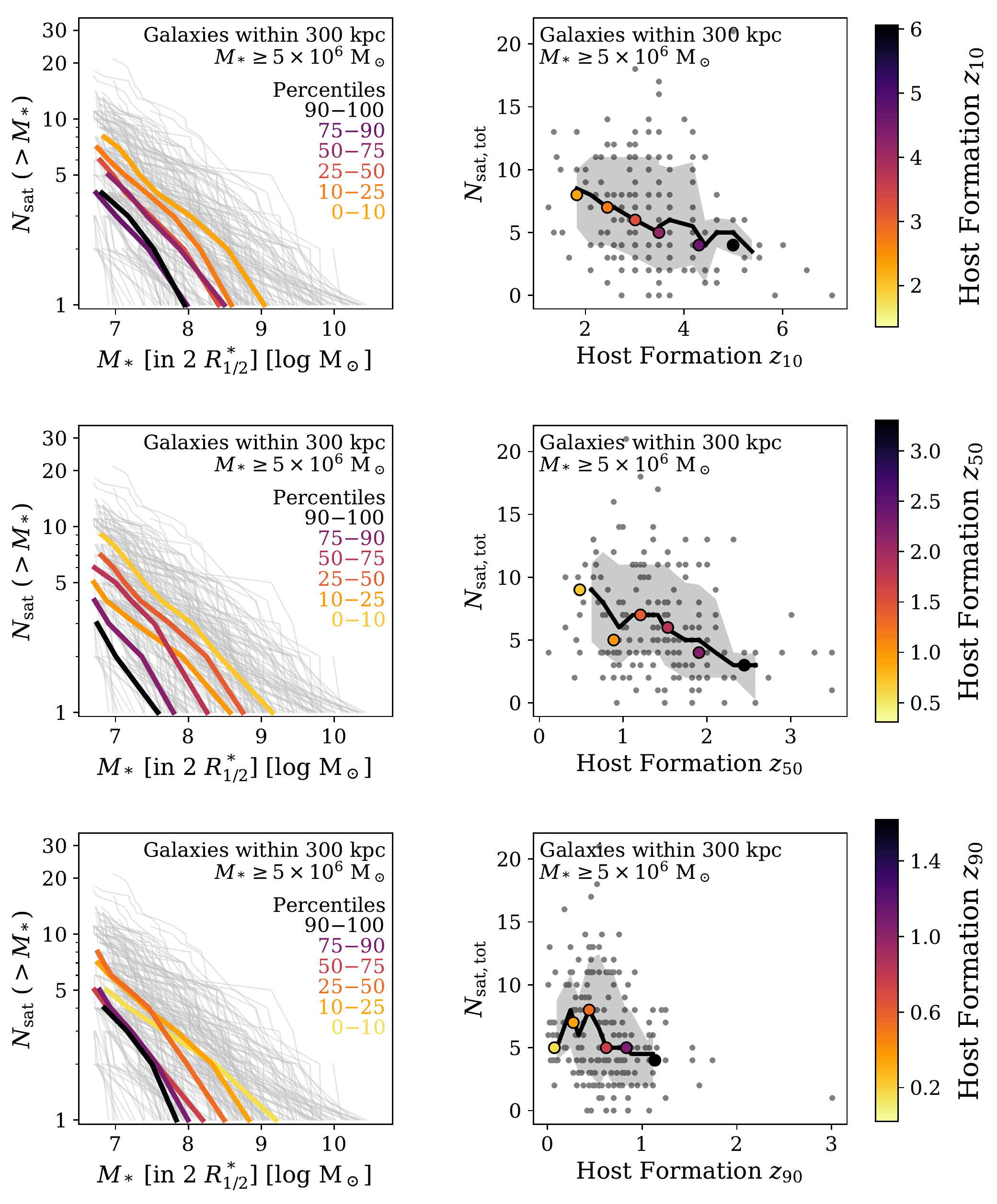}
    \caption{{\bf Dependence of satellite abundance on different definitions of host halo formation time} for satellites within $300~\rmn{kpc}$ (3D) of their MW/M31-like host and with a stellar mass of at least $5 \times 10^6~\rmn{M}_\odot$. Each row investigates a different stage of halo assembly: $z_{10}$, $z_{50}$, and $z_{90}$ (from top to bottom), i.e. the redshifts at which the host halo had assembled 10, 50, or 90 per cent of its present-day total mass. \textit{Left panels:} median satellite stellar mass functions in various percentiles of the assembly time in question (thick, yellow to black curves). The thin, grey curves in the background denote satellite stellar mass functions of individual TNG50 MW/M31-like hosts as a reference. \textit{Right panels:} total number of satellites as a function host properties for the percentiles (yellow to black circles), all TNG50 MW/M31-like galaxies (grey circles), as well as their running median (black curves) and scatter (grey shaded area, $16^\rmn{th}$ and $84^\rmn{th}$ percentiles).}
    \label{fig:haloAssembly}
\end{figure*}

We examine the abundance of satellites around MW/M31-like galaxies in TNG50 and the dependence on different stages in host halo assembly in Figure~\ref{fig:haloAssembly}. Here, we cover the hosts' early, intermediate, and late time formation using $z_{10}$, $z_{50}$, and $z_{90}$ (from top to bottom), i.e. the redshifts at which the host halo had assembled 10, 50, or 90 per cent of its present-day total mass. The results given below represent an update based on a hydrodynamical galaxy-formation simulation of previous analyses based on DM-only calculations \citep{Gao2011, Mao2015}.

The trends on satellite populations are illustrated using percentiles in stellar mass functions in terms of the assembly time in question (left panels), as well as the total number of satellites as a function of host assembly (right panels). The specific colours of percentiles vary slightly depending on the distribution of the respective host formation time. While the total number of satellites exhibits the clearest correlation at early assembly $z_{10}$ -- hosts with a more quiet early assembly tend to have more satellites at the present-day -- this trend becomes less pronounced when considering later assembly times with $z_\rmn{50}$ and $z_\rmn{90}$. We find (albeit we do not show) that, for lower mass hosts, earlier formation times are more clearly correlated with total host mass than later formation times. However, there is no correlation of formation time and host mass for massive hosts. While the connection of formation time and host mass influences the trend of total satellite abundance with different halo formation stages, there still seems to be an intrinsic correlation. Furthermore, we find a secondary, less pronounced correlation with the slopes of the satellite stellar mass functions, which we do not show in Figure~\ref{fig:haloAssembly}. Stellar mass function slopes exhibit the reversed development: while there is no discernible trend with early assembly $z_\rmn{10}$, intermediate and late assembly $z_\rmn{50}$ and $z_\rmn{90}$ exhibit distinct correlations.

\section{Dependence of subhalo abundance on host halo properties}
\label{sec:subMF_haloProps}

\begin{figure*}
    \centering
    \includegraphics[width=.75\textwidth]{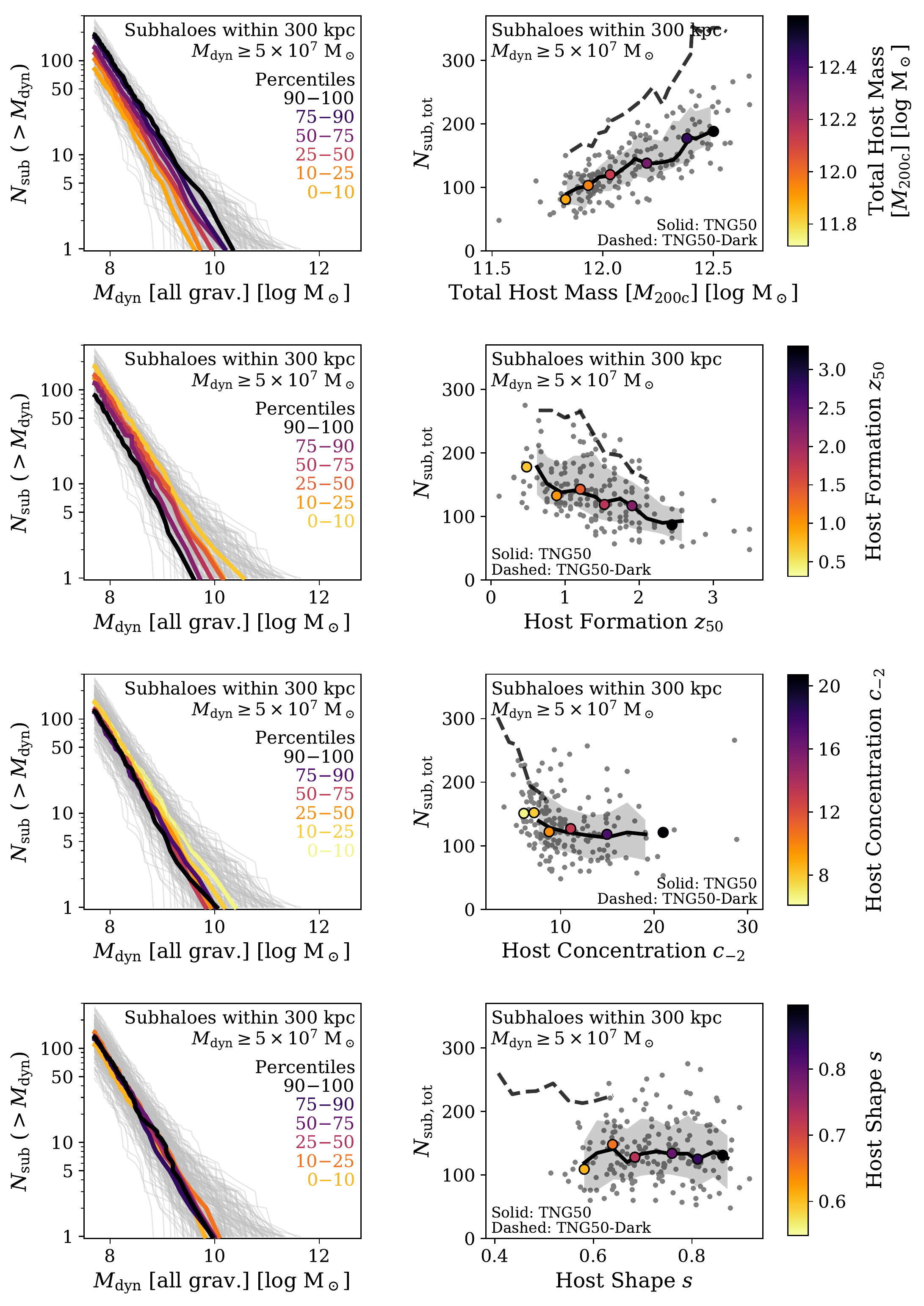}
    \caption{{\bf Dependence of {\it subhalo abundance} (instead of satellite abundance as in Figure~\ref{fig:satMF_haloProps}) on host halo properties} for subhaloes within $300~\rmn{kpc}$ (3D) of their MW/M31-like host and with a dynamical mass of at least $5 \times 10^7~\rmn{M}_\odot$. Each row investigates a different host property (from top to bottom): total mass $M_\rmn{200c}$, halo assembly time $z_{50}$, i.e. the redshift at which the MW/M31-like host had assembled 50 per cent of its mass, halo concentration $c_{-2}$, as well as halo shape $s$ as minor-to-major axis ratio. \textit{Left panels:} median subhalo dynamical mass functions in various percentiles of the host property in question (thick, yellow to black curves). The thin, grey curves in the background denote subhalo dynamical mass functions of individual TNG50 MW/M31-like hosts as a reference. \textit{Right panels:} total number of subhaloes as a function host properties for the percentiles (yellow to black circles), all TNG50 MW/M31-like galaxies (grey circles), as well as their running median (solid, black curves) and scatter (grey shaded area, $16^\rmn{th}$ and $84^\rmn{th}$ percentiles). Dashed, black curves denote the corresponding medians from the DM-only analogue simulation TNG50-Dark. Baryonic processes reduce the strength of the correlations between subhalo number and host properties -- particularly with host halo concentration.}
    \label{fig:haloProps_subhaloes}
\end{figure*}

To connect to earlier results based on DM-only calculations \citep[e.g.][and references therein]{Boylan-Kolchin2010}, we examine correlations of host halo properties and subhalo populations in Figure~\ref{fig:haloProps_subhaloes}, as opposed to satellite galaxy populations in Section~\ref{sec:hostProps} and Figure~\ref{fig:satMF_haloProps}. Trends with subhalo abundance are illustrated using percentile stellar mass functions of the halo property in question (left panels, yellow to black curves), as well as the total number of subhaloes as a function of host halo properties (right panels). Each row of panels presents the dependence on another halo property (from top to bottom): total halo mass $M_\rmn{200c}$, halo assembly time $z_{50}$, i.e. the redshift at which the host halo had assembled 50~per~cent of its present-day total mass, halo concentration $c_{-2}$, and halo shape as its minor-to-major axis ratio $s$. Overall, we find the same trends for subhalo abundances as for satellite galaxies -- albeit more pronounced than in Figure~\ref{fig:satMF_haloProps}. More massive host haloes and those that formed 50 per cent of their present-day total mass later in time tend to have a larger number of surviving subhaloes at $z=0$. There are only slight trends with host halo concentration: while less concentrated MW/M31-like hosts have somewhat more subhaloes, this correlation decreases and flattens towards higher concentrations. Furthermore, there are no significant trends with the host halo's shape. As in Section~\ref{sec:hostProps}, we did check for trends with the slope of the subhalo mass function, however, we recover no significant correlations.

Finally, we compare the correlations in the right panels to their analogues from the DM-only run TNG50-Dark (dashed, black curves). Consistent with our findings in Section~\ref{sec:bary_vs_dmo} and Figure~\ref{fig:satMF_acc}, we find overall larger subhalo abundances around MW/M31-like hosts in TNG50-Dark than in TNG50. We find the same qualitative trends in terms of total host halo mass and host formation time~$z_{50}$ -- namely, more massive host haloes and those that formed later in time have a higher total number of subhaloes. In this respect, TNG50-Dark hosts cover a similar range in host properties as their baryonic counterparts. However, we find significant differences with host concentration~$c_{-2}$ and shape~$s$. Hosts in TNG50-Dark are generally less concentrated and have smaller minor-to-major axis ratios: this is qualitatively consistent with the effects of baryons due to galaxy formation processes in Illustris \citep{Chua2017, Chua2019}. While subhalo abundances exhibit a clear correlation with host concentration -- more subhaloes tend to reside in less concentrated host haloes -- there is still no significant trend with host shape. As in Illustris \citep{Chua2017}, TNG host concentration correlates more strongly with subhalo abundance in DM-only than in baryonic simulations of MW/M31-like hosts.

\bsp	
\label{lastpage}
\end{document}